\documentclass[twocolumn,tighten,astrosymb]{aastex631}

\usepackage{amsmath}
\usepackage{enumerate}
\usepackage{multirow}
\usepackage{CJKutf8}

\newcommand{\PypeIt}{\texttt{PypeIt}}
\newcommand{\thePayne}{\texttt{the Payne}}

\newcommand{\atlas}{\texttt{ATLAS12}}
\newcommand{\synthe}{\texttt{SYNTHE}}
\newcommand{\flambdatext}{$f_\lambda(\theta_*)$}
\newcommand{\flambda}{f_\lambda(\theta_*)}
\newcommand{\tefftext}{$T_\text{eff}$}
\newcommand{\teff}{T_\text{eff}}
\newcommand{\loggtext}{$\log g$}
\newcommand{\logg}{\log g}
\newcommand{\vmicrotext}{$v_\text{micro}$}
\newcommand{\vmicro}{v_\text{micro}}
\newcommand{\vmacrotext}{$v_\text{macro}$}
\newcommand{\vmacro}{v_\text{macro}}
\newcommand{\wave}[1]{$\lambda$#1}

\newcommand{\perpix}{pixel$^{-1}$}

\submitjournal{ApJS}

\shorttitle{Multi-Resolution Spectroscopy}
\shortauthors{Sandford et al.}
\graphicspath{{./}{figures/}}

\begin{document}
\begin{CJK*}{UTF8}{gbsn}
\title{Validating Stellar Abundance Measurements from Multi-Resolution Spectroscopy}

\correspondingauthor{Nathan Sandford}
\email{nathan\_sandford@berkeley.edu}

\author[0000-0002-7393-3595]{Nathan R. Sandford}
\affiliation{Department of Astronomy, University of California Berkeley, Berkeley, CA 94720, USA}

\author[0000-0002-6442-6030]{Daniel R. Weisz}
\affiliation{Department of Astronomy, University of California Berkeley, Berkeley, CA 94720, USA}

\author[0000-0001-5082-9536]{Yuan-Sen Ting (丁源森)}
\affiliation{Research School of Astronomy and Astrophysics, Australia National University, Cotter Road, ACT 2611, Canberra, Australia}

\begin{abstract}  
Large-scale surveys will provide spectroscopy for $\sim$50 million resolved stars in the Milky Way and Local Group. However, these data will have a high degree of heterogeneity and most will be low-resolution ($R<10000$), posing challenges to measuring consistent and reliable stellar labels. Here, we introduce a framework for identifying and remedying these issues. By simultaneously fitting the full spectrum and \textit{Gaia} photometry with \thePayne\, we measure $\sim$40 abundances for 8 red giants in M15. From degraded quality Keck/HIRES spectra, we evaluate trends with resolution and S/N and find that 
(i) $\sim$20 abundances are recovered consistently within $\lesssim$0.1 dex agreement and with $\lesssim$0.05--0.15~dex systematic uncertainties from $10000\lesssim R\lesssim80000$; 
(ii) for 9 elements (C, Mg, Ca, Sc, Ti, Fe, Ni, Y, Nd), this systematic precision and accuracy extends down to $R\sim2500$; and 
(iii) while most elements do not exhibit strong S/N-dependent systematics, there are non-negligible biases for 4 elements (C, Mg, Ca, and Dy) below $\text{S/N}\sim10$~\perpix.
We compare statistical uncertainties from MCMC sampling to the easier-to-compute Cram\'er-Rao bounds and find that they agree for $\sim$75\% of elements, indicating the latter to be a reliable and faster way to estimate uncertainties.
Our analysis illustrates the great promise of low-resolution spectroscopy for stellar chemical abundance work, and ongoing improvements to stellar models (e.g., 3D-NLTE physics) will only further extend its viability to more elements and to higher precision and accuracy.
\end{abstract}


\section{Introduction}
\label{sec:introduction}
\end{CJK*}
Astronomy is in the midst of a multi-decade golden era of stellar spectroscopy. 
Large spectroscopic surveys (e.g., APOGEE; \citealt{majewski:2017}, GALAH; \citealt{desilva:2015}, LAMOST; \citealt{cui:2012}, Gaia-ESO; \citealt{gilmore:2012}, Gaia-RVS; \citealt{recio-blanco:2022}, DESI; \citealt{cooper:2022}), are mapping the detailed chemical abundance patterns of millions of stars across the Milky Way (MW), and in doing so have ushered in a renaissance of chemodynamical studies seeking to piece together the complex formation history of the MW and its satellite system.
Meanwhile, deep observations with $6+$ meter telescopes have pushed the limits of resolved star spectroscopy beyond the MW and have begun unveiling the chemical evolution of other Local Group (LG) galaxies \citep[e.g.,][]{kirby:2018, escala:2019, gilbert:2019}, including those that are relics from the early universe \citep[e.g.,][and references therein]{tolstoy:2009, simon:2019a}.

\begin{figure}[ht!]
    \epsscale{1.25}
    \plotone{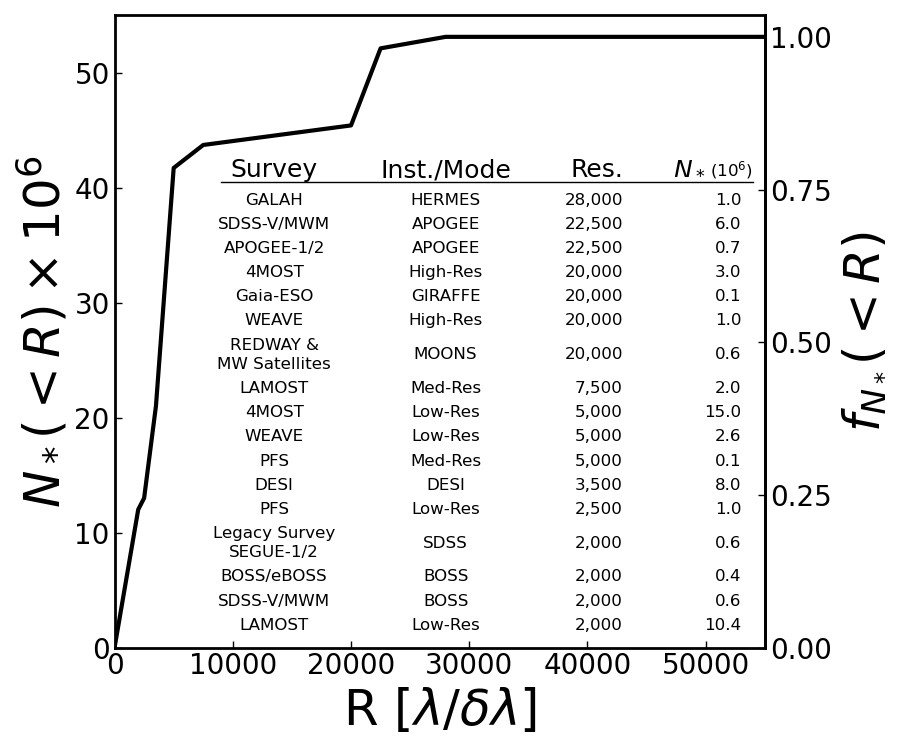}
    \caption{
    Forecasted number of stars observed by large spectroscopic surveys by $\sim$2030 as a function of spectral resolving power. Surveys with very limited wavelength coverage suitable (e.g., RAVE, Gaia-RVS, H3) are excluded. Surveys with fewer than $10^5$ stars 
    are also excluded as they contribute to the figure imperceptibly. Survey overlap is not considered. In 2030, $\sim$75\% of $>$50 million observed stellar spectra in the MW and LG will be taken at $R<10000$. \label{fig:N_vs_R}}
\end{figure}

Over the course of the coming decade, the next iteration of ambitious stellar spectroscopic surveys (e.g., WEAVE; \citealt{dalton:2016}, SDSS-V; \citealt{kollmeier:2017}, PFS; \citealt{tamura:2018}, MOONS; \citealt{taylor:2018}, 4MOST; \citealt{dejong:2019},  FOBOS; \citealt{bundy:2019}) will deliver an order-of-magnitude gain in the number of stars for which detailed chemical abundance patterns can be measured. By $\sim$2030, stellar spectra will be acquired for roughly 50 million resolved stars throughout the MW and LG (Figure \ref{fig:N_vs_R}). 
Spectrographs on next-generation large-aperture space- and ground-based telescopes (e.g., JWST; \citealt{gardner:2006}, GMT; \citealt{fanson:2020}, TMT; \citealt{skidmore:2015}, E-ELT \citealt{gilmozzi:2007}) will further supplement these surveys; their unparalleled sensitivity and light-collecting power enabling spectroscopic observations out to several Mpc, far beyond the capabilities of existing ground-based facilities \citep{sandford:2020b}.

However, the vast increase in data volume and availability made possible by these past, present, and future observations also pose newfound technical challenges. The combination of these large and numerous spectroscopic datasets will feature a high degree of heterogeneity across wavelength regime, signal/noise (S/N), and spectral resolving power ($R \equiv \lambda/\delta\lambda$), all of which can introduce complications in deriving consistent and reliable stellar chemical abundance measurements \citep[][and references therein]{jofre:2019}.

As can be seen in Figure \ref{fig:N_vs_R}, the majority (75\%) of the resolved star spectra acquired in the next decade will be obtained at ``low-resolution" ($R < 10000$), where lower dispersion, higher throughput, and improved multiplexing provide both better observational efficiency and access to fainter and more distant stars. 
For these same reasons, the relative prolificity of low-resolution stellar spectroscopy becomes more pronounced with increasing distance---very few stars beyond a few hundred kpc will have high-resolution spectroscopy of modest or higher S/N ($\gtrsim$40 \perpix) available. 
The trade-off is that low-resolution stellar spectroscopy suffers from severe blending of absorption features, which necessitates full spectral modeling and robust synthetic stellar spectra to precisely and accurately measure detailed chemical abundance patterns.

While the combination of low-resolution spectroscopy and full spectral fitting has lead to enormous scientific gains \citep[e.g.,][]{kirby:2009, kirby:2010, ting:2017b, ting:2018a, kirby:2018, xiang:2019, wang:2022}, a variety of questions remain about the fidelity of abundance recovery in the low-resolution regime given their heavy reliance on synthetic stellar models. Namely, a major concern is that most spectral models used for full-spectrum fitting do not or do not fully capture the 3D and non-local thermodynamic equilibrium (NLTE) effects of the stellar atmosphere on line formation. Similarly, despite ongoing and sustained efforts \citep[e.g.,][to just name a few contributions]{lawler:2013, ryabchikova:2015, denhartog:2019, smith:2021}, there are many atomic and molecular transitions that are missing or imperfectly calibrated in the linelists employed by these spectral models. 

For high-resolution observations, imperfections in the spectral model can be sidestepped by simply ignoring problematic features. But for low-resolution observations, poorly modeled spectral features become blended and inseparable from neighboring features and may introduce systematic biases and uncertainties into the measured chemical abundances if they are not handled carefully \citep{nissen:2018}. 
Given the ongoing proliferation of low-resolution stellar spectroscopy and the crucial role that low-resolution observations will play in extragalactic chemical abundance measurements, quantifying (and addressing) the systematics incurred as a function of resolution will be of the utmost importance. Without a firm grasp of these systematics, it will be difficult to draw firm conclusions across the disparate datasets, especially between the high-resolution studies that define our understanding of the MW and the low-resolution studies that provide our only window into galaxies beyond 1 Mpc.

It is relatively common practice in low-resolution stellar chemical abundance studies to correct for systematic biases, quantify systematic uncertainties, or otherwise validate the fidelity of low-resolution measurements by comparing these measurements with high-resolution literature measurements for a subset of stars \citep[e.g.,][]{kirby:2010}. 
%
%
In may cases, however, these cross-validations are themselves quite heterogeneous, featuring measurements made with both full-spectrum fitting techniques and classical equivalent width (EW) fitting techniques, which frequently employ a great diversity of model atmospheres, spectral synthesis codes, and line lists \citep[e.g., see Table 9 of][]{kirby:2010}.
While many studies \citep[e.g.,][]{bedell:2014, hinkel:2016, jofre:2017, blanco-cuaresma:2019, arentsen:2022} have attempted to quantify methodological, instrumental, or model-oriented systematics, we are aware of no studies to date, which perform a comparison of abundance measurements as a function of resolving power using solely full-spectrum fitting techniques.

It is worth taking a moment to mention that for some scientific purposes, namely kinematic studies, high-resolution low-S/N ($\gtrsim$5 \perpix) spectra is sufficient. In these instances, multi-element abundance measurements are not attempted as historically, only high-resolution spectra with moderate to high S/N ($\gtrsim 40$ \perpix) has been deemed necessary \citep{jofre:2019}. In large part, this is because EWs are challenging to measure precisely in noisy spectra and can lead to biased results \citep[e.g.,][]{smiljanic:2014, heiter:2014}. Consequently, high-resolution spectroscopy, even with large 10-m telescopes like Keck, has been limited to relatively bright stars ($r<19.5$), excluding all but the brightest RGB stars in nearby dwarf galaxies \citep{simon:2019a}. 
Full spectrum fitting techniques, however, are predicted to better leverage the information content of low S/N spectra---even if a single noisy absorption line is only weakly informative, the ensemble of all spectral features should still provide strong constraints on the chemical abundances of a star \citep{ting:2017a, sandford:2020b}. While applications of full spectrum fitting to high-resolution stellar spectroscopy are becoming more common place, most are concerned with bright MW stars for which acquiring high S/N spectra is relatively easy. The utility of low-S/N high-resolution spectra for chemical abundance measurements, especially for extragalactic metal-poor stars, has yet to be thoroughly demonstrated.

In this paper, we quantify the systematic biases and uncertainties in stellar chemical abundance measurements as a function of resolution and S/N by applying self-consistent full-spectrum fitting techniques to initially exquisite Keck/HIRES spectra ($R > 50000$, $\text{S/N} > 100$ \perpix) that we have artificially degraded to lower resolution and S/N ($R\sim2500$; $\text{S/N}\sim5$ \perpix). 
By fitting real observations from a single instrument, as opposed to mock spectra or observations from multiple spectrographs, we capture the impact of model inaccuracies on stellar label recovery when propagated to lower resolutions,
while reducing complicating factors associated with different instruments, reduction pipelines, observing conditions, and stellar models.
Our sample consists of 8 red giant branch (RGB) stars in MW globular cluster M15 with a rich history of previous study on which we validate our measurements. 

This paper is structured as follows. In \S\ref{sec:observations}, we describe the archival data and their degradation to lower resolution and S/N. We present our full-spectrum fitting techniques in \S\ref{sec:methods}. In \S\ref{sec:results}, we present our results
as a function of resolution and as a function of S/N. We discuss our primary findings in \S\ref{sec:discussion}, and present our conclusions in \S\ref{sec:conclusion}.

\section{Observations}
\label{sec:observations}

\subsection{Archival Data}
\label{sec:archival_data}
We use publicly available archival spectra from the Keck Observatory Archive\footnote{\url{https://koa.ipac.caltech.edu/}} taken with the HIRES instrument on the Keck {\sc I} Telescope \citep{vogt:1994}. In total, we analyze 40 individual spectra of 8 RGB stars in the M15 globular cluster. Observations span the wavelength range 3160--8370 \AA\ and provide nominal resolving powers ($R=\lambda/\delta\lambda$) from $37500$ to $86600$. 
In addition to archival Keck/HIRES spectroscopy, we also employ Gaia DR3 photometry \citep{gaia_dr3:2022} to better constrain stellar fundamental parameters (e.g., \tefftext, \loggtext). We apply extinction corrections to this photometry using the \citet{schlafly:2011} dust map, the Gaia extinction coefficients from \citet{gaia_dr2:2018}, and adopting $R_V=3.1$.

\begin{figure}[ht!]
    \epsscale{1.15}
    \plotone{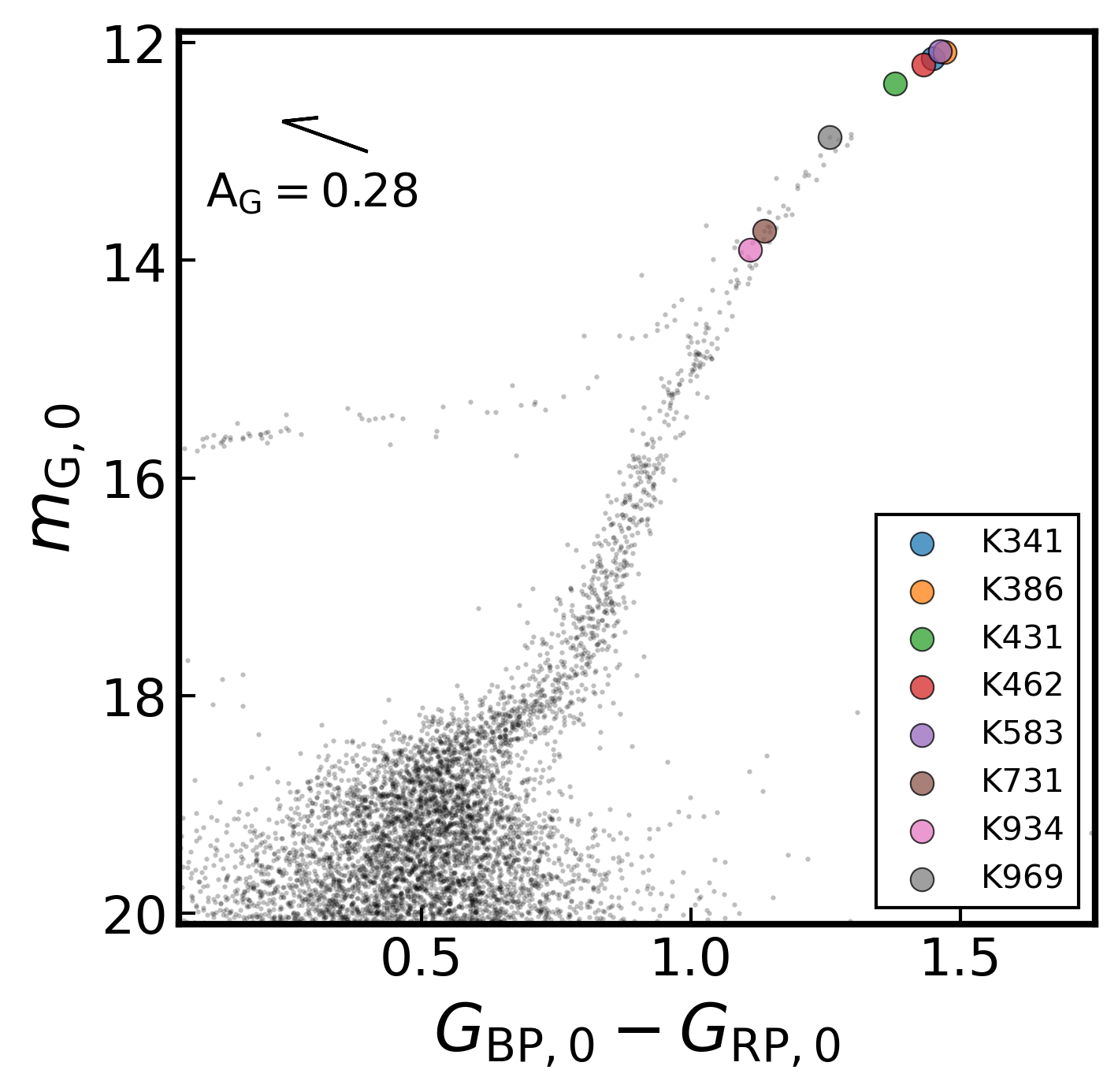}
    \caption{
        Gaia DR3 color-magnitude diagram of likely M15 members as identified by \citet{vasiliev:2021}. Stars analyzed in this work are represented by filled circles, which are all located on the upper part of the RGB. The median extinction correction applied to the sample is denoted by the arrow in the upper left-hand corner of the figure. \label{fig:m15_cmd}
    }
\end{figure}

Table \ref{tab:stars} provides a list of the stars analyzed in this work, and Table \ref{tab:observations} provides a summary of the spectroscopic observations. 
Figure \ref{fig:m15_cmd} shows the location of these stars on the Gaia DR3 color-magnitude diagram of probable M15 members as determined by \citet{vasiliev:2021}. All of the stars considered in this study reside on the upper portion of the RGB.

\begin{deluxetable*}{ccccc}
	\centerwidetable
	\caption{M15 Stars Analyzed in this Work}
	\label{tab:stars}
	\tablehead{
	    \colhead{Kustner ID}        &
	    \colhead{2MASS ID}          & 
	    \colhead{Other IDs}         &
	    \colhead{$m_\text{G,0}$}                 &
	    \colhead{$\text{G}_\text{BP,0} - \text{G}_\text{RP,0}$}
	    \vspace{-2mm}\\
	    \colhead{\citep{kustner:1921}}      &
	    \colhead{\citep{skrutskie:2006}}    &
	    \colhead{}                          &
	    \colhead{}                          &
	    \colhead{}                          
	}
	\startdata
	    K341 & J21295492+1213225 & CBG 4099 & 12.39 & 1.59\\
	    K386 & J21295562+1210455 & CBG 40825 & 12.32 &  1.62\\
	    K431 & J21295618+1212337 & S1 & 12.62 & 1.53 \\
	    K462 & J21295666+1209463 &  \nodata & 12.45 & 1.58\\
	    K583 & J21295856+1209214 &  \nodata & 12.32 & 1.61\\
	    K731 & J21300053+1211369 & ARP I-63, CBG 45062  & 13.99 & 1.29\\
	    K934 & J21300480+1211469 & ARP I-62  & 14.17 & 1.26\\
	    K969 & J21300637+1206592 & S8  & 13.11 & 1.40\\
	\enddata
	\tablecomments{
	    For brevity, we will refer to stars throughout this work using their Kustner IDs. Alternative identifiers are as follows: ARP = \citet{arp:1955}, CBG = \citet{carretta:2009b}, and S = \citet{sandage:1970}. G-band magnitudes and BP-RP colors are from Gaia DR3 and corrected for extinction \citep{gaia_dr3:2022}.
	}
\end{deluxetable*}

\begin{deluxetable*}{ccccccc}
	\centerwidetable
	\caption{HIRES Observations of M15 Stars}
	\label{tab:observations}
	\tablehead{
	    \colhead{Kustner ID}                    & \colhead{Wavelength} &
	    \colhead{Resolution}                    &
	    \colhead{Date}                          & \colhead{Program ID} &
	    \colhead{Program PI}                    & \colhead{Exposures}  \vspace{-2mm}\\
	    \colhead{\citep{kustner:1921}} & \colhead{Range (\AA)} & 
	    \colhead{($\lambda/\delta\lambda$)}     &  
	    \colhead{(DD-MM-YYYY)}                  &  &  & 
	}
	\startdata
	    K341 & 3650--5200 & 45000 & 03-09-1997 & U09H & R.\ Kraft & $5\times1800$s \\
	    K386 & 3650--5200 & 45000 & 04-09-1997 & U09H & R.\ Kraft & $7\times1800$s \\
	    K431 & 3840--8370 & 86600 & 09-09-2011 & C316Hr & E.\ Kirby & $4\times1770$s \\
	    K431 & 3840--8370 & 86600 & 17-09-2011 & C316Hr & E.\ Kirby & $1500$s \\
	    K462 & 3650--5200 & 45000 & 03-09-1997 & U09H & R.\ Kraft & $8\times1800$s \\
	    K583 & 3650--5200 & 45000 & 04-09-1997 & U09H & R.\ Kraft & $6\times1800$s \\
	    K731 & 3840--8370 & 37500 & 10-06-2008 & C147Hr & J.\ Cohen & $1000$s \\
	    K731 & 3840--8370 & 37500 & 11-06-2008 & C147Hr & J.\ Cohen & $1000$s \\
	    K934 & 3840--8370 & 37500 & 11-06-2008 & C147Hr & J.\ Cohen & $800$s \\
	    K969 & 3840--8370 & 86600 & 09-09-2011 & C316Hr & E.\ Kirby & $3\times1725$s,  $3\times1475$s
	\enddata
	\tablecomments{
	    Summary of archival observations analyzed in this work. All raw data are available on the Keck Observatory Archive. Several archival HIRES observations of M15 stars are omitted from this study because they lack suitable flat-field exposures for \texttt{PypeIt} reductions and/or lack Gaia photometry.
	}
\end{deluxetable*}

\subsection{Data Reduction}
\label{sec:reduction}
All archival data were reduced using version 1.3.1 of the \PypeIt\ data reduction pipeline \citep{pypeit:joss}\footnote{
    \url{https://pypeit.readthedocs.io}
}.
At the time of reduction, \PypeIt\ did not support Keck/HIRES data, so a few minor alterations to the reduction code were necessary, which we summarize below.

Echelle orders were manually identified for each observational setup by matching preliminary wavelength solutions to the HIRES Echelle Format Simulator\footnote{\url{https://www2.keck.hawaii.edu/realpublic/inst/hires/HIRES-efs-master/efs.html}}. Spectral orders were discarded if $\gtrsim$50\% of their extent fell off or between detectors---no attempt was made to stitch together orders that spanned multiple detectors. As a result, 
order 67 (5280--5370 \AA) was discarded from the C147Hr and C316Hr programs.

Wavelength calibrations were performed using \PypeIt's ``reidentify" method, in which the observed arc spectra are cross-correlated against archival arc spectra. Appropriate archival spectra for each setup were adopted from the \texttt{MAKEE} data reduction package\footnote{\url{https://sites.astro.caltech.edu/~tb/makee/}}.

Default \PypeIt\ methods and algorithms were employed for bias subtraction, flat-fielding, flexure correction, cosmic ray rejection, sky subtraction, and object extraction. After extraction, the stellar spectra were velocity corrected into the Heliocentric reference frame using the default \texttt{astropy}\footnote{https://www.astropy.org/} Solar System ephemeris. To minimize information loss, repeat observations of the same star are not stacked, but fit individually. A ``stacked" measurement is obtained by combining the posteriors of fits to individual exposure using a hierarchical model (see \S\ref{sec:methods_stack}).

We do not formally flux calibrate the 1D extracted spectra but rather fit for the star's pseudo-continuum simultaneously with its atmospheric parameters and elemental abundances (see \S\ref{sec:continuum}). As a part of the pseudo-continuum fitting, we define a scaled blaze function for each order, which we extract from the combined flat-field calibration frame and scale to the flux of each observed spectral order.

\begin{figure*}[ht!]
    \epsscale{1.15}
    \plotone{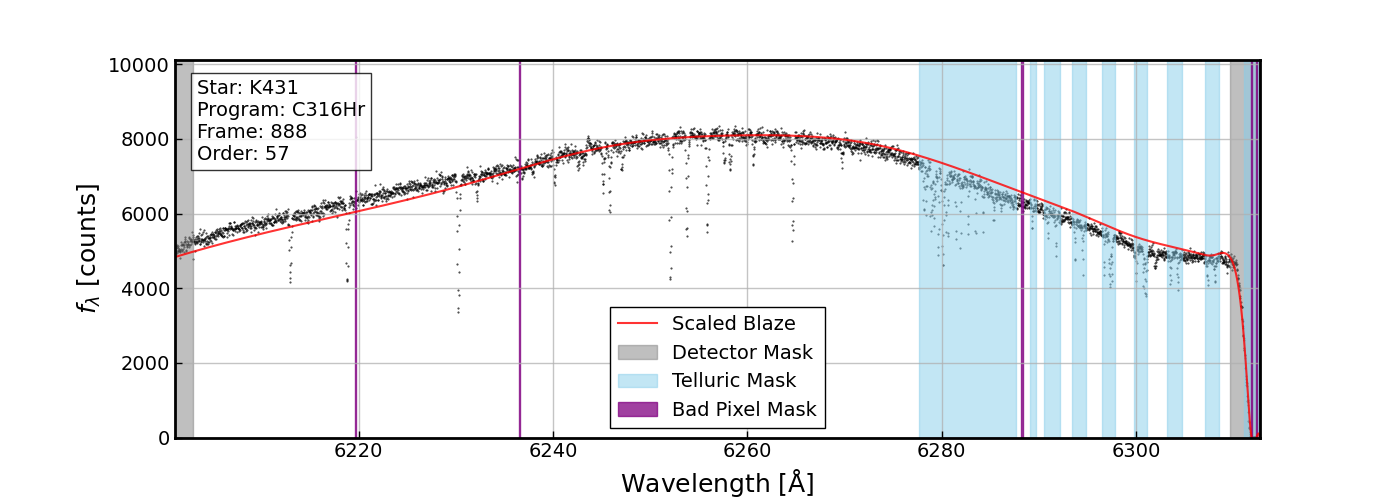}
    \caption{A sample order from one reduced archival observations (black points) illustrating the types of masks we apply to the data. The solid red line represents the scaled blaze function, which we use for the zeroth-order continuum determination. Deviations from the observed continuum are accounted for using a polynomial as described in \S\ref{sec:continuum}. The gray, blue, and purple shaded regions represent the detector boundary mask, the telluric mask, and the bad pixel mask respectively. Pixels that lie within these observational masks are ignored in the spectral fitting analysis. \label{fig:obs_mask}}
\end{figure*}

In Figure \ref{fig:obs_mask}, we present a sample order from one of the reduced archival observations. The scaled blaze function for the order is over-plotted in red, and the adopted observational masks (described in \S\ref{sec:obs_masking}) are included as vertical shaded bands. A complete library of the reduced spectra analyzed in this work can be made available upon request.

\subsection{Observational Masks}
\label{sec:obs_masking}

In Figure \ref{fig:obs_mask}, we illustrate the three types of observational masks adopted to flag pixels with large observational artifacts or uncertainties and exclude them in our spectral fitting analysis. The telluric absorption mask (blue shaded regions), includes all pixels that contain strong telluric contamination, as identified in the ``List of Telluric Lines" provided by \texttt{MAKEE}\footnote{
    \url{https://www2.keck.hawaii.edu/inst/common/makeewww/Atmosphere/atmabs.txt}
}. The detector boundary mask (gray shaded regions) includes the first 64 and last 128 pixels of every order in the 
C147Hr and C316Hr programs, which exhibit strongly non-linear response functions that bias polynomial fits to the spectral continuum\footnote{The Older U09H program observations do not exhibit strong non-linear effects near the detector boundaries, so no detector boundary mask is necessary.}. Lastly, the bad pixel mask (purple shaded regions) includes all hot pixels, improperly subtracted sky lines, and cosmic rays as identified automatically with \PypeIt\ or by visual inspection.

\subsection{Post-Processing Observations}
\label{sec:methods_obs}
A primary goal of this paper is to self-consistently test the robustness of stellar spectroscopic label recovery as a function of spectral resolving power and S/N using real (as opposed to mock) data. 
Specifically, we consider stellar label recovery along two axes: i) as a function of resolution at fixed integration time and ii) as a function of S/N at fixed resolution.
In order to satisfy these conditions using archival data from only one spectrograph, we apply several post-processing operations to the data (e.g., to degrade resolution or S/N), which we now describe.

\subsubsection{Varying Resolution at Fixed Integration Time}
\label{sec:methods_fixed_t}
\begin{figure*}[ht!]
    \epsscale{1.15}
    \plotone{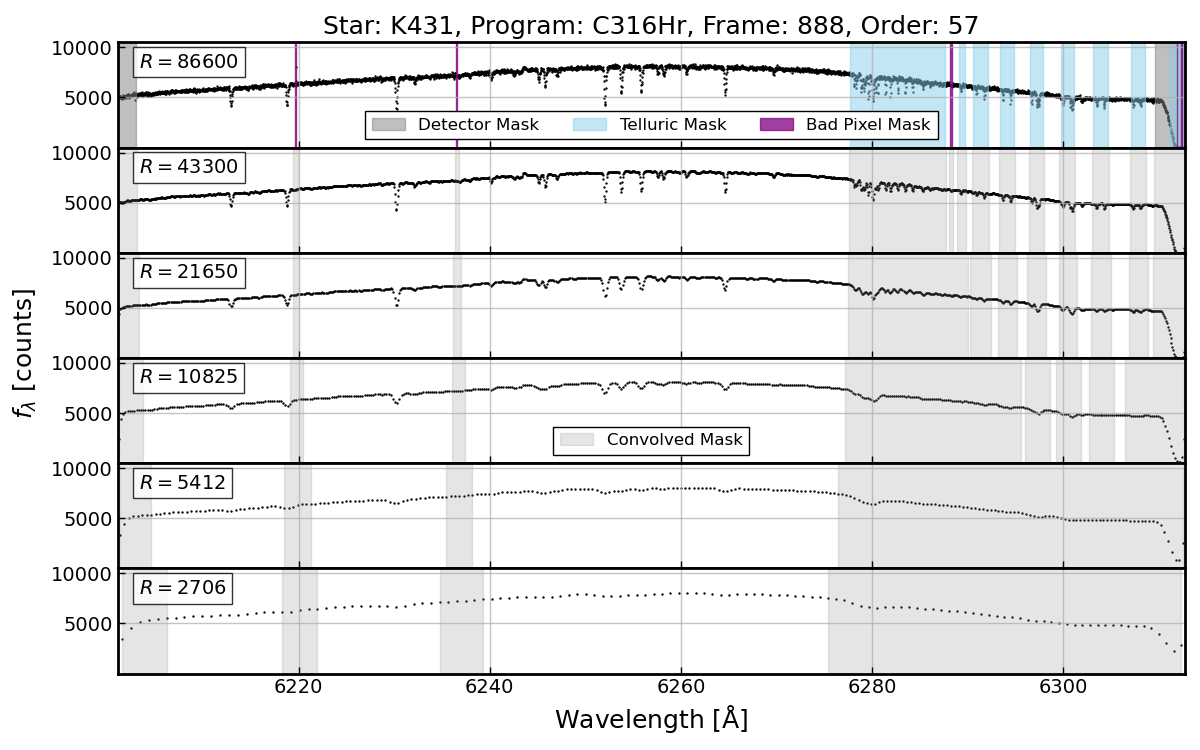}
    \caption{An illustration of the effects of varying spectral resolution on the observational masks using the same sample order and observational masks from Figure \ref{fig:obs_mask} (top). Lower panels depict the observed order convolved to lower resolutions by successive factors of 2. As the spectral resolving power decreases, the observational masks (light grey bands) grow to include pixels impacted by the broadening of masked features. The spectrum is also re-binned as it is convolved to lower resolution to maintain a constant $N_\text{pix}/\text{FWHM}$. The S/N of the spectrum scales with $R^{-1/2}$ as a result of this re-binning. \label{fig:convolve}}
\end{figure*}

Because the archival spectra are all taken at high resolution, testing stellar label recovery at lower resolution requires that we artificially degrade the resolving power of the archival spectrum and repeat our analysis at each resolution. We perform this degradation by convolving each archival spectrum to successively halved resolving powers down to $R\sim2500$---a factor of 16-32 lower than the native instrumental resolution. The convolution of a sample order from one reduced archival spectrum is presented in Figure \ref{fig:convolve}.

Here, and throughout this paper, we perform spectral convolutions assuming that the instrumental broadening kernel, $\mathcal{F}_v^\text{inst}$, is well-described by a zero-mean Gaussian with constant width, $\sigma_\text{inst}=1/2.355R$, where $R$ is the spectral resolving power of the instrumental configuration used in the observation. We also assume that $R$ is constant as a function of wavelength though this is not strictly true in practice.
Given an observation's initial resolving power, $R_0$, we achieve the desired resolving power, $R$, by convolving each order of the initial spectrum with a Gaussian kernel of width
\begin{equation}
    \sigma_\text{inst} = \left[\left(2.355R\right)^{-2} - \left(2.355R_0\right)^{-2}\right]^{1/2}.
\end{equation}
We perform these convolutions via  multiplication of the spectrum and the broadening kernel in Fourier-space which increases computational efficiency and better preserves spectral information. An identical convolution is applied to the flux uncertainty of each order. 

However, convolving observational data has several unavoidable consequences that must be handled properly for a self-consistent analysis. First, by convolving the spectra on their native wavelength grid results in spectra that are over-sampled (i.e., $N_\text{pix}/\text{FWHM}\gtrsim3$). For example, a spectrum with $N_\text{pix}/\text{FWHM}\sim3$ at $R=40000$ would have   $N_\text{pix}/\text{FWHM}\sim6$ at $R=20000$ and $N_\text{pix}/\text{FWHM}\sim48$ at $R=2500$, which is unrealistically over-sampled. 
Instead, to more realistically emulate low resolution observations, we down-sample the spectra by a factor of $R_0/R$ to maintain constant $N_\text{samp}\sim3$ pixels/FWHM. This downsampling is performed using the using the \texttt{SpectRes}\footnote{\url{https://spectres.readthedocs.io/en/latest/}} Python package \citep{carnall:2017}. Importantly, \texttt{SpectRes} re-bins the spectra and its uncertainties in a manner that conserves flux, resulting in the S/N of the convolved and down-sampled spectra increasing as the resolution is decreased according to $\text{S/N}\propto R^{-1/2}$.

Second, convolution also complicates the use of the observational masks described in \S\ref{sec:obs_masking}. The convolution kernel not only broadens spectral features, but also sky lines, detector artifacts, and bad pixels, causing them to ``spill out" from the existing masks. Our solution for this is to treat our masks as binary arrays with 0's corresponding to masked pixels and 1's corresponding to unmasked pixels. We then broaden and interpolate these masks in the same manner as the observed spectrum and expand them to include any pixels where the convolved mask is $<$0.99---that is, any region where a masked pixel contributes $>$1\% of its flux. 
For bad pixels with extremely outlying values, this can still lead to substantial contributions to unmasked pixels. To mitigate this, we replace all bad pixels with the mean value of the nearest non-masked pixel prior to convolution. Broadened observational masks are represented in Figure \ref{fig:convolve} by light grey vertical bands.

A third complication is potential edge effects.  To illustrate the issue, consider the pathological example of a strong absorption line with a central wavelength that lies just outside the range of an observed spectral order. At high resolution, the absorption from this line might be completely excluded from the observed order. But at low resolution, the line might be broadened to the point where its wings bleed into the observed order. Convolving the observed spectrum artificially as we do in this study, would completely omit the contribution of this broadened line, introducing additional systematic error into the analysis. For spectra from the 
C147Hr and C316Hr programs, the detector boundary masks are sufficient to exclude any edge effects. For spectra from the U09H program, we implement a one pixel mask at each end of each order and proceed with the mask convolution procedure described above. This will similarly exclude any potential edge effects.

As a result of expanding the observational masks, a greater fraction of the spectrum is masked at lower resolution. For example, in the C316Hr observations $\sim$10\% of the pixels are masked at $R\sim80000$ vs. $\sim$25\% at $R\sim2500$, and in the U09H observations $\sim$1\% of the pixels are masked at $R\sim45000$ vs. $\sim$7\% at $R\sim2500$. While larger contamination from telluric lines is to be expected at lower resolution, it is not typically the case for cosmic rays, hot/dead pixels, and detector edge effects. This is a minor, but necessary, trade-off in our choice to use the same exposures at multiple resolutions.  We believe the value in using real data (as opposed to synthetic spectra) greatly outweighs these minor complications.

\subsubsection{Varying S/N at Fixed Resolution}
\label{sec:methods_fixed_r}

\begin{figure}[ht!]
    \epsscale{1.15}
    \plotone{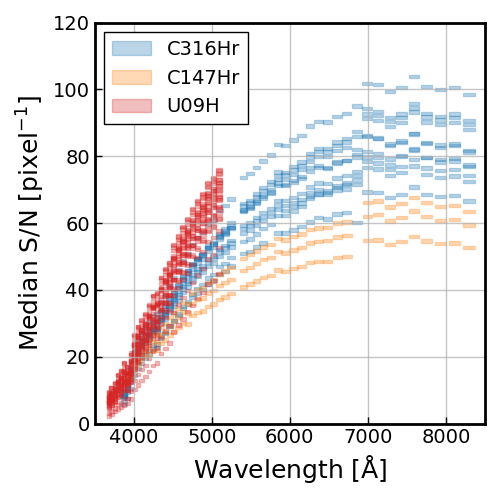}
    \caption{
        Median S/N per pixel of each echelle order in each exposure analyzed in this study before the quality of the data is degraded. The width of the horizontal bars represent the wavelength coverage spanned by the order. The colors denote the observing programs outlined in Table \ref{tab:observations}.
        \label{fig:nat_obs_snr}
    }
\end{figure}

The archival spectra was taken with specific science goals in mind, which translate to minimal S/N requirements (e.g., $\text{S/N}\gtrsim 40$ \perpix\ at \wave{5000}). This is illustrated in Figure \ref{fig:nat_obs_snr}, which presents the median S/N of each individual echelle order analyzed in this study. 

In order to test the robustness of stellar label recovery as a function of S/N, we add artificial white noise to the reduced spectra in order to decrease the median S/N by factors of 2 down to $\text{S/N}\sim5$ \perpix. For this analysis, we consider only spectra convolved to $R\sim10000$ as we expect the results at moderately lower and higher resolutions to be similar.

\begin{figure*}[ht!]
    \epsscale{1.15}
    \plotone{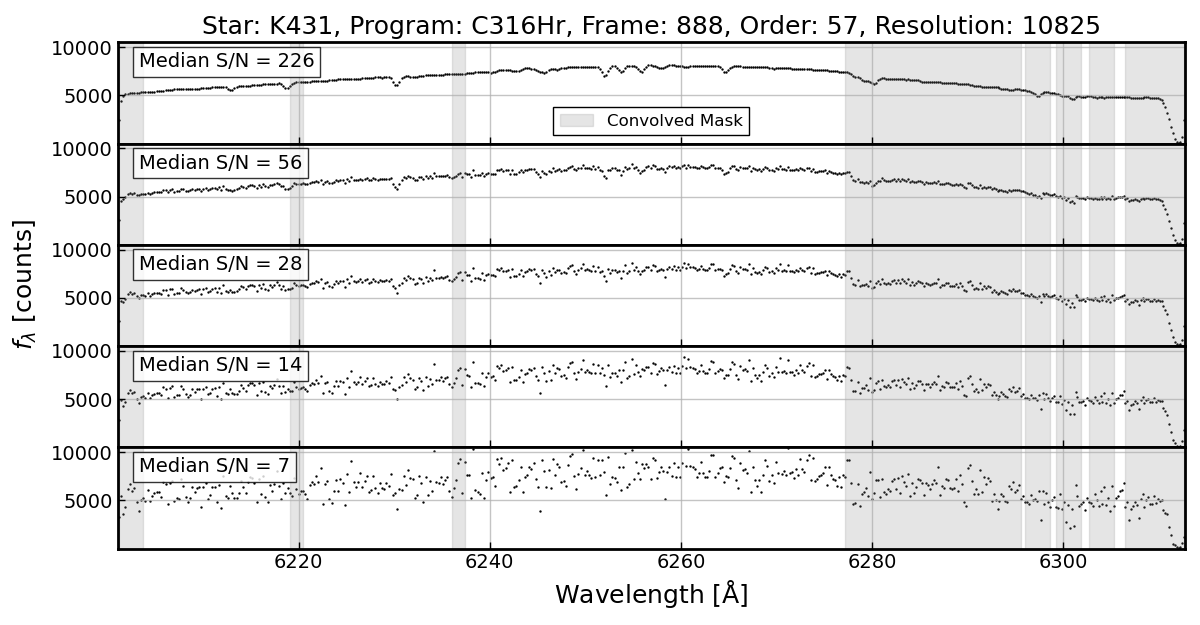}
    \caption{
        The same sample order and observational mask from Figure \ref{fig:obs_mask} convolved to $R\sim10000$ (top). The lower panels depict the observed order noised up by factors of 4, 8, 16, and 32 respectively. While very little information appears to remain at the lowest S/N, this is only a small portion of the full stellar spectrum.
        \label{fig:noise}
    }
\end{figure*}

For a reduced spectrum, $D_0$, with flux errors, $\sigma_{D_0}$, reported from the \PypeIt\ reduction pipeline,  we add Gaussian noise to the spectrum as follows:
\begin{equation}
    D = D_{0} + \mathcal{N}(D_{0}, ~\sigma),
\end{equation}
where $\sigma$ satisfies the condition that the resulting flux uncertainties,
\begin{equation}
    \sigma_D = \sqrt{(\sigma)^2+(\sigma_{D_0})^2},
\end{equation}
yield the desired median S/N,
\begin{equation}
    \text{Med(S/N)} = \text{Med}\left(\frac{D}{\sigma_D}\right).
\end{equation}
Figure \ref{fig:noise} illustrates an example spectral order degraded to lower S/N values. 

\section{Spectral Fitting Analysis}
\label{sec:methods}
In this section, we describe our framework for fitting stellar spectra.
The overarching structure of our analysis (and this section) is as follows. We begin in \S\ref{sec:emulator} by generating a normalized synthetic spectrum for a set of stellar labels using \thePayne, a fast neural-network spectral emulator. Then in \S\ref{sec:forward}, this model spectrum is forward-modelled into the observational domain given additional parameters describing various spectral broadening effects, the star's radial velocity, and the spectrum's continuum. Lastly in \S\ref{sec:evaluation}, the model spectrum is compared directly to the observed spectrum on the pixel-by-pixel level and a posterior probability is calculated. The best-fit stellar (and nuisance) parameters are found by maximizing the posterior using both optimization techniques and Markov chain Monte Carlo (MCMC) sampling.
Throughout this section, we borrow much of our notation from \S2 of \citet{czekala:2015}, which we found to be a clear, illustrative, and mathematically rigorous presentation of forward-modelling stellar spectra. The code used to perform the described spectral analysis is made public in the \texttt{PayneOptuna} Github repository\footnote{https://github.com/NathanSandford/PayneOptuna}.

\subsection{Generating Model Spectra with The Payne}
\label{sec:emulator}
At the core of most full-spectrum fitting techniques is a model that can generate a realistic stellar spectrum, $f_\lambda$, from a set of stellar parameters or labels, $\theta_*$.
Because generating \flambdatext\ on the fly from stellar atmosphere and radiative transfer codes is computationally prohibitive, we employ \texttt{the Payne} \citep{ting:2019}, a powerful tool for spectral emulation that has been successfully used in a number of spectroscopic studies \citep[e.g.,][]{el-badry:2018a, ting:2019, xiang:2019, kovalev:2019, xiang:2022, straumit:2022}. At its core, \thePayne\ is a fully-connected neural network that is trained to efficiently and accurately interpolate a high-dimensional grid of ab initio stellar spectra. Because \thePayne\ is trained on synthetic spectra, it avoids confusing astrophysical correlation between elemental abundances (like bulk $\alpha$-enhancements) with real spectroscopic abundance information \citep[e.g.,][]{ting:2017b, xiang:2019}.

In short, we generate a grid of $\mathcal{O}(10^{4})$ stellar labels, $\theta_* = \{\teff, ~\logg, ~\vmicro, ~\text{[X/H]}\}$, where X includes 36 elements (C, N, O, Na, Mg, Al, Si, K, Ca, Sc, Ti, V, Cr, Mn, Fe, Co, Ni, Cu, Zn, Ga, Sr, Y, Zr, Ba, La, Ce, Pr, Nd, Sm, Eu, Gd, Dy, Ho, Er, Os, and Th). For each $\theta_*$, we compute a continuum-normalized $R=300000$ ab initio spectrum with the 1D LTE stellar atmosphere and radiative transfer codes, \atlas\ and \synthe\ \citep{kurucz:1970, kurucz:1981, kurucz:1993, kurucz:2013, kurucz:2017}.
These spectra are convolved and sub-sampled down to the highest spectral resolution and wavelength sampling present in our archival data ($R=86600$; d$v=1.17$km/s \perpix). \texttt{The Payne} is then trained on this grid of convolved spectra.
A detailed technical description of \thePayne's architecture, training, and accuracy, is provided in Appendix \ref{app:payne}.

\subsubsection{Model Uncertainties}
\label{sec:model_uncertainties}

\begin{figure*}[ht!]
    \epsscale{1.15}
    \plotone{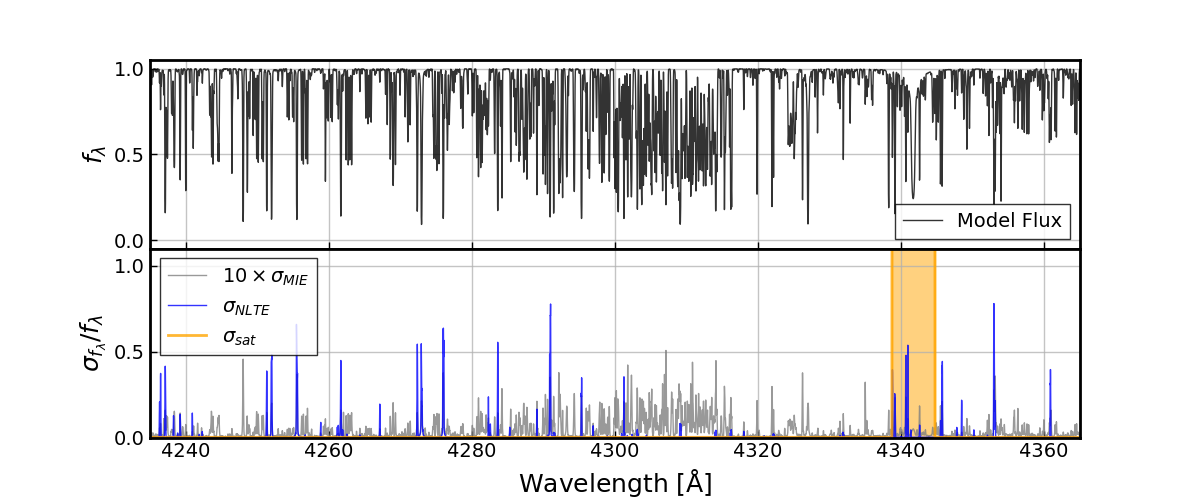}
    \caption{A portion of a synthetic spectrum generated wit \thePayne\ (top) and its fractional flux uncertainty (bottom). The total model uncertainty is the quadrature sum of the three components displayed here: the MIE of \thePayne (gray), NLTE effects (blue), and saturated lines (yellow). For visibility, the MIE has been inflated by a factor of 10 in this figure. The saturated line masked from this portion of the spectrum is the H$\gamma$ line at $\lambda$4341.7.
    \label{fig:model_uncertainties}}
\end{figure*}

In addition to the flux uncertainty of the observations, we also incorporate the flux uncertainty of our models. Specifically, we include three sources of model uncertainty: interpolation errors of \texttt{the Payne}, NLTE effects, and saturated lines. These are illustrated in Figure \ref{fig:model_uncertainties}.

The first source of uncertainty captures how well our spectral model, \flambdatext, can reproduce the ab initio spectra generated directly with \texttt{ATLAS12} and \texttt{SYNTHE}. Even a well-trained model has non-zero interpolation errors, which can vary as a function of wavelength and stellar labels. We adopt the median interpolation error (MIE), $\sigma_\text{MIE}$, as the fundamental flux uncertainty of our model spectra (gray line in Figure \ref{fig:model_uncertainties}).
On the whole, $\sigma_\text{MIE}$ is small---the median value across the entire spectrum is $\sim4\times10^{-4}$. There are portions of the spectrum, however, that exhibit larger interpolation errors---roughly 1\% of the model spectrum has $\sigma_\text{MIE}\gtrsim10^{-2}$. This is predominantly the case for strong lines and complicated molecular features like the CH molecular band at $\lambda$4300 seen in Figure \ref{fig:model_uncertainties}.
For simplicity, we assume that $\sigma_\text{MIE}$ is independent of stellar labels, though we find it to be larger for spectra with $\text{[Fe/H]} > -2$. Fortuately, the stars considered in this study are all found to have $\text{[Fe/H]} \lesssim -2.4$. For more details on the MIE, see App \ref{app:payne_acc}.

The second source of uncertainty is introduced by the 1D LTE assumptions of our model atmosphere and radiative transfer codes. Many stellar absorption lines are known to be sensitive to NLTE effects, which will be poorly modelled by \flambdatext\ \citep[e.g.,][and references therein]{asplund:2005}. Instead of simply masking out NLTE lines as is standard in 1D LTE analyses, we attempt to mitigate the impact of our 1D LTE assumptions by including an additional source of uncertainty, $\sigma_\text{NLTE}$. We define this to be the difference in normalized flux expected from LTE and NLTE treatments:
\begin{equation}
    \sigma_{\text{NLTE}} = |f_{\lambda,\text{LTE}} - f_{\lambda,\text{NLTE}}|
\end{equation}
(blue line in Figure \ref{fig:model_uncertainties}).
To calculate $\sigma_{\text{NLTE}}$, we use the NLTE Abundance Correction tool\footnote{\url{http://nlte.mpia.de/gui-siuAC_secE.php}} developed and maintained by M.\ Koval\"ev, which includes NLTE effects for lines of O, Mg, Si, Ca, Ti, Cr, Mn, Fe, and Co as calculated by \citet{mashonkina:2007, sitnova:2013, bergemann:2008, bergemann:2010a, bergemann:2010b, bergemann:2011, bergemann:2012, bergemann:2013}, and \citet{bergemann:2017}. This is, of course, a far from complete accounting of the NLTE effects present in real spectra, but should nevertheless substantially reduce the impact of the LTE assumptions made throughout this study.

Third and finally, a few strong spectral features, notably the Ca H\&K and the Hydrogen Balmer lines, in our observations are strongly saturated and thus poorly modelled by $f_\lambda(\theta_*)$. We mask these lines out with
\begin{equation}
    \sigma_\text{sat}=
    \begin{cases}
    1, & |\lambda-\lambda_0| < \delta\lambda\\
    0, & \text{otherwise}\\
    \end{cases},
\end{equation}
where $\lambda_0$ is the line center of the saturated feature and $\delta\lambda$ is chosen to generously encompass the width of the line (yellow region in Figure \ref{fig:model_uncertainties}). We provide $\lambda_\text{center}$ and $\delta\lambda$ for these lines in Table \ref{tab:masked_lines}.

\begin{deluxetable}{lcc}
	\centerwidetable
	\caption{Masked Saturated Lines}
	\label{tab:masked_lines}
	\tablehead{
	    \colhead{Line} & \colhead{$\lambda_0$ [\AA]} & \colhead{$\delta\lambda$ [\AA]} 
	}
	\startdata
	    Ca H & 3969.6 & 20 \\
	    Ca K & 3934.8 & 20 \\
	    H$\alpha$ & 6564.6 & 3 \\
	    H$\beta$ & 4862.7 & 3 \\
	    H$\gamma$ & 4341.7 & 3 \\
	    H$\delta$ & 4102.9 & 3 \\
	    H$\epsilon$ & 3971.2 & 3 \\
	    H$\zeta$ & 3890.2 & 3 \\
	    H$\eta$ & 3836.5 & 3 
	\enddata
	\tablecomments{
	    All line centers are given in vacuum wavelengths.
	}
\end{deluxetable}

Under the reasonable assumption that these three sources of uncertainty are largely uncorrelated, the total model uncertainty is then their quadrature sum,
\begin{equation} \label{eq:model_unc}
    \sigma_{f_\lambda} = \sqrt{\sigma_{\text{MIE}}^2 + \sigma_{\text{NLTE}}^2 + \sigma_{\text{sat}}^2}.
\end{equation}

\subsection{Forward Modelling}
\label{sec:forward}
By construction, \thePayne\ emulates the normalized spectra generated by the \atlas\ and \synthe\ models and, as is, omits important observational and instrumental effects. As a result, it is necessary to incorporate these effects via forward modelling of the synthetic spectra before it can be compared directly to real data. This forward modelling is done in three steps, which are described below. In each step, the model flux and the model flux uncertainties are operated on identically.

\subsubsection{Radial Velocity and Broadening Kernels}
\label{sec:rv_broadening}
In the first forward modelling step, we account for observational and instrumental effects that alter the stellar spectrum along its wavelength dimension. We implement broadening from two sources, the instrument's line spread function (LSF) and macroturbulent motion in the star's photosphere. We also Doppler shift the spectrum according to the star's radial velocity. Each of these can be characterized by a kernel that modifies the line-of-sight velocity distribution function of \flambdatext. 

For the instrumental broadening kernel, $\mathcal{F}^\text{inst}_v$, we adopt a zero-mean Gaussian with constant-width parameterized by the instrumental resolving power, $R$, as previously described in \S\ref{sec:methods_fixed_t}. 
For computational efficiency, we also adopt a zero-mean Gaussian for the macroturbulent broadening kernel, $\mathcal{F}^\text{turb}_v$, which we parameterize with the macroturbulent velocity, \vmacrotext\footnote{The ``radial-tangential" model described in \citet[Equation 17.15 of][]{gray:2021} would be more accurate, but adopting a Gaussian kernel for both the instrumental and macroturbulent broadening kernels allows the two broadening steps to be easily combined.}.
Lastly, the Doppler shift is implemented with a delta function kernel, $\mathcal{F}^\text{dop}_v=\delta(v-v_r)$, centered at the star's radial velocity, $v_r$. 

\flambdatext\ and $\sigma_{f_\lambda}$ are modified via a convolution with these kernels in velocity space, i.e., 
\begin{equation}
    f_{\lambda}(\theta_*, \theta_v) = 
    \flambda * \mathcal{F}^\text{dop}_v * \mathcal{F}^\text{inst}_v * \mathcal{F}^\text{turb}_v
\end{equation}
and
\begin{equation}
    \sigma_{f_{\lambda}}(\theta_v) =
    \sigma_{f_{\lambda}} * \mathcal{F}^\text{dop}_v * \mathcal{F}^\text{inst}_v * \mathcal{F}^\text{turb}_v
\end{equation}
respectively, where $\theta_v=\{R,~\vmacro,~v_r\}$ includes the additional model parameters characterizing each kernel.
These convolutions are performed by multiplying the spectra with the kernels in Fourier space.

We note two velocity-related convolutions that are excluded from this post-processing: microturbulent broadening and rotational broadening. Microturbulent broadening is excluded here because it is already incorporated into the model spectra generation as part of $\theta_*$ passed to \synthe. Rotational velocity is excluded because the stars in our sample are most likely slow-rotating low-mass giant stars, whose spectra do not typically exhibit substantial rotational broadening \citep{carlberg:2011}.

In practice, we hold $R$ fixed as we expect it to be very degenerate with measurements of \vmacrotext\ and other stellar parameters, especially at low resolution. Moreover, $R$ is typically a well-known characteristic of the spectroscopic observing configuration. 

\subsubsection{Wavelength Interpolation}
\label{sec:interpolation}
At this point in the post-processing, the convolved and Doppler shifted model spectrum is highly oversampled compared to real observations. It is thus necessary to resample the model flux and its uncertainties onto the discrete wavelengths corresponding to each pixel of each order, $o$, in the observed spectrum, i.e.,
\begin{equation}
    f_{\lambda}(\theta_*, \theta_v) \mapsto M_{o}(\theta_*, \theta_v)
\end{equation}
and
\begin{equation}
    \sigma_{f_{\lambda}}(\theta_*, \theta_v) \mapsto \sigma_{M_{o}}(\theta_*, \theta_v).
\end{equation}
This resampling is performed via linear interpolation of $f_{\lambda}(\theta_*, \theta_v)$ and $\sigma_{f_{\lambda}}$.

\subsubsection{Stellar Continuum and Detector Response}
\label{sec:continuum}
This forward modelling step addresses the fact that the model we have established thus far, $M_{o}(\theta_*, \theta_v)$, generates a normalized stellar spectrum.  However, the shape of the observed spectra is that of the stellar continuum modulated by the instrumental response function.
To incorporate a realistic continuum into the normalized model spectra, we apply a two-part continuum scaling. The first operation captures the spectrograph's response function both within and across spectral orders as well as the star's large-scale spectral energy distribution. To do this, we multiply each order of the model spectrum by that order's blaze function, which we have extracted from the combined flat-field calibration frame and scaled to the observations (see \S\ref{sec:reduction}).

To account for any deviations that remain, we multiply each order of the model spectrum by a low-order $n$th degree polynomial function, $P_o$. This polynomial function can be described by a set of $n+1$ coefficients for each order, $\phi_\text{P}=\left\{c_{o,n}\right\}$. To improve the stability of this correction while fitting, we evaluate each polynomial not as a function of $\lambda$ but of a scaled wavelength
\begin{equation}
    \lambda_o^{\prime} = \frac{2a}{\lambda_{o,\text{max}}-\lambda_{o,\text{min}}}
    \left(\lambda_o - \lambda_{o,\text{mean}}\right),
\end{equation}
where $\lambda_{o,\text{max}}$, $\lambda_{o,\text{min}}$, and $\lambda_{o,\text{mean}}$ are the maximum, minimum, and mean wavelengths of each order respectively, and $-a < \lambda_o^{\prime} < a$.

The resulting continuum-corrected and fully post-processed spectrum can then be written as:
\begin{align}
        M(\Theta) &= \left\{M_o(\theta_*, \theta_v) \times B_o P_o\right\} \\
        &= \left\{M_o(\theta_*, \theta_v) \times B_o\sum_{n=0}^{N_\text{deg}}c_{o,n}(\lambda_o^{\prime})^{n}\right\},\label{eq:final_M}
\end{align}
where $\Theta = \{\theta_*, \theta_v, \phi_\text{P}\}$ represents all physical and nuisance parameters of the model. 

In summary, each model spectrum is described by 39 stellar labels (3 atmospheric parameters and 36 elemental abundances), 3 labels describing spectral broadening and Doppler shift ($R$, \vmacrotext, and $v_r$), and $N_\text{ord}\times(N_\text{deg}+1)$ continuum coefficients, where $N_\text{ord}$ is the number of orders in the spectrum and $N_\text{deg}$ is the degree of the continuum correction polynomial. We find $N_\text{deg}=4$ is suitable for most HIRES observations.

\subsection{Model Evaluation and Spectral Fitting}
\label{sec:evaluation}
With spectral model $M(\Theta)$ now defined (Equation \ref{eq:final_M}), we can infer the physical (and nuisance) parameters, $\Theta$, that best reproduce an observed spectrum, $D$, by maximizing the posterior probability 
\begin{equation}
    \ln P(\Theta|D) = \ln L(D|\Theta) + \ln\Pi(\Theta), \label{eq:posterior}
\end{equation}
where $\ln L(D|\Theta)$ is the log-likelihood of the data given the model parameters and $\ln\Pi(\Theta)$ is the log-prior on the model parameter. For each observed spectrum, we first use an optimization algorithm to recover the maximum \textit{a posteriori} value of $\Theta$. Then we use MCMC to sample directly from $P(\Theta|D)$, validating the results of the optimizer and providing uncertainties and covariances for the recovered labels. A technical description of each method is provided in Appendix \ref{sec:fitting}.

For both optimization and MCMC sampling, we adopt a Gaussian log-likelihood function for $\ln L(D|\Theta)$ in Equation \ref{eq:posterior}:
\begin{equation}
    \ln L(D|M) = -\frac{1}{2}\sum^{N_\text{ord}}\sum^{N_\text{pix}}\left[\ln(2\pi\sigma_\text{tot}^2) + (R/2\sigma_\text{tot})^{2}\right],
\end{equation}
where
\begin{equation}
    R \equiv R(\Theta) \equiv D - M(\Theta)
\end{equation}
is the residual spectrum and 
\begin{equation}
    \sigma_\text{tot} = \sqrt{\sigma_{M}^2 + \sigma_D^2}
\end{equation}
is the combined flux uncertainty of the model and the data. The total log-likelihood is the summation of the individual pixel log-likelihoods over all spectral orders excluding those pixels ignored by the observational masks.

\subsubsection{Fitting \texorpdfstring{\tefftext}{Teff} and \texorpdfstring{\loggtext}{logg}}
\label{photometric_priors}
In practice, \tefftext\ and \loggtext\ are often determined independent of the spectral analysis using photometry.In most cases these photometrically determined values are held fixed or coarsely iterated over during the abundance determination \citep[e.g.,][]{kirby:2010}. This approach is frequently taken for 1D LTE analysis of low-metallicity RGB stars where ``overionization" departures from LTE become increasingly important \citep[e.g.,][and references therein]{asplund:2005}. We find, as have previous studies \citep[e.g.,][]{sneden:2000a, sobeck:2006}, that attempting to fit \tefftext\ and \loggtext\ from spectroscopy alone frequently results in surface gravities that are $>$0.3 dex too small or stars occupying completely unphysical parts of the Kiel diagram.

Here, we recover \tefftext\ and \loggtext\ deterministically and simultaneously with the spectral analysis by interpolating MIST isochrones using the star's extinction-corrected Gaia photometry and [Fe/H] abundance. That is
\begin{align}
   \teff, \logg &= f_\text{Iso}(m_\text{G,0}, G_\text{BP,0}-G_\text{RP,0},~\text{[Fe/H]}),
\end{align}
where $f_\text{Iso}$ is the interpolation function for the MIST isochrone.
To convert from apparent to absolute magnitudes, we adopt a distance modulus to M15 of $\mu_\text{M15}=10.71$ from \citet{baumgardt:2021}.
Because [Fe/H] is itself a free parameter, \tefftext\ and \loggtext\ are updated iteratively with each step of the optimizer and MCMC walker. This is similar to, though less sophisticated than
the techniques employed in the \texttt{MINESWEEPER} spectral fitting code \citep{cargile:2020}.

\subsubsection{Priors}
\label{sec:methods_priors}
With a few exceptions, we adopt the same priors when optimizing and sampling $P(\Theta|D)$. These priors are specified below. The total log-prior is the sum of the log-priors for each label's individual log-prior, $\Pi(\Theta)=\Pi(\teff)+\Pi(\logg)+...+\Pi(c_{n,o})$.

As described in \S\ref{photometric_priors}, Gaia photometry is used to essentially impose a delta-function prior on \tefftext\ and \loggtext\ given $m_{\text{G}, 0}$, $\text{G}_{\text{BP}, 0}-\text{G}_{\text{RP}, 0}$, and [Fe/H]. For the remaining stellar labels (\vmicrotext\ and all elemental abundances), we adopt uniform priors over the range of values included in the spectral training grid (see Table \ref{tab:sampling}):
\begin{align}
    v_\text{micro} &\sim \mathcal{U}(1.2, ~2.5) \nonumber\\
    [\text{Fe/H}] &\sim \mathcal{U}(-4.00, ~-1.00) \nonumber\\
    [\text{X}_1\text{/Fe}] &\sim \mathcal{U}(-1.00, ~1.00) \nonumber\\
    [\text{X}_2\text{/Fe}] &\sim \mathcal{U}(-0.50, ~0.50) \nonumber\\
    [\text{X}_3\text{/Fe}] &\sim \mathcal{U}(-0.25, ~1.00), \nonumber
\end{align}
where $\text{X}_1 =~$C, N, and O; $\text{X}_2 =~$Na, Sc, V, Cr, Mn, Co, Ni, Cu, Zn, Ga, Sr, Y, Zr, Ba, and La; and $\text{X}_3 =~$Mg, Al, Si, K, Ca, Ti, Ce, Pr, Nd, Sm, Eu, Gd, Dy, Ho, Er, Os, and Th.

Though the resolving power, $R$, is a parameter in our model, we simply adopt the observatory-provided resolutions (and subsequent artificial reductions). This effectively imposes a delta-function prior on $R$,
\begin{align}
    R_\text{inst} &\sim \delta(R_\text{obs}). \nonumber
\end{align}
We impose a uniform prior on the log macroturbulent velocity, $\log_{10}v_\text{macro}$, of from -1.0 to 1.3 
\begin{align}
    \log_{10}v_\text{macro} &\sim \mathcal{U}(-1.0, ~1.3), \nonumber
\end{align}
which is equivalent to bounding \vmacrotext\ in linear units from 0.1 to 20 km/s.
We adopt a broad uniform prior on the radial velocity from $-300$ to 300 km/s
\begin{align}
    \mathcal{U}(-300~\text{km/s}, ~300~\text{km/s}). \nonumber
\end{align}

Because it is difficult to predict the appropriate range of values for the continuum polynomial coefficients, $c_{n,o}$, \textit{a priori}, we adopt infinitely broad uniform priors on $c_{n,o}$ during optimization. Unfortunately, the large number of coefficients ($n\times N_\text{ord}$) make including all $c_{n,o}$ as free parameters in the MCMC sampling computationally infeasible. Future work with Hamiltonian Monte Carlo and/or nested sampling methods may eventually make this tractable, but for now we fix all $c_{n,o}$ to the best fit optimization values with a delta function prior:
\begin{align}
    c_{o,n} &\sim 
    \begin{cases}
        \mathcal{U}(-\infty,\infty)   & \text{Optimizer} \\
        \delta\left(c_{o,n}^\text{(Opt)}\right)     & \text{MCMC}
    \end{cases}. \nonumber
\end{align}

\subsubsection{Reparameterization}
To aid in the optimization and sampling of our posteriors, we find it advantageous to reparameterize a subset of our model parameters so that they share a similar dynamic range. Instead of fitting \vmacrotext\ in linear units, we fit for $\log v_{\text{macro}}$. The radial velocity, $v_{r}$, is scaled by a factor of 100 so that it has units of 100 km/s. The stellar labels, $\theta_{*}$, are scaled in the same manner as they are for the training of the \thePayne\ to be between $-0.5$ and $0.5$ (see Appendix \ref{app:payne}). The priors for these reparameterized labels are transformed accordingly.

\subsubsection{Fitting to Multiple Exposures}
\label{sec:methods_stack}
There are several approaches to handling the extra constraining power enabled by multiple exposures of the same star. The simplest and most common approach involves co-adding the spectra from individual exposures to create a ``stacked" spectrum with a higher S/N than from the individual exposures. This approach is limited, however, in that it hides potential observational systematics introduced at the inter-exposure level---it is impossible to say how each exposure impacts the stacked fit. 

A second approach and the one we adopt in this study is to treat each exposure of the same star as an independent observation of that star.
The joint log-likelihood for the $N_\text{exp}$ exposures is then just the sum of each individual exposure's log-likelihood,
\begin{equation}
   \ln L(D|\Theta) = \sum^{N_\text{exp}} \ln L(D_i|\Theta),
   \label{eq:joint_like}
\end{equation}
and the ``stacked" posterior of the multiple exposures is 
\begin{equation}
    \ln P(\Theta|D) = \ln \Pi(\Theta) + \sum^{N_\text{exp}} \ln L(D_i|\Theta),
    \label{eq:stacked_posterior}
\end{equation}
where $\ln \Pi(\Theta)$ are the log-priors described in \S\ref{sec:methods_priors}.

While we can calculate the joint likelihood by fitting all exposures simultaneously, we choose to construct it after first sampling the posteriors of the individual exposure fits. We then fit these marginalized posteriors assuming they are well-described by 1-dimensional Gaussian distributions truncated at the bounds of the uniform priors. With functional forms of the posterior distributions in hand, we convert them into likelihood functions (a task made trivial by the use of uniform priors), and combine them into a joint likelihood function. Re-introducing priors results in the stacked posterior distribution function given in Equation \ref{eq:stacked_posterior}, which we also fit assuming 1-dimensional truncated Gaussian distributions. We take the mean and standard deviation of these distributions as the best-fit value and $1\sigma$ statistical uncertainty of the stellar label except when the best-fit value is $<1\sigma$ from the uniform priors bounds adopted in \S\ref{sec:methods_priors}. In these instances, we instead adopt the 95\% upper/lower limit in lieu of the mean and standard deviation.

\begin{figure*}[ht!]
    \plotone{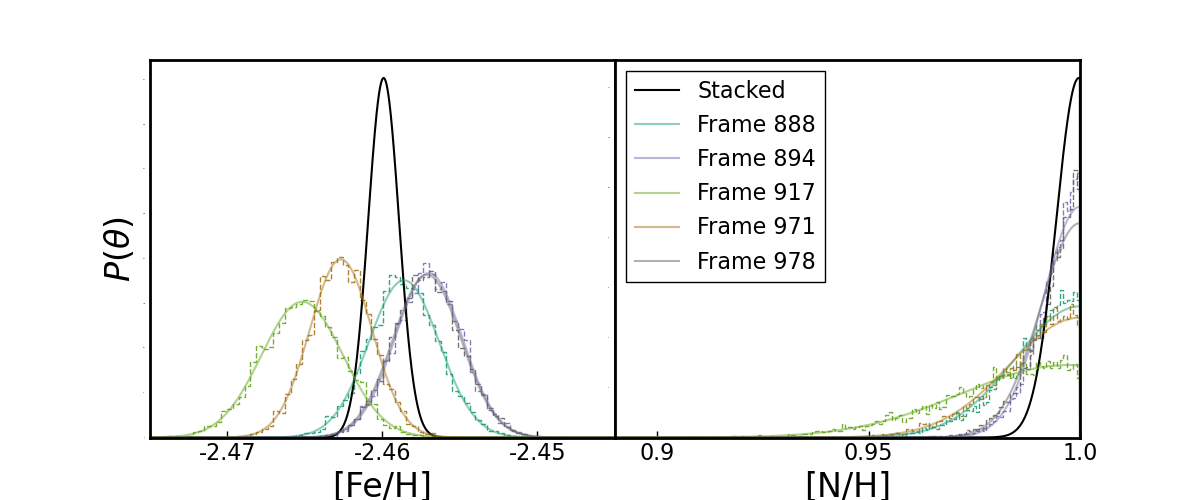}
    \epsscale{1.15}
    \caption{
        Marginalized posteriors for [Fe/H] (left) and [N/Fe] (right) for K431 observed in the C316Hr program at the degraded resolution of $R\sim20000$. Posterior samples and the best fit truncated normal distribution for the 5 individual exposures are plotted in the thin colored dashed histograms and solid curves respectively. The stacked posterior recovered when combining the individual likelihoods is plotted in the thick black line. In the case of [N/Fe], the best fit value is at the boundary of our priors (set by the extent of our training grid), necessitating the use of a truncated distribution.
    } \label{fig:3_posterior_stack}
\end{figure*}

The left panel of Figure \ref{fig:3_posterior_stack} illustrates an example stacked posterior for [Fe/H] (black curve) that is recovered when the five individual exposure posteriors (colored curves) are combined. The right panel illustrates the same for [N/Fe] and demonstrates the importance of using truncated distributions.

\section{Results}
\label{sec:results}
In this section, we present the results of our spectral fitting. We begin in \S\ref{sec:Abundance_v_Resolution} with the recovery of stellar labels as a function of resolving power and conclude in \S\ref{sec:Abundance_v_SNR} with the recovery of stellar labels as a function of S/N. For a comparison of the stellar labels we measure from un-altered (i.e., default resolution and S/N) spectra to literature values, see Appendix \ref{sec:literature}.

\subsection{Label Recovery as a Function of Resolution}
\label{sec:Abundance_v_Resolution}
At each resolution and for each star in our sample, we calculate the change in stellar labels, $\delta\theta$, relative to the recovered labels at the highest available resolution for that star
\begin{equation}
\delta\theta_{R} = \theta_{R} - \theta_{R_{0}}.
\end{equation}
Taken together, the trends of $\delta\theta$ vs.\ resolution for our sample provide a coarse marginalization over the spectroscopic configurations (e.g., wavelength coverage) and stellar parameters (e.g., \tefftext, \loggtext, [Fe/H]) presented in this work. We summarize the collective trend for each stellar label with two quantities: a resolution-dependent systematic bias, $\Delta\theta$, and a resolution-dependent systematic uncertainty, $\sigma_\text{syst}$. The systematic bias captures how much a stellar label is likely to be over/underestimated when measured at a lower resolution, while the systematic uncertainty captures the dispersion in $\delta\theta$ found across the programs and stars analyzed. We define $\Delta\theta$ to be the median and $\pm\sigma_\text{syst}$ to be the 16th and 84th percentiles of $\delta\theta$ at each resolution. We omit from these calculations any poorly constrained fits for which the statistical uncertainty $>$0.5 dex.

As described in \S\ref{sec:methods_stack}, some stellar label fits result in the recovery of upper or lower limits. While $\Delta\theta$ and $\sigma_\text{syst}$ are robust to the presence of a few upper and lower limits, if a large enough fraction of the measurements of a stellar label at a given resolution are upper/lower limits, the 16th and 84th percentiles of $\delta\theta$---and thus $\pm\sigma_\text{syst}$---may correspond to a limit. In these instances, the systematic uncertainty will be underestimated. In rare cases, $\Delta\theta$ may also correspond to a limit and be similarly underestimated.

In Figures \ref{fig:4_atm1_v_res}-\ref{fig:4_neutron3_v_res}, we present these systematic biases (solid black line) and uncertainties (gray shaded region) as a function of resolution. Solid red lines at the edge of the gray shaded region denote regions where the bias and/or uncertainty may be underestimated due to the limitations imposed by our training set and priors. For a few elements, measurements for each of the individual stars in the sample are included, color-coded by their observing program, to highlight instances where substantially different trends are exhibited from one archival dataset to the next. In these cases the U09H, C147Hr, and C316Hr programs are indicated with red squares, orange triangles, and blue circles respectively. Individual star measurements for all elements can be made available upon request.
In Table \ref{tab:abundance_v_resolution}, we provide $\Delta\theta$ and $\sigma_\text{syst}$ for each element at resolutions of $R\sim2500$, 5000, 10000, 20000, and 40000. In the remainder of this section, we discuss the resolution-dependent recovery of each stellar label individually. For clarity, we organize our discussion of each element in groups loosely based on shared nucleosynthetic origin.

\begin{deluxetable*}{cCCCCC}
	\centerwidetable
	\caption{Trends in Stellar Label Recovery with Resolution}
	\label{tab:abundance_v_resolution}
	\tablehead{
         &
        \colhead{$R\sim2500$} &
        \colhead{$R\sim5000$} &
        \colhead{$R\sim10000$} &
        \colhead{$R\sim20000$} &
        \colhead{$R\sim40000$} \\
        \colhead{$\theta$} &
        \colhead{$\Delta\theta\pm_{\sigma_\text{16th}}^{\sigma_\text{84th}}$} &
        \colhead{$\Delta\theta\pm_{\sigma_\text{16th}}^{\sigma_\text{84th}}$} &
        \colhead{$\Delta\theta\pm_{\sigma_\text{16th}}^{\sigma_\text{84th}}$} &
        \colhead{$\Delta\theta\pm_{\sigma_\text{16th}}^{\sigma_\text{84th}}$} &
        \colhead{$\Delta\theta\pm_{\sigma_\text{16th}}^{\sigma_\text{84th}}$}
	}
	\startdata		
		\tefftext   & 0.78\pm_{0.65}^{0.80}                 & 0.68\pm_{0.64}^{0.33}                 & 0.31\pm_{0.30}^{0.60}                 & 0.16\pm_{0.18}^{0.17}                 & 0.17\pm_{0.03}^{0.03}                 \\
		\loggtext   & -0.01\pm_{0.00}^{0.00}                & -0.00\pm_{0.00}^{0.00}                & -0.00\pm_{0.00}^{0.00}                & -0.00\pm_{0.00}^{0.00}                & -0.00\pm_{0.00}^{0.00}                \\
		\vmicrotext & 0.15\pm_{0.07}^{0.08}                 & 0.26\pm_{0.07}^{0.08}                 & 0.25\pm_{0.10}^{0.11}                 & 0.14\pm_{0.02}^{0.03}                 & 0.08\pm_{0.00}^{0.00}                 \\
		\vmacrotext & 8.26^{*}\pm_{7.50}^{11.23^{*}}        & 5.96\pm_{5.42}^{8.67}                 & 2.52\pm_{2.18}^{5.02}                 & 0.97\pm_{0.82}^{2.59}                 & 0.63\pm_{0.11}^{0.11}                 \\
		$v_r$       & 0.34\pm_{1.03}^{0.37}                 & -0.09\pm_{0.17}^{0.28}                & -0.07\pm_{0.06}^{0.21}                & -0.03\pm_{0.05}^{0.14}                & -0.03\pm_{0.01}^{0.01}                \\
		{[}C/H{]}   & 0.06\pm_{0.10}^{0.02}                 & 0.06\pm_{0.10}^{0.03}                 & 0.05\pm_{0.01}^{0.01}                 & 0.02\pm_{0.01}^{0.01}                 & 0.01\pm_{0.00}^{0.00}                 \\
		{[}N/H{]}   & -0.00\pm_{0.01}^{0.01^{*}}            & 0.00^{*}\pm_{0.00}^{0.01^{*}}         & 0.01^{*}\pm_{0.01}^{0.01^{*}}         & 0.00^{*}\pm_{0.00}^{0.02^{*}}         & 0.00^{*}\pm_{0.00}^{0.00^{*}}         \\
		{[}O/H{]}   & -0.17\pm_{0.51}^{0.17^{*}}            & -0.03\pm_{0.37}^{0.03^{*}}            & -0.00^{*}\pm_{0.05}^{0.00^{*}}        & -0.00^{*}\pm_{0.02}^{0.00^{*}}        & 0.00^{*}\pm_{0.00}^{0.00^{*}}         \\
		{[}Na/H{]}  & -0.34\pm_{0.26^{*}}^{0.22^{*}}        & -0.24\pm_{0.69^{*}}^{0.23^{*}}        & -0.17\pm_{0.63^{*}}^{0.17^{*}}        & -0.07\pm_{0.25}^{0.06^{*}}            & -0.15\pm_{0.02}^{0.02}                \\
		{[}Mg/H{]}  & 0.02\pm_{0.10^{*}}^{0.16}             & 0.05\pm_{0.11}^{0.05}                 & 0.00\pm_{0.08}^{0.06}                 & -0.01\pm_{0.02^{*}}^{0.02}            & -0.03\pm_{0.03}^{0.03}                \\
		{[}Al/H{]}  & 0.13^{*}\pm_{0.36^{*}}^{0.38}         & 0.14\pm_{0.14^{*}}^{0.43}             & 0.00\pm_{0.18^{*}}^{0.48}             & 0.00\pm_{0.32^{*}}^{0.12}             & 0.13^{*}\pm_{0.09^{*}}^{0.09}         \\
		{[}Si/H{]}  & 0.14\pm_{0.05}^{0.22}                 & 0.11\pm_{0.02}^{0.19}                 & 0.08\pm_{0.04}^{0.22}                 & 0.02\pm_{0.04}^{0.19}                 & 0.12\pm_{0.06}^{0.06}                 \\
		{[}K/H{]}   & \nodata                               & \nodata                               & \nodata                               & -0.00^{*}\pm_{0.22^{*}}^{0.00^{*}}    & -0.54^{*}\pm_{0.37}^{0.37^{*}}        \\
		{[}Ca/H{]}  & -0.02\pm_{0.12}^{0.10}                & 0.02\pm_{0.08}^{0.04}                 & -0.00\pm_{0.03}^{0.05}                & 0.00\pm_{0.02}^{0.02}                 & -0.00\pm_{0.00}^{0.00}                \\
		{[}Sc/H{]}  & -0.05\pm_{0.10}^{0.13}                & -0.04\pm_{0.01}^{0.13}                & 0.01\pm_{0.03}^{0.08}                 & 0.02\pm_{0.03}^{0.03}                 & 0.03\pm_{0.00}^{0.00}                 \\
		{[}Ti/H{]}  & -0.03\pm_{0.03}^{0.09}                & -0.01\pm_{0.01}^{0.04}                & 0.00\pm_{0.02}^{0.04}                 & 0.01\pm_{0.03}^{0.01}                 & 0.01\pm_{0.00}^{0.00}                 \\
		{[}V/H{]}   & 0.07\pm_{0.22}^{0.10}                 & 0.06\pm_{0.13}^{0.04}                 & 0.02\pm_{0.03}^{0.03}                 & 0.02\pm_{0.02}^{0.01}                 & 0.01\pm_{0.00}^{0.00}                 \\
		{[}Cr/H{]}  & 0.13\pm_{0.41^{*}}^{0.13}             & 0.07\pm_{0.30}^{0.04}                 & 0.04\pm_{0.12}^{0.02}                 & 0.02\pm_{0.04}^{0.01}                 & 0.02\pm_{0.01}^{0.01}                 \\
		{[}Mn/H{]}  & 0.00^{*}\pm_{0.00^{*}}^{0.00}         & 0.00^{*}\pm_{0.00^{*}}^{0.00}         & 0.04\pm_{0.04^{*}}^{0.08}             & 0.00^{*}\pm_{0.00^{*}}^{0.02}         & 0.14^{*}\pm_{0.09^{*}}^{0.09}         \\
		{[}Fe/H{]}  & -0.02\pm_{0.01}^{0.01}                & -0.02\pm_{0.01}^{0.01}                & -0.01\pm_{0.01}^{0.01}                & -0.00\pm_{0.01}^{0.01}                & -0.00\pm_{0.00}^{0.00}                \\
		{[}Co/H{]}  & -0.03\pm_{0.31}^{0.07}                & 0.06\pm_{0.19}^{0.02}                 & 0.04\pm_{0.10}^{0.02}                 & -0.00\pm_{0.03}^{0.01}                & -0.02\pm_{0.01}^{0.01}                \\
		{[}Ni/H{]}  & 0.01\pm_{0.07}^{0.09}                 & -0.00\pm_{0.02}^{0.04}                & -0.02\pm_{0.04}^{0.04}                & 0.00\pm_{0.02}^{0.01}                 & -0.01\pm_{0.01}^{0.01}                \\
		{[}Cu/H{]}  & 0.28^{*}\pm_{0.28^{*}}^{0.57^{*}}     & 0.26^{*}\pm_{0.26^{*}}^{0.59^{*}}     & 0.10^{*}\pm_{0.10^{*}}^{0.26}         & 0.07^{*}\pm_{0.07^{*}}^{0.12}         & 0.02\pm_{0.03}^{0.03}                 \\
		{[}Zn/H{]}  & 0.32\pm_{0.53}^{0.44^{*}}             & 0.30\pm_{0.29}^{0.32}                 & 0.14\pm_{0.06}^{0.20}                 & 0.07\pm_{0.02}^{0.05}                 & 0.06\pm_{0.02}^{0.02}                 \\
		{[}Ga/H{]}  & \nodata                               & \nodata                               & \nodata                               & \nodata                               & \nodata                               \\
		{[}Sr/H{]}  & -0.03\pm_{0.19}^{0.06}                & 0.11\pm_{0.11}^{0.06}                 & 0.10\pm_{0.14}^{0.07}                 & 0.02\pm_{0.07}^{0.09}                 & 0.07\pm_{0.03}^{0.03}                 \\
		{[}Y/H{]}   & 0.07\pm_{0.13}^{0.19}                 & -0.01\pm_{0.03}^{0.14}                & -0.04\pm_{0.02}^{0.07}                & -0.02\pm_{0.01}^{0.03}                & -0.01\pm_{0.01}^{0.01}                \\
		{[}Zr/H{]}  & -0.12\pm_{0.37}^{0.08}                & -0.02\pm_{0.15}^{0.06}                & 0.02\pm_{0.08}^{0.03}                 & 0.02\pm_{0.04}^{0.02}                 & 0.02\pm_{0.00}^{0.00}                 \\
		{[}Ba/H{]}  & -0.00^{*}\pm_{0.54}^{0.00^{*}}        & -0.13\pm_{0.12}^{0.13^{*}}            & -0.04^{*}\pm_{0.17}^{0.04^{*}}        & -0.01^{*}\pm_{0.07}^{0.01^{*}}        & -0.01^{*}\pm_{0.01}^{0.01^{*}}        \\
		{[}La/H{]}  & 0.01^{*}\pm_{0.01}^{0.17^{*}}         & 0.02^{*}\pm_{0.02}^{0.16^{*}}         & 0.03\pm_{0.03}^{0.06^{*}}             & 0.02\pm_{0.02}^{0.02^{*}}             & 0.02^{*}\pm_{0.01}^{0.01^{*}}         \\
		{[}Ce/H{]}  & 0.25\pm_{0.09}^{0.07}                 & 0.08\pm_{0.01}^{0.11}                 & 0.05\pm_{0.04}^{0.06}                 & 0.02\pm_{0.02}^{0.04}                 & 0.04\pm_{0.02}^{0.02}                 \\
		{[}Pr/H{]}  & 0.08\pm_{0.06}^{0.03}                 & 0.11\pm_{0.01}^{0.03}                 & 0.08\pm_{0.01}^{0.04}                 & 0.05\pm_{0.01}^{0.02}                 & 0.03\pm_{0.01}^{0.01}                 \\
		{[}Nd/H{]}  & -0.04\pm_{0.13}^{0.07}                & 0.02\pm_{0.08}^{0.06}                 & 0.02\pm_{0.04}^{0.04}                 & 0.02\pm_{0.02}^{0.01}                 & 0.02\pm_{0.00}^{0.00}                 \\
		{[}Sm/H{]}  & -0.28\pm_{0.14}^{0.13}                & -0.07\pm_{0.02}^{0.06}                & 0.03\pm_{0.03}^{0.03^{*}}             & 0.02\pm_{0.02}^{0.02}                 & 0.02\pm_{0.00}^{0.00}                 \\
		{[}Eu/H{]}  & -0.12\pm_{0.43}^{0.07}                & -0.00\pm_{0.10}^{0.01^{*}}            & 0.01^{*}\pm_{0.01}^{0.03^{*}}         & 0.00^{*}\pm_{0.00}^{0.01^{*}}         & 0.00^{*}\pm_{0.00}^{0.00^{*}}         \\
		{[}Gd/H{]}  & 0.14\pm_{0.51}^{0.29}                 & 0.10\pm_{0.16}^{0.05}                 & 0.04\pm_{0.05}^{0.03}                 & 0.01\pm_{0.03}^{0.02}                 & 0.01\pm_{0.01}^{0.01}                 \\
		{[}Dy/H{]}  & 0.10\pm_{0.47}^{0.14^{*}}             & 0.03\pm_{0.23}^{0.15}                 & 0.05\pm_{0.07}^{0.07}                 & 0.05\pm_{0.01}^{0.01}                 & 0.04\pm_{0.01}^{0.01}                 \\
		{[}Ho/H{]}  & \nodata                               & \nodata                               & -0.00^{*}\pm_{0.22}^{0.03^{*}}        & -0.00^{*}\pm_{0.13}^{0.03^{*}}        & 0.13^{*}\pm_{0.09}^{0.09^{*}}         \\
		{[}Er/H{]}  & 0.33^{*}\pm_{0.19}^{0.27^{*}}         & 0.17\pm_{0.02}^{0.11^{*}}             & 0.13\pm_{0.06}^{0.14}                 & 0.05\pm_{0.02}^{0.05}                 & 0.10\pm_{0.04}^{0.04}                 \\
		{[}Os/H{]}  & \nodata                               & \nodata                               & \nodata                               & \nodata                               & \nodata                               \\
		{[}Th/H{]}  & -0.41                                 & -0.18\pm_{0.04}^{0.04}                & -0.02\pm_{0.07}^{0.02}                & 0.00\pm_{0.05}^{0.00}                 & -0.04                 
	\enddata
	\tablecomments{
	    Asterisks denote instances where the reported quantities are impacted by the imposed boundaries of the training set. If no systematic uncertainty is provided, there was only one measurement for which the statistical uncertainty was $<$0.5 dex.
	}
\end{deluxetable*}

\subsubsection{Atmospheric Parameters}
\label{sec:rec_vs_res_atm}
\begin{figure}[ht!]
    \epsscale{1.20}
    \plotone{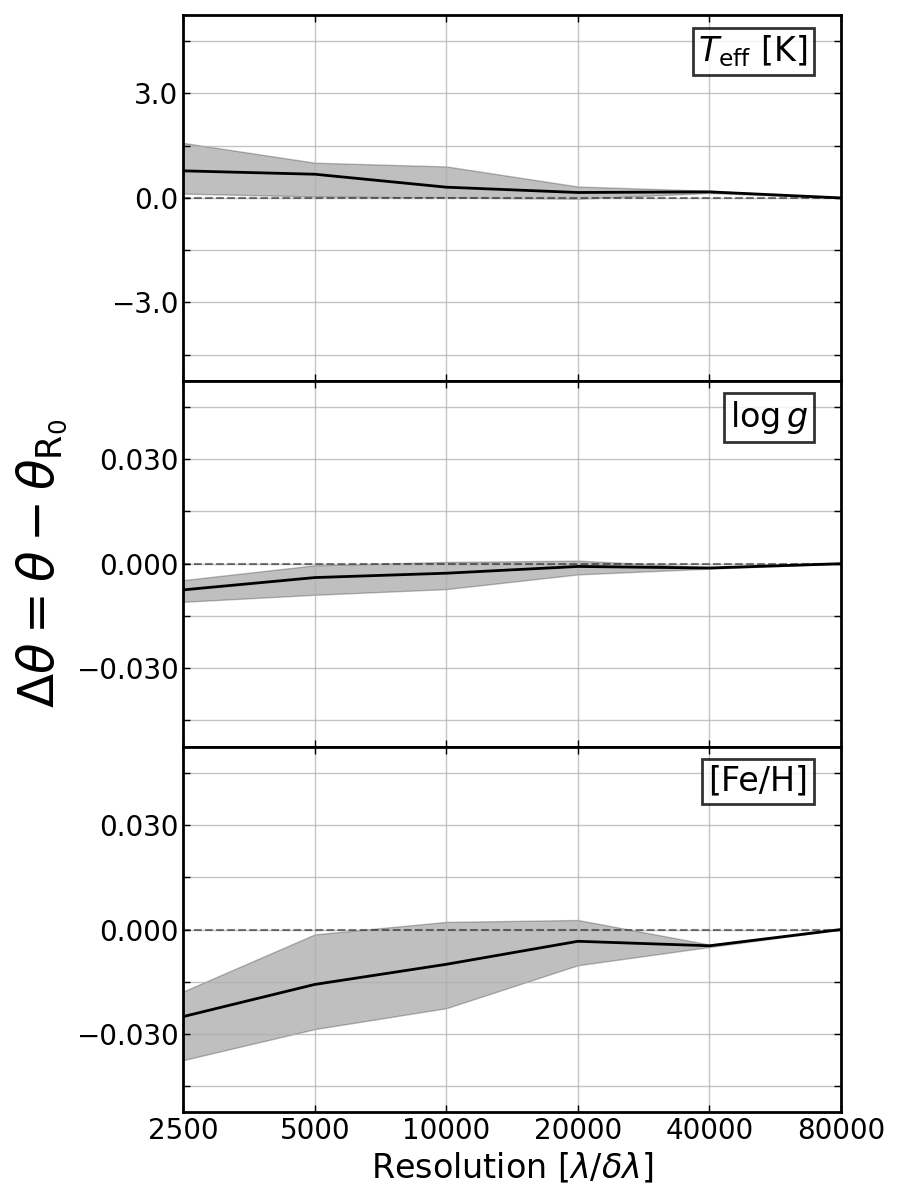}
    \caption{
        Systematic biases (solid black lines) and 1-$\sigma$ systematic uncertainties (gray shaded regions) in the recovery of \tefftext\ (top), \loggtext\ (middle), and [Fe/H] (bottom) as a function of resolution. All three labels are recovered with only very minimal differences ($+1$ K, $-0.01$ dex, and $-0.03$ dex) across the entire range of resolutions analyzed. Systematic uncertainties are similarly small.
        \label{fig:4_atm1_v_res}
    }
\end{figure}

\paragraph{Effective Temperature, Surface Gravity, and [Fe/H]}
In Figure \ref{fig:4_atm1_v_res}, we present the change in recovered atmospheric parameters \tefftext, \loggtext, and [Fe/H] as a function of resolution. Only very minimal differences are found between high-resolution and low-resolution measurements. At $R\sim2500$, \tefftext, \loggtext, and [Fe/H] only differ by approximately $+1$ K, $-0.01$ dex, and $-0.02$ dex respectively from the measurements made at $R\sim40000$--80000. The systematic uncertainties are similarly small.

The robust recovery of [Fe/H] at all resolutions is reassuring, albeit unsurprising given the abundance of well calibrated Fe absorption lines and a long history of reliable low-resolution (and photometric) stellar metallicity measurements. The wealth of well-modelled Fe lines minimizes the impact of blending with imperfectly modelled lines at low resolution.
%
%
The similarity in trend between \tefftext, \loggtext, and [Fe/H] is a direct result of the strong covariance between these labels, which is introduced by the determination of \tefftext\ and \loggtext\ from isochrones dependent on the star's photometry and [Fe/H].

\begin{figure}[ht!]
    \epsscale{1.20}
    \plotone{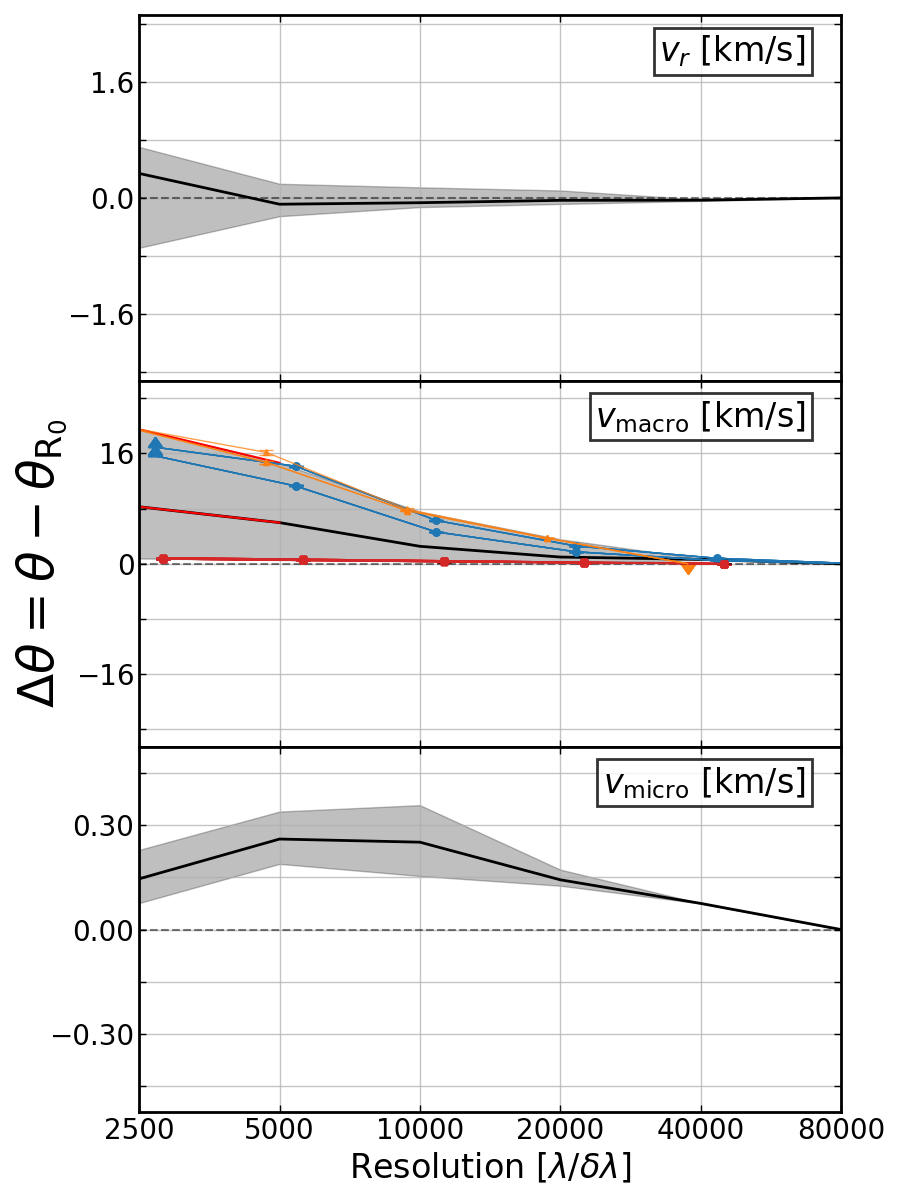}
    \caption{
        Same as Figure \ref{fig:4_atm1_v_res} except for velocity-based atmospheric parameters $v_r$ (top), \vmacrotext\ (middle), and \vmicrotext\ (bottom). 
        $v_r$ is recovered consistently at all resolutions with small systematic uncertainties ($\lesssim$0.5 km/s for $R\gtrsim5000$ and $\sim$1.5 km/s for $R\sim2500$). 
        \vmacrotext\ exhibits distinct trends between the U09H observations (red squares) and the 147Hr and C316Hr observations (orange triangles and blue circles respectively). The large systematic offsets seen at low resolution for the latter observations is attributed to incorrectly specified instrumental broadening.
        We recover $\vmicro\lesssim0.5$ km/s higher at lower resolutions consistent with its correlation with [Fe/H] and the small trend seen for [Fe/H] in Figure \ref{fig:4_atm1_v_res}.
        \label{fig:4_atm2_v_res}
    }
\end{figure}

\paragraph{Radial, Macroturbulent, and Microturbulent Velocities}
Figure \ref{fig:4_atm2_v_res} shows the changes in the recovered velocity-related parameters \vmicrotext, \vmacrotext, and $v_r$.
As expected, the recovery of radial velocity across the sample shows little trend with resolving power. The systematic uncertainty in $v_r$ is $\lesssim$0.25 km/s for $R\gtrsim5000$ and $\sim$1 km/s for $R\sim2500$. These spreads are large compared to the formal measurement uncertainties (0.1--0.6 km/s), but on par with expectations for low- and medium-resolution surveys \citep[e.g.,][]{xiong:2021}.

For \vmacrotext, we find two distinct trends, one for the older U09H observations (red squares) and one for the post-upgrade C147Hr and C316Hr observations (orange triangles and blue circles respectively). For the newer observations, the measured value of \vmacrotext\ increases by up to 16 km/s as the resolution is decreased to $R\sim2500$, while for the older observations the resolution dependence is much weaker with an offset of only $\sim$1 km/s at $R\sim2500$. This suggests that the observed trend is driven by an observational systematic present in the C147Hr and C316Hr data, most likely a mismatch between the assumed and true default spectroscopic resolution. Because both macroturbulent and instrumental broadening are implemented with Gaussian kernels, \vmacrotext\ and $R$ are entirely degenerate. As a result of not fitting for $R$, \vmacrotext\ compensates for this mismatch. We do not find evidence that \vmacrotext\ is correlated in any meaningful way with stellar chemical abundances, which are the primary concern of this study. As such, we simply treat \vmacrotext\ as a nuisance parameter that characterizes the instrumental LSF.

For \vmicrotext\, a more moderate trend with resolution is seen with measurements $\sim$0.1--0.3 km/s larger at $R\lesssim20000$ than the measurements made at $R\gtrsim40000$. Most, if not all, of this offset can be attributed to the correlation of \vmicrotext\ and [Fe/H], which we find to be the two most highly correlated stellar labels in our analysis (with the exception of \tefftext\ and \loggtext). The $-0.03$ dex $\Delta$[Fe/H] seen in Figure \ref{fig:4_atm1_v_res} alone can explain $\Delta\vmicro\sim0.15$. The growth of spectral masks with decreasing resolution (see \S\ref{sec:model_uncertainties}) may also impact the fitting of extended line profiles, which could introduce systematics into the measurement of \vmicrotext.

\subsubsection{Iron-Peak Element Abundances}
\label{sec:rec_vs_res_fepeak}

\begin{figure}[ht!]
    \epsscale{1.20}
    \plotone{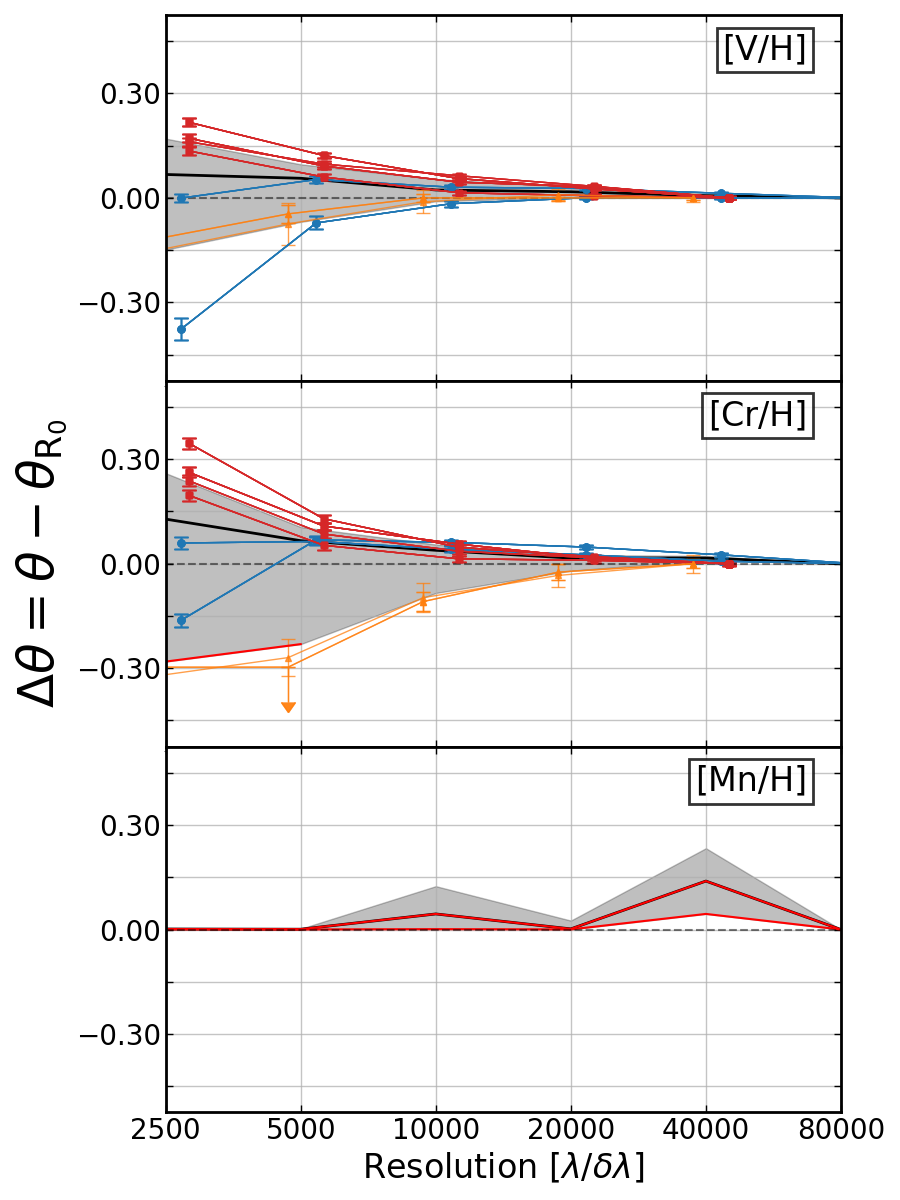}
    \caption{
        Same as Figure \ref{fig:4_atm1_v_res} except for iron-peak elements  V (top), Cr (middle), and Mn (bottom). [V/H] and [Cr/H] are recovered consistently down to $R\sim5000$ with gradually increasing systematic uncertainties and a slight bias towards higher values at the lowest resolution. These systematic trends are driven by a combination of blending of imperfectly modeled lines in the blue and imperfectly modeled continuum regions in the red. Upper limits on [Mn/H] are recovered at all resolutions.
        \label{fig:4_fepeak1_v_res}
    }
\end{figure}

\begin{figure}[ht!]
    \epsscale{1.20}
    \plotone{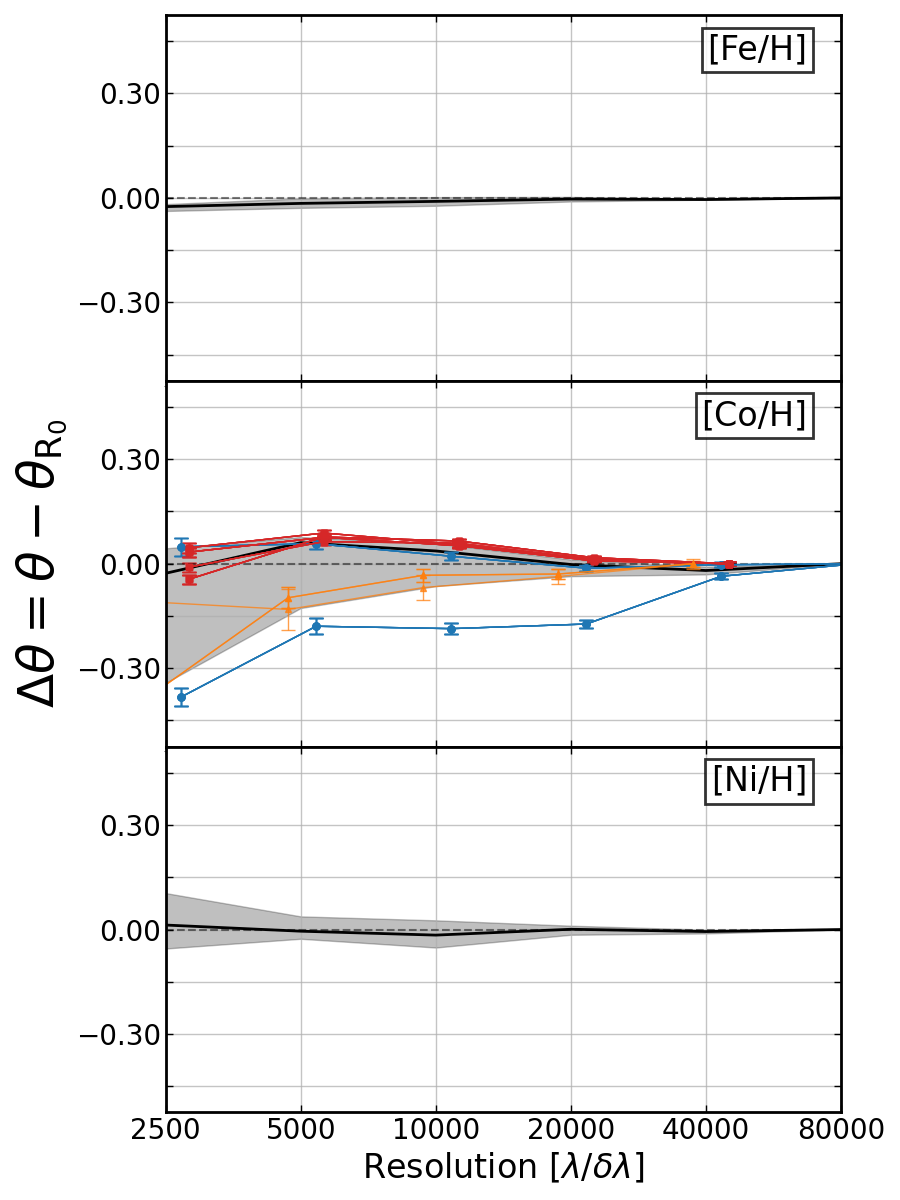}
    \caption{
        Same as Figure \ref{fig:4_atm1_v_res} except for iron-peak elements  Fe (top), Co (middle), and Ni (bottom). We find very small ($\lesssim0.03$  dex) systematic effects for [Fe/H]. [Co/H] and [Ni/H] are also recovered consistently at all resolutions, though [Co/H] exhibits a substantial $\sim$0.3 dex systematic uncertainty at the lowest resolutions. 
        \label{fig:4_fepeak2_v_res}
    }
\end{figure}

\begin{figure}[ht!]
    \epsscale{1.20}
    \plotone{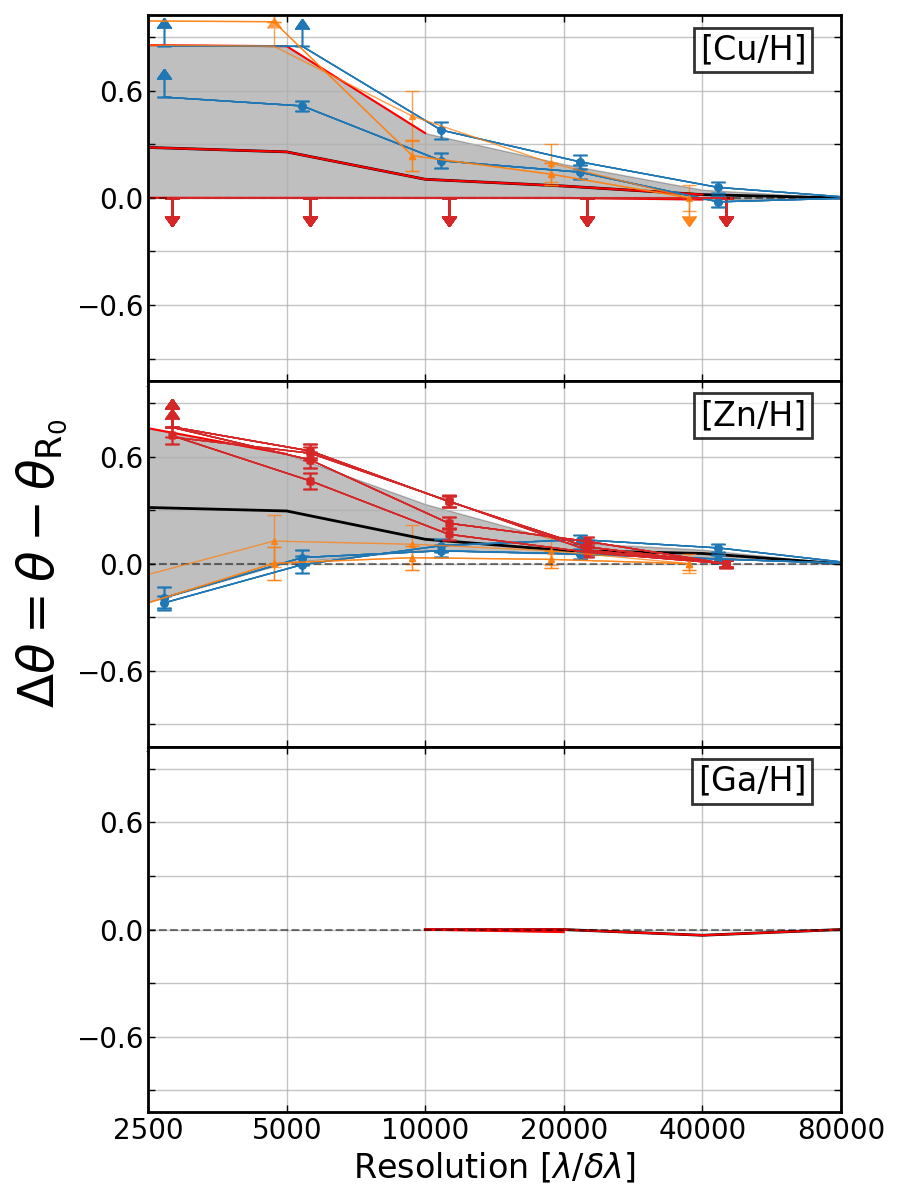}
    \caption{
        Same as Figure \ref{fig:4_atm1_v_res} except for iron-peak elements  Cu (top), Zn (middle), and Ga (bottom). We find the recovery of [Cu/H] and [Zn/H] to be strongly resolution dependent with both large biases and uncertainties. These systematic effects are driven by the reliance on only a handful of lines which are easily impacted by the modeling of neighboring lines at lower resolutions. Due to the paucity of Ga lines, we do not recover [Ga/H] within the bounds of the training set at any resolution.
        \label{fig:4_fepeak3_v_res}
    }
\end{figure}

In Figures \ref{fig:4_fepeak1_v_res}, \ref{fig:4_fepeak2_v_res}, and \ref{fig:4_fepeak3_v_res}, we present the systematic bias and uncertainty of iron-peak elements (V, Cr, Mn, Fe, Co, Ni, Cu, Zn, and Ga) as a function of resolution. In summary, we find that [V/H], [Cr/H], [Fe/H], [Co/H], and [Ni/H] are recovered consistently at all resolutions, though [V/H] and [Cr/H] display small biases towards higher abundances at low resolution. [Cu/H] and [Zn/H] exhibit large systematic biases and uncertainties at low resolution, and Ga proves challenging to recover at all. For [Mn/H] the boundary of the training grid limit a complete picture of the bias and spread in abundance measurement as a function of resolution. We describe the results for each element in detail below. 

\paragraph{Vanadium}
We recover [V/H] consistently and with small systematic uncertainties ($<$0.05 dex) at all resolutions higher than $R\gtrsim10000$. At lower resolutions, a small bias towards higher [V/H] develops and grows to 0.07 dex, but increasing systematic uncertainties maintain $1\sigma$ consistency with the $R\sim40000$--80000 measurements. The increasing systematic uncertainty is driven by diverging trends between the older blue-only U09H observations (red squares), which trend higher as the resolution is decreased, and the newer full-optical C147Hr and C316Hr observations (orange triangles and  blue circles respectively), which trend lower as the resolution is decreased. These trends are driven by heavy blending of mis-modeled lines in the blue ($\lambda<4500$ \AA) and mis-fit continuum regions coupled with weak lines in the red ($\lambda\sim6000$ \AA) respectively.

\paragraph{Chromium}
The recovery of [Cr/H] as a function of resolution resembles that of [V/H] discussed above. As the resolution is decreased to $R\sim2500$, the systematic bias and uncertainty increase to $\sim$0.13 and $\sim$0.3 dex respectively. As with [V/H], the increasing systematic uncertainty is driven by diverging behavior between the U09H, C147Hr, and C316Hr observations. The same underlying causes can be attributed as well. At $R<5000$, only upper limits on [Cr/H] are recovered for the C147Hr observations, leading to the lower uncertainty interval being underestimated. 


\paragraph{Manganese}
For nearly all observations and resolutions we recover [Mn/H] that are near or at the training set's lower bound ($\text{[Mn/Fe]} = -0.5$), which precludes any robust quantification of the systematic bias and uncertainty. As we discuss in Appendix \ref{sec:lit_by_element}, this is in general agreement with previous LTE measurements from \citet{sobeck:2006} and \citet{sobeck:2011} who measure [Mn/Fe] for these stars in the range of $-0.3$ to $-0.6$ dex. Fits to the $\sim$50 Mn lines in the spectra appear qualitatively reasonable, suggesting that the [Mn/Fe] value we would recover with an extended training set is  not too far beyond the currently imposed limits.


\paragraph{Iron}
As discussed previously in \S\ref{sec:rec_vs_res_atm}, [Fe/H] is recovered with only small ($\lesssim$0.02 dex) systematic biases and uncertainties across the full range in resolutions analyzed. 

\paragraph{Cobalt}
We find that [Co/H] is generally recovered consistently from $R\sim2500$--80000. [Co/H] recovery exhibits a small bias of $+$0.04--0.06 dex at $10000\gtrsim R\gtrsim5000$, but none at $R\sim2500$. A negatively skewed systematic uncertainty increases gradually and grows to $\sim$0.3 dex at the lowest resolution, primarily driven by large negative biases in the measurements from C147Hr and C316Hr observations. 

Upon deeper investigation, we determine that these biases can be traced to two sources: the poorly-modeled CN band at \wave{3883} in K731, K934, and K969 and the reliance on weak red-optical Co I lines, which are biased by the presence of correlated noise in the low-resolution, high-S/N regime. 
With their bluer wavelength coverage, the U09H observations contain approximately three times as many Co lines, leading to more consistent [Co/H] recovery.

\paragraph{Nickle}
We recover [Ni/H] consistently  across the full range of resolutions analyzed. Systematic uncertainties increase gradually with decreasing resolution to 0.08 dex at $R\sim2500$.

\paragraph{Copper}
We find the resolution-dependent recovery of [Cu/H] to be strongly dependent on the observational setup. For U09H observations, we recover only upper bounds ($\text{[Cu/Fe]} < -0.5$ dex), while for C147Hr and C316Hr observations, we measure [Cu/H] values that steadily rise by nearly 1 dex and become lower limits ($\text{[Cu/Fe]} > 0.5$ dex) as the resolution is decreased to $R\sim2500$. The result is a very large (0.3--0.8 dex) systematic uncertainty that is likely still underestimated due to the limits imposed by our priors. The systematic bias, nominally 0.3 dex at $R\sim2500$, is likely also underestimated.

For the U09H observations, constraints on [Cu/H] come predominantly from two weak Cu I lines (\wave\wave{4063.8, 5107.0}), which are both underestimated. The C316Hr and C147Hr observations also include the \wave{5783.7} Cu I line, which is located next to the edge of an order making it quite sensitive to the continuum determination. Indeed, we find that the trend towards higher [Cu/Fe] with decreasing resolution is caused by an increasingly poor fit to the continuum in  the region of this line.

\paragraph{Zinc}
The recovery of [Zn/H] as a function of resolution resembles a more extreme case of the systematic biases and uncertainties seen for [V/H] and [Cr/H]. As we decrease the resolution below $R\lesssim10000$, a $\sim$0.3 dex positive bias develops and the systematic uncertainty grows to $\sim$0.5 dex. We find that this trend is predominantly driven by the recovery of [Zn/H] from U09H measurements, which are $\sim$0.7 dex larger at $R\sim2500$ than at the default resolution of $R\sim40000$. [Zn/H] measurements from C147Hr and C316Hr observations are largely consistent across all resolutions.

Because the measurement of [Zn/H] relies on only three Zn I lines (\wave\wave{4681.4,4723.5,4811.9}), the measurement is quite sensitive to systematics. In the U09H observations, the \wave{4723.5} line falls near the edge of the detector and is partially lost as the resolution is decreased. This further increases the impact of blending with poorly-modeled lines on the remaining two lines.

\paragraph{Gallium}
Owing to the lack of good Ga absorption lines in the archival spectra, we struggle at all resolutions to recover [Ga/H] within the bounds of our training set ($-0.5\leq\text{[Ga/Fe]}\leq0.5$). The only two lines accessible to us are $\lambda\lambda$4034.1 and 4173.2, both of which are quite weak, heavily blended with adjacent lines, and fall within NLTE masks. At lower resolutions, these issues are further exacerbated. As a result, we cannot quantify the dependence of [Ga/H] recovery on resolution.

\subsubsection{\texorpdfstring{$\alpha$}{alpha} Element Abundances}
\label{sec:rec_vs_res_alpha}
\begin{figure}[ht!]
    \epsscale{1.20}
    \plotone{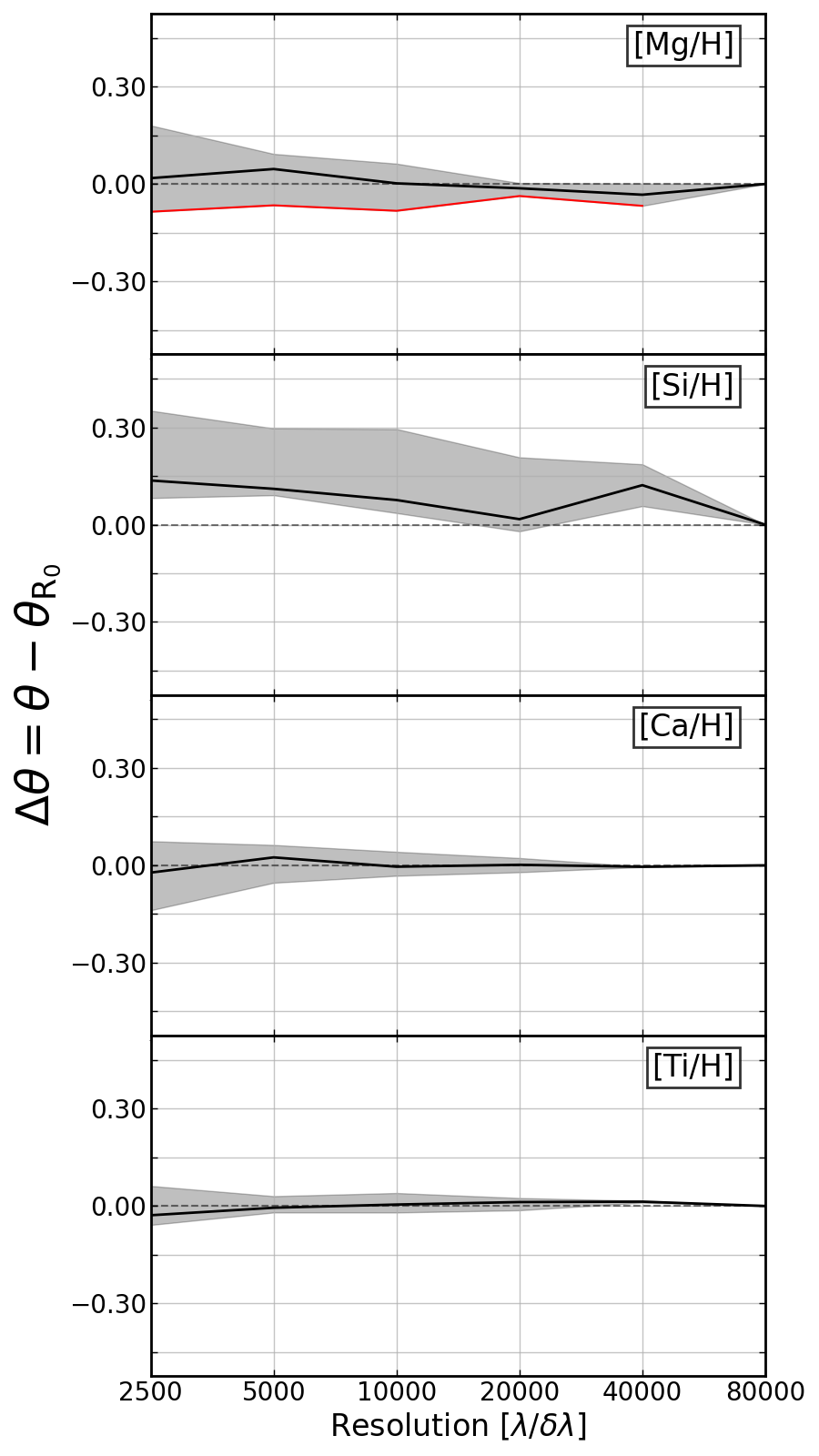}
    \caption{
        Same as Figure \ref{fig:4_atm1_v_res} except for $\alpha$ elements Mg, Si, Ca, and Ti (from top to bottom). We find the recovery of [Mg/H], [Ca/H], and [Ti/H] to be consistent as a function of resolution down to $R\sim2500$. Larger uncertainties on [Mg/H] are due to the masking of NLTE-sensitive lines. [Si/H] displays a substantial bias with resolution and large systematic uncertainties as a result of its strong dependence on the stellar atmospheric structure.
        \label{fig:4_alpha_v_res}
    }
\end{figure}

In Figure \ref{fig:4_alpha_v_res}, we present the change in the recovered abundance of $\alpha$ elements (Mg, Si, Ca, and Ti) as a function of resolution. In summary, we find that [Mg/H], [Ca/H], and [Ti/H] are recovered consistently with small to modest systematic uncertainties across all resolutions, while [Si/H] displays a substantial bias towards higher abundances at nearly all resolutions. We describe the results for each element in detail below. 

\paragraph{Magnesium}
We recover [Mg/H] consistently at all resolutions and find that the systematic uncertainty gradually increases to $\sim$0.15 dex as the resolution is decreased to $R\sim2500$. Systematic uncertainties on [Mg/H] are slightly underestimated for $R<40000$ measurements, because we recover only upper bounds ($\text{[Mg/H]}<-0.25$) for stars K731 and K969. 
We find sizeable scatter ($\sim$0.1--0.2 dex) between the [Mg/H] measured from repeat observation, which contributes to the systematic uncertainty seen for the stacked measurements. This is due to the fact that most of the strong Mg lines in the spectrum exhibit substantial NLTE effects and are masked or down-weighted in the fit. As a result, the measurement of [Mg/H] relies more heavily on weaker Mg lines and indirect information scattered throughout the spectrum. 

\paragraph{Silicon}
We recover [Si/H] to be 0.1--0.15 dex larger at nearly all resolutions smaller than the default resolution. Similarly sized systematic uncertainties are present as well, though they are positively skewed to even higher [Si/H] abundances. Combined with the mixed agreement to literature [Si/H] measurements (see Appendix \ref{sec:lit_by_element}), this suggests that substantial model inaccuracies exist. Indeed, much of the spectral information for Si is indirectly accessible through absorption lines of other elements, which are not modeled sufficiently accurately in this work. As with [Mg/H], this reliance on indirect spectral features also adds $\sim$0.1--0.2 dex scatter between repeat observations of the same star.

\paragraph{Calcium}
We recover [Ca/H] consistently across the full range of resolutions analyzed. Systematic uncertainties increase gradually with decreasing resolution to $\sim$0.1 dex at $R\sim2500$. 

\paragraph{Titanium}
We recover [Ti/H] consistently across the full range of resolutions analyzed. Systematic uncertainties increase gradually with decreasing resolution to $\sim$0.05 dex at $R\sim2500$.

\subsubsection{C, N, O Abundances}
\label{sec:rec_vs_res_cno}
\begin{figure}[ht!]
    \epsscale{1.20}
    \plotone{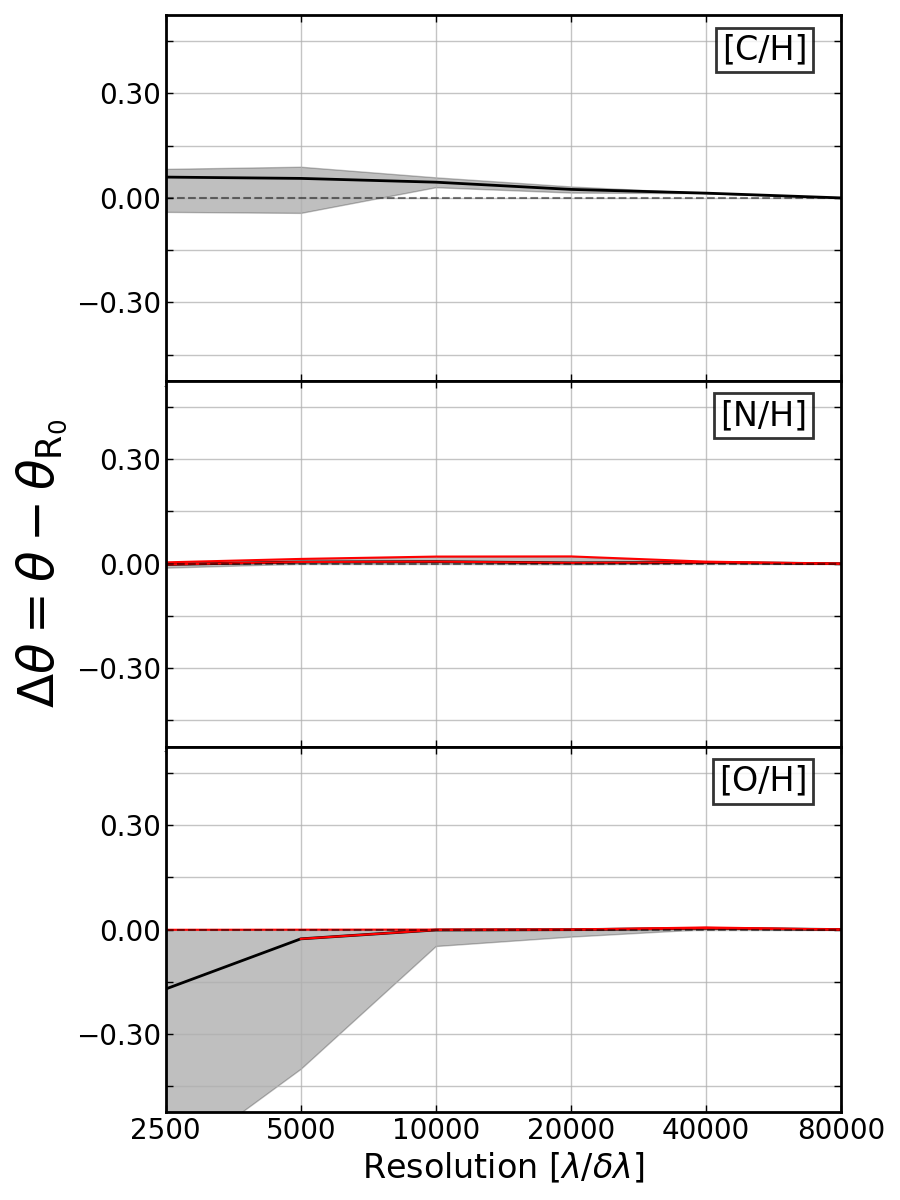}
    \caption{
        Same as Figure \ref{fig:4_atm1_v_res} except for C (top), N (middle), and O (bottom). We recover [C/H] consistently with small uncertainties at all resolutions. Resolution-dependent systematics are challenging to quantify for [N/H] and [O/H] due to the measurement of lower limits. For the U09H observations, the measurement of [N/H] appears consistent as a function of resolution. The measurement of [O/H] from these spectra is particularly challenging as most of the information comes indirectly from O's impact on the atmospheric structure.
        \label{fig:4_cno_v_res}
    }
\end{figure}

In Figure \ref{fig:4_cno_v_res}, we present the change in the recovered abundance of the light elements C, N, and O as a function of resolution. In summary, we find [C/H] and [N/H] to be recovered robustly and consistently at all resolutions, while the consistent recovery of [O/H] is more challenging. For [N/H] and [O/H] the boundary's of the training grid limit a complete picture of the bias and spread in abundance as a function of resolution. The recovery of each element is described in more detail below. 

\paragraph{Carbon}
While [C/H] recovery exhibits a small $\lesssim$0.05 dex positive bias for $R\lesssim10000$, it is largely consistent across the full range of resolutions. Systematic uncertainties increase gradually with decreasing resolution to $\sim$0.05 dex at $R\sim2500$. The small positive bias may be related to the strong negative correlation we find between [C/H] and [Fe/H] (see \S\ref{sec:crlb}). Despite their complicated nature, the C molecular features are fit well at all resolutions. This is reassuring given the large number of low-resolution searches for C-enhanced metal poor stars \citep[e.g.,][and references therein]{arentsen:2022}.

\paragraph{Nitrogen}
Because we recover lower limits on [N/H] ($\text{[N/H}>1.0$) for most of the stars in our sample, it is difficult to robustly quantify any resolution-dependent  systematic effects. For 4 stars with blue-optical U09H observations, K341, K386, K462, and K934, we do obtain constraints on [N/H] (i.e., not lower limits) at all resolutions. For all of these but K462, we recover [N/H] consistently (to better than $<0.05$ dex). In the case of K462, we recover [N/H] to be $\sim$0.15 dex lower at $R\sim2500$ than at the default resolution of $R\sim45000$, though the cause of this bias is challenging to diagnose. Given the presence of so many lower limits, we cannot rule out the presence of a positive bias, nor can we quantify a positive systematic uncertainty.

\paragraph{Oxygen}
Similar to [N/H], we recover only lower limits on [O/H] ($\text{[O/H}>1.0$) for the majority of stars, and thus cannot fully quantify the nature of resolution-dependent systematics on the measurement of [O/H]. Large scatter in the U09H observations towards lower [O/H] at low resolution lead to large ($>$0.3 dex) negative systematic uncertainties and a $\sim$0.15 dex bias below $R\sim10000$. This is likely because the vast majority of the O information is present only through indirect effects on C molecular features and changes to the atmospheric structure \citep[see][]{ting:2018b}. In the C147Hr and C316Hr observations, two O I lines are accessible at \wave\wave{6302.0, 6365.5}, but the former falls in a telluric mask and the later is very weak. Our inability to make conclusive statements regarding the recovery of [O/H] speaks to the challenge of measuring oxygen abundances from optical spectra---even at high resolution. 

\subsubsection{Light-Odd Element Abundances}
\label{sec:rec_vs_res_lightodd}
\begin{figure}[ht!]
    \epsscale{1.20}
    \plotone{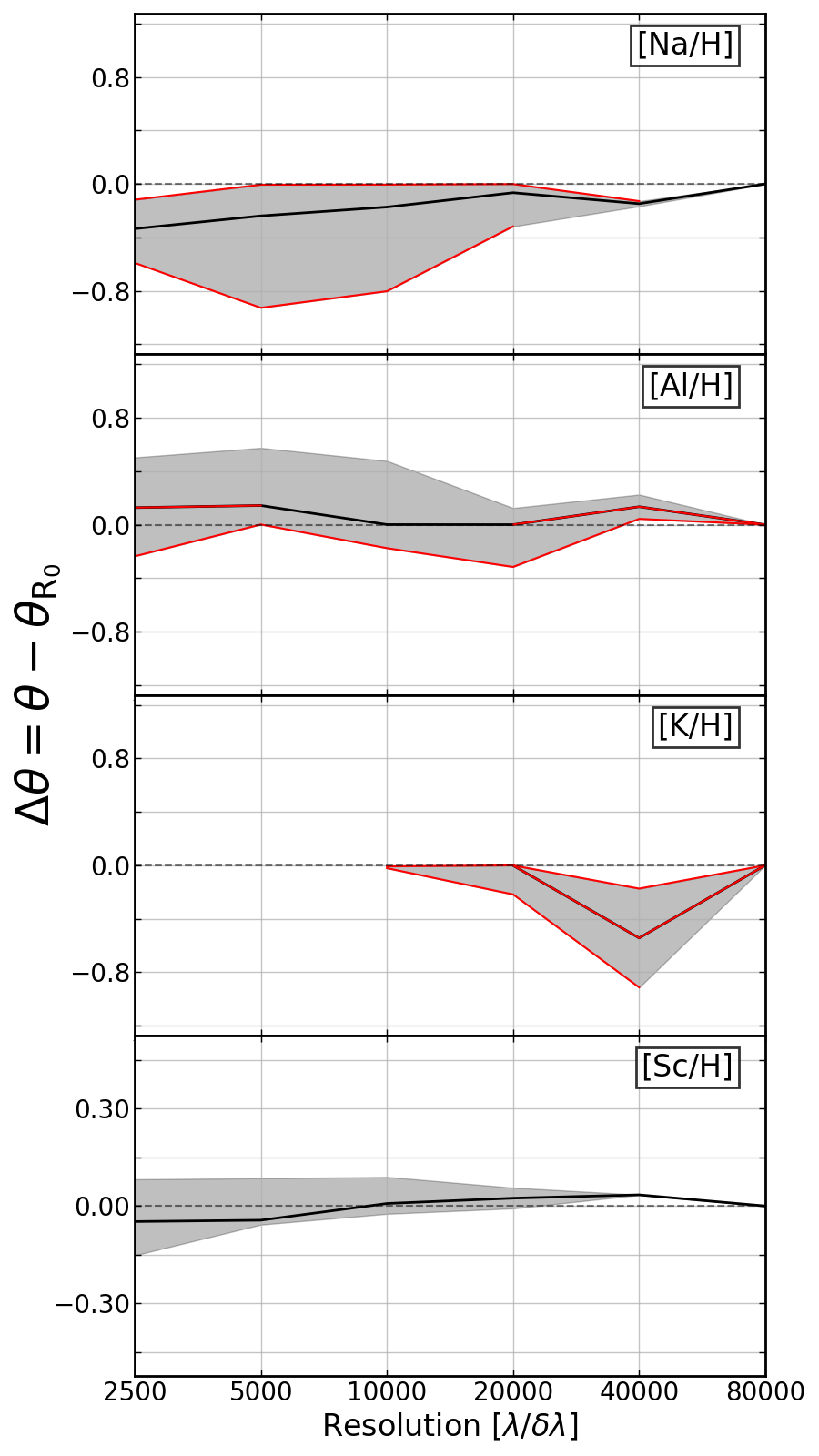}
    \caption{
        Same as Figure \ref{fig:4_atm1_v_res} except for Na, Al, K, and Sc (from top to bottom). We find poor and inconsistent recovery of [Na/H], [Al/H], and [K/H] at nearly all resolutions owing to the sparsity of absorption features. [Sc/H] is recovered consistently as a function of resolution with modest systematic uncertainties. 
        \label{fig:4_lightodd_v_res}
    }
\end{figure}

In Figure \ref{fig:4_lightodd_v_res}, we present the change in the recovered abundance of light-odd elements (Na, Al, K, and Sc) as a function of resolution. With the exception of [Sc/H], we find that light-odd elements are recovered quite poorly and inconsistently at nearly all resolutions. The recovery of each element is described in more detail below. 

\paragraph{Sodium}
We struggle to recover [Na/H] consistently at nearly all resolutions. The systematic bias towards lower values of [Na/H] brings some measurements into better agreement with the literature (e.g., K341 and K431) but also worsens the agreement of others (e.g., K462; see Appendix \ref{sec:lit_by_element}). The $\gtrsim0.4$ dex systematic uncertainty seen at $R\lesssim10000$ is characteristic of the large scatter seen in literature measurements for [Na/H], but the presence of both lower and upper limits in our sample mean that these already large systematic uncertainties are likely underestimated.
The challenge in recovering consistent [Na/H] is driven largely by the lack of good Na lines in the spectrum. The two strongest Na feature, the Na doublet at \wave\wave{5891.6, 5897.6}, falls entirely within telluric masks, and the three next-strongest lines at \wave\wave{4979.9, 4984.2, 5684.2, 5689.8} are all fairly weak. Two of these lines, those at \wave\wave{4979.9, 4984.2}, contribute strongly to the negative bias as they are in close proximity to a handful of poorly-fit Fe I lines.

\paragraph{Aluminum}
As with [Na/H], we find [Al/H] to be challenging to measure consistently for nearly all observations at all resolutions. The presence of both lower and upper limits in our sample mean that both the large systematic uncertainties ($\gtrsim$0.3 dex) and the $\sim$0.1 dex systematic bias are likely underestimated. Like Na, the challenge in recovering consistent [Al/H] is due to the lack of good Al lines in the data. The two strongest Al features, the Al I lines at \wave\wave{3945.1, 3962.6}, are lost to the Ca H\&K mask. The remaining Al information is either indirect (mainly through the CN bands) or a handful of very weak Al I lines.

\paragraph{Potassium}
The recovery of [K/H], like the recovery of [Na/H] and [Al/H], is challenging at all resolutions. Very few measurements fall within the bounds of our training set ($-0.25\leq\text{[K/Fe]}\leq1.0$), making it impossible to quantify the true impact of resolution-dependent systematics. This is due to the extreme paucity of K lines in the observed spectra. The most prominent potassium feature, the K I line at \wave{7701.1}, is a part of the telluric mask. The remaining K features at \wave\wave{4045.3, 4048.4} are very weak, and are further down-weighted by NLTE masks. This lack of K lines prevents any measurement of [K/H] to better precision than 0.5 dex below $R<10000$.

\paragraph{Scandium}
Unlike for the other light-odd elements, we recover [Sc/H] consistently down to $R\sim10000$ and with only a small ($\sim$0.05 dex) systematic negative bias at lower resolutions. The systematic uncertainty gradually increases with decreasing resolution to $\sim$0.1 dex at $R\sim2500$. In contrast to the dearth of Na, Al, and K lines, there are $\sim$40 Sc lines contained in the archival spectra, enabling robust [Sc/H] measurements at all resolutions. Blending with neighboring imperfectly modelled lines is responsible for the small systematic uncertainty and bias at low resolution. 

\subsubsection{Neutron-Capture Element Abundances}
\label{sec:rec_vs_res_neutron}
\begin{figure}[ht!]
    \epsscale{1.20}
    \plotone{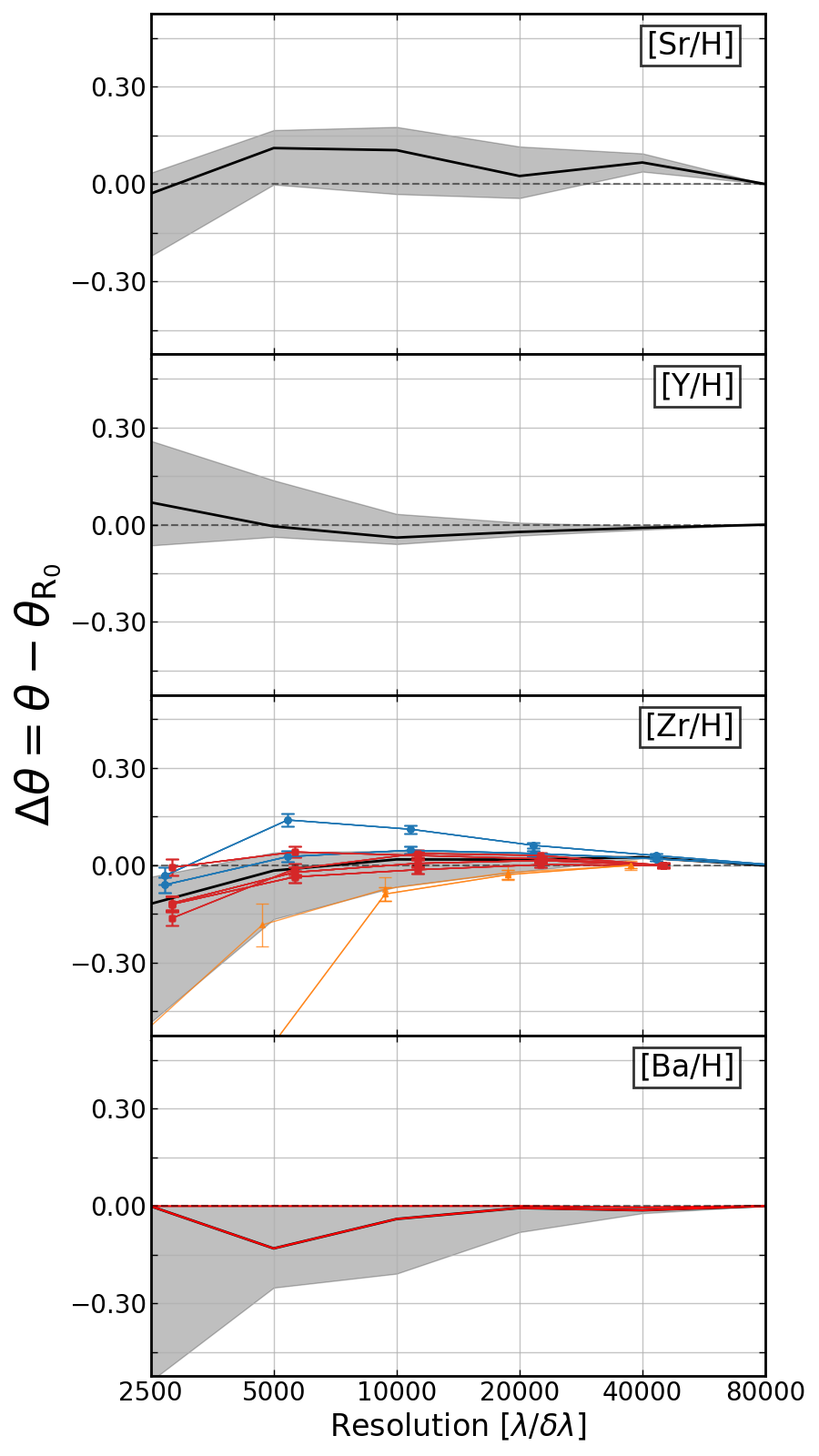}
    \caption{
        Same as Figure \ref{fig:4_atm1_v_res} except for neutron-capture elements Sr, Y, Zr, and Ba (from top to bottom). We recover [Y/H] and [Zr/H] consistently down to $R\sim5000$ and with small positive and negative biases respectively at $R\sim2500$. We recover [Sr/H] with a slight positive bias at lower resolution. The presence of lower limits prevents the robust quantification of systematics for [Ba/H]. The measurement of both [Sr/H] and [Ba/H] suffer from substantial NLTE effects and hyperfine splitting in their strong resonance lines.
        \label{fig:4_neutron1_v_res}
    }
\end{figure}

\begin{figure}[ht!]
    \epsscale{1.20}
    \plotone{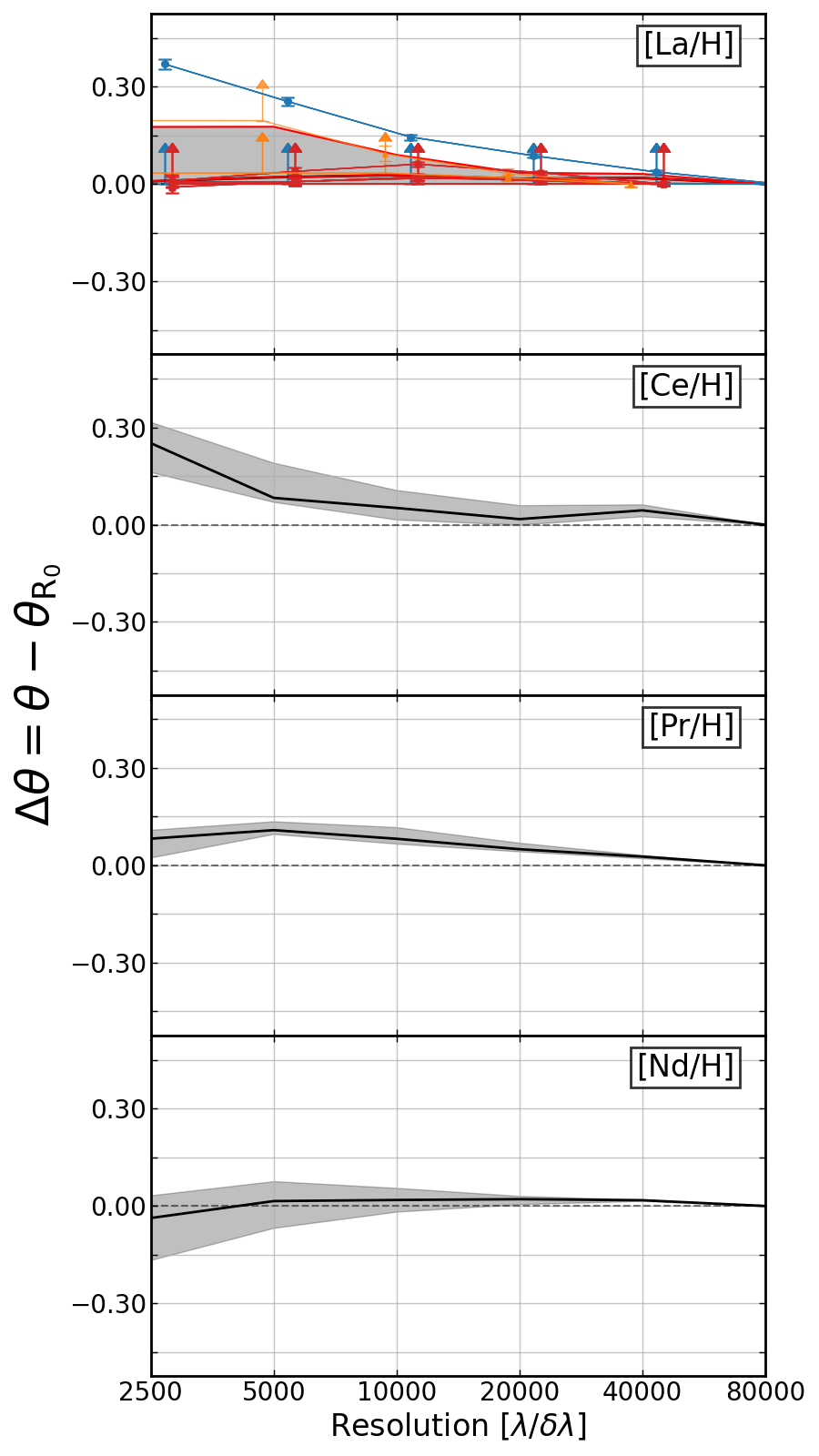}
    \caption{
        Same as Figure \ref{fig:4_atm1_v_res} except for neutron-capture elements La, Ce, Pr, and Nd (from top to bottom). We find a strong resolution dependence for the recovery of [Ce/H] and [Pr/H] as a result of blending in the feature-dense region around 4000 \AA. The recovery of [La/H] appears similarly biased though the presence of lower limits prevents robust quantification of the systematic effects. [Nd/H] is recovered consistently as a function of resolution with modest systematic uncertainties. 
        \label{fig:4_neutron2_v_res}
    }
\end{figure}

\begin{figure}[ht!]
    \epsscale{1.20}
    \plotone{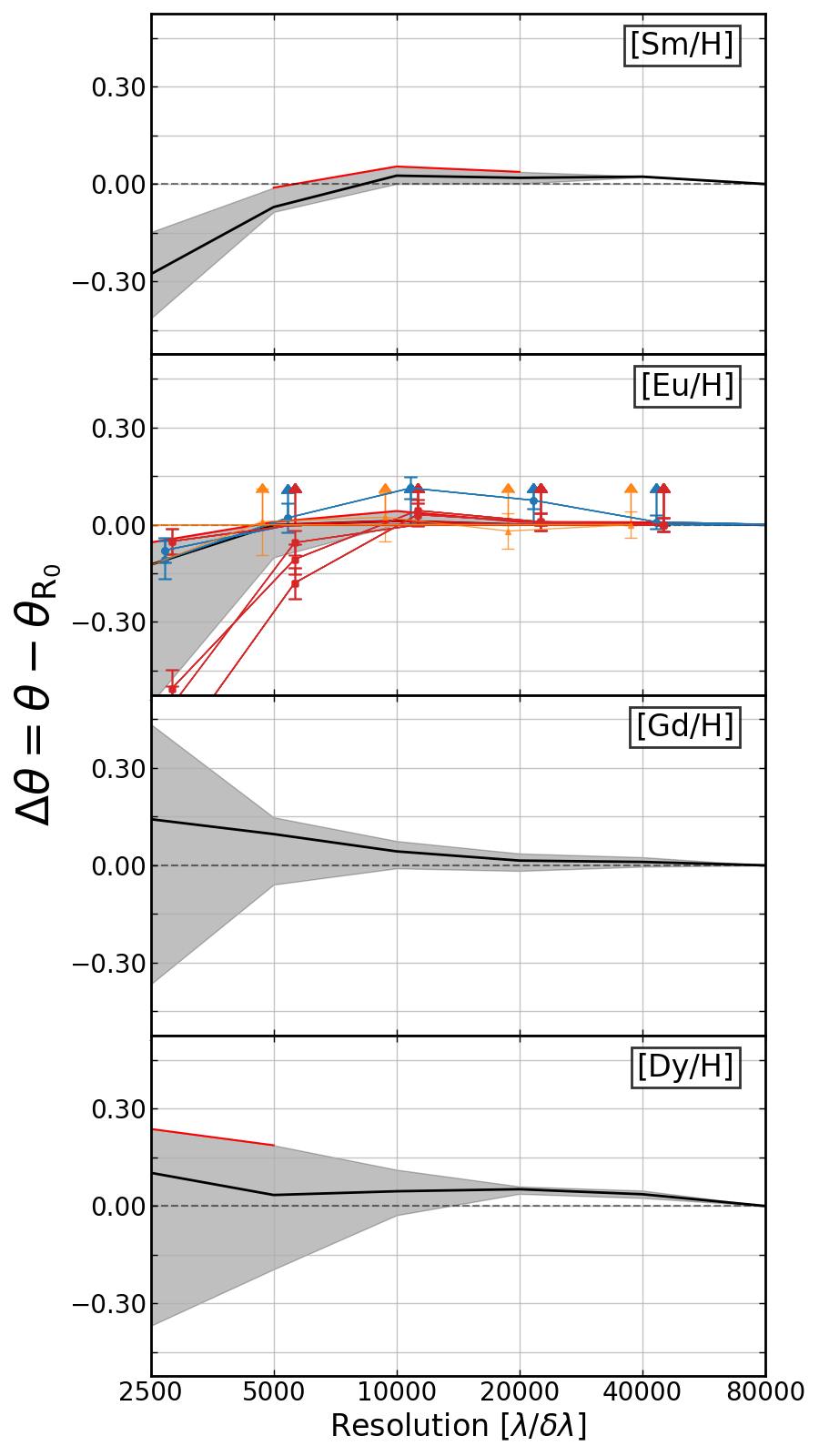}
    \caption{
        Same as Figure \ref{fig:4_atm1_v_res} except for neutron-capture elements Sm, Eu, Gd, and Dy (from top to bottom). The recovery of [Gd/H] and [Dy/H] appears consistent, albeit with rapidly increasing uncertainties down to $R\sim2500$. We find that [Sm/H] is well recovered down to $R\sim10000$, but exhibits increasing bias at lower resolutions. The measurement of [Gd/H], [Dy/H], and [Sm/H] are all characterized by fitting many very weak lines in crowded regions of the stellar spectrum. The presence of lower limits for [Eu/H] prevents robust quantification of the resolution-dependent systematic effects on its recovery. [Eu/H] also suffers from substantial NLTE effects and hyperfine splitting in their strong resonance lines.
        \label{fig:4_neutron3_v_res}
    }
\end{figure}

In Figures \ref{fig:4_neutron1_v_res}--\ref{fig:4_neutron3_v_res}, we present the change in the recovered abundance of neutron-capture elements (Sr, Y, Zr, Ba, La, Ce, Pr, Nd, Sm, Eu, Gd, Dy, Ho, Er, Os, and Th) as a function of resolution. In  summary, we find [Sr/H], [Y/H], [Zr/H], [Nd/H], [Sm/H], [Gd/H], [Dy/H], and [Th/H] to be recovered consistently down to at least $R\sim10000$, though a subset show large uncertainties or noticeable biases at the lowest resolutions. We find substantial biases with resolution for [Ce/H], [Pr/H], and [Er/H]. For [Ba/H], [La/H], and [Eu/H], the boundaries of the training grid limit a complete picture of the resolution-dependent systematic bias and uncertainties. [Ho/H] and [Os/H] prove challenging to recover at all. The recovery of each element is described in more detail below. 

\paragraph{Strontium}
While the recovery of [Sr/H] is slightly biased by $\sim$0.1 dex at a few individual resolutions, it it consistent to within the 0.1--0.2 dex systematic uncertainties for $R\lesssim20000$. Constraints on [Sr/H] come primarily from the two strong Sr II resonance lines at \wave\wave{4078.9, 4216.7} and secondarily from two weak lines at \wave\wave{4163.0, 4608.6}. Blending, saturation, and NLTE effects in the strong lines all contribute to sizeable ($\sim$0.2 dex) scatter in the measured [Sr/H] from repeat observations at all resolutions.

\paragraph{Yttrium}
We find the recovery of [Y/H] to be largely consistent as a function of resolution. A small $\lesssim$0.05 (0.10) dex negative (positive) bias is seen for $R\sim10000$ (2500), but this is within the systematic uncertainty, which grows gradually as the resolution is decreased to $\sim$0.15 dex at $R\sim2500$. Roughly 40 Y lines between 4100 and 5500 \AA\ contribute to the robust measurement of [Y/H]. Blending of these lines with neighboring lines with small errors is responsible for the small systematic uncertainty and bias at low resolution.

\paragraph{Zirconium}
We recover [Zr/H] consistently for all resolutions $R\gtrsim5000$ and biased by $\sim$0.1 dex lower values at $R\sim2500$. Positive systematic uncertainties grow gradually to $\sim$0.1 dex at $R\sim2500$, while negative systematic uncertainties grow gradually to $\sim$0.3 dex. Both the negative bias and substantially larger negative systematic uncertainties are driven by the measurements from the C147Hr observations, for which we recover [Zr/H] to be as much as $\sim$0.6 dex smaller at $R\sim2500$ than at the default resolution. The bias in the C147Hr measurements appears to be driven by a combination of blending and a poorly approximated continuum shape.

\paragraph{Barium}
For over half of the stars in our sample, we recover lower limits on [Ba/H] ($\text{[Ba/Fe]}>0.5$), which obfuscate the complete picture of resolution-dependent systematics. As we decrease the resolution, we do see increasingly large negative uncertainties---up to 0.45 dex at $R\sim2500$. This result, along with the poor agreement with literature [Ba/H] measurements (see Appendix \ref{sec:lit_by_element}) suggest that there are some quite substantial inaccuracies in our model spectrum's Ba features. Indeed, line saturation, NLTE effects, and hyperfine splitting are all at play in the strongest optical Ba lines \citep[e.g.,][]{eitner:2019}.

\paragraph{Lanthanum}
The recovery of [La/H] displays a biased towards larger values at lower resolutions, though the extent of this bias is unknown due to the boundary of our model grid. Lower limits ($\text{[La/Fe]} > 0.5$ dex) are recovered for the majority of stars. Similarly, the systematic uncertainty on [La/H] grows to at least $\sim$0.15 dex as we decrease the resolution to $R\sim2500$.
This bias is predominantly driven by measurements made with the C147Hr and C316Hr observations, which are biased by as much $0.35$ dex as a result of blending in a few important lines at longer wavelengths (\wave\wave{4663.8, 4922.4, 4923.2, 5124.4, 6392.3}). The U09H observations, on the other, yield quite consistent (to $\lesssim$0.05 dex) [La/H] measurements across all resolutions.

\paragraph{Cerium}
We find a growing systematic bias towards higher [Ce/H] as the resolution is decreased. At $R\sim2500$ ($R\sim5000$), we measure [Ce/H] $\sim$0.25 (0.08) dex higher than we do at the default resolutions. Despite the substantial bias, systematic uncertainties remain small ($\lesssim0.05$ dex for $R\gtrsim10000$ and $\sim$0.1 dex for $R\lesssim5000$). While one would expect the large number (100's--1000's) of Ce lines present in these spectra to yield robust [Ce/H] measurements regardless of resolution, a closer inspection of the spectra reveals that the majority of these Ce lines reside between 3800 and 4600 \AA\ among a high density of other lines, including complex molecular absorption bands. As a result, the impact of blending in this portion of the spectrum is especially large. When the resolution decreases, [Ce/H] increases to compensate for missing and underestimated lines.

\paragraph{Praseodymium}
We find a similar, albeit smaller, systematic bias with resolution for [Pr/H] recovery as we do for [Ce/H]. At $R\sim20000$, we recover [Pr/H] to be 0.05 dex larger than at higher resolutions. At lower resolutions, this bias increases slightly to $\sim$0.1 dex. Systematic uncertainties remain small ($\lesssim$0.05 dex) across all resolutions. As with Ce, blending, especially in the region between 4000 and 4100 \AA, is source of the systematic bias.

\paragraph{Neodymium}
We recover [Nd/H] consistently  across the full range of resolutions analyzed. Systematic uncertainties increase gradually with decreasing resolution to $\sim$0.1 dex at $R\sim2500$. While Nd has a similar number of lines as Ce and Pr, these lines are more broadly distributed throughout the spectrum. As a result, the recovery of [Nd/H] is less susceptible to the impact of blending in the blue-optical.

\paragraph{Samarium}
We recover [Sm/H] consistently at resolutions above $R\gtrsim10000$. At lower resolutions, a systematic negative bias develops and grows to $0.25$ dex at $R\sim2500$. Systematic uncertainties grow to as large as $\sim$0.2 dex, though these may be underestimated due to lower limits recovered at our model grid boundary ($\text{[Sm/Fe]} > 1$). 
Upon visual inspection, the ability of our spectral model to fit the many Sm lines present in the data is quite mixed---some are fit well, others are overestimated, and others still are underestimated. We believe the source of the bias at the lowest resolutions is due to the dominance of a few of the stronger lines, namely at \wave\wave{4069.5, 4108.4, 4156.4, 4204.2, 4468.6}, which are overestimated at the default resolution. 

\paragraph{Europium}
The recovery of [Eu/H] as a function of resolution, like that of [Sm/H], is obscured by lower limits at the boundary of our model grid ($\text{[Eu/Fe]}>1$). At the lowest resolution, we find a systematic $\sim$0.15 dex bias towards lower values of [Eu/H] and a large $\sim$0.4 dex systematic uncertainty. This bias is most pronounced for measurements of [Eu/H] from U09H observations. This result, along with the poor agreement with literature [Eu/H] measurements (see Appendix \ref{sec:lit_by_element}) suggest that there are some quite substantial inaccuracies in our model spectrum's Eu features. As with Ba, line saturation, NLTE effects, and hyperfine splitting are all at play in the strongest optical Eu lines \citep{Mashonkina:2000}.

\paragraph{Gadolinium}
We recover [Gd/H] consistently at all resolutions when the systematic uncertainty is taken into account. As the resolution is decreased to $R\sim5000$ (2500), a systematic bias towards higher [Gd/H] grows to $\sim$0.1 (0.25) dex, while the systematic uncertainty grows at a larger rate, reaching $\sim$0.15 (0.55) dex. The increasing systematic uncertainty is driven by blending of complicated and imperfectly modeled absorption features with the $\sim$100 weak Gd lines present bluewards of 4500 \AA.

\paragraph{Dysprosium}
The recovery of [Dy/H] is largely consistent across all resolutions, though a small positive bias of $\sim$0.05 dex is seen at intermediate resolutions ($5000\lesssim R\lesssim40000$). The systematic uncertainty in [Dy/H] recovery grows steadily with decreasing resolution to $\sim$0.4 dex at $R\sim2500$. As for Gd, the increasing systematic uncertainty is driven by blending of complicated and imperfectly modeled absorption features with the $\sim$100 weak Dy lines present bluewards of 4500 \AA.

\begin{figure}[ht!]
    \epsscale{1.20}
    \plotone{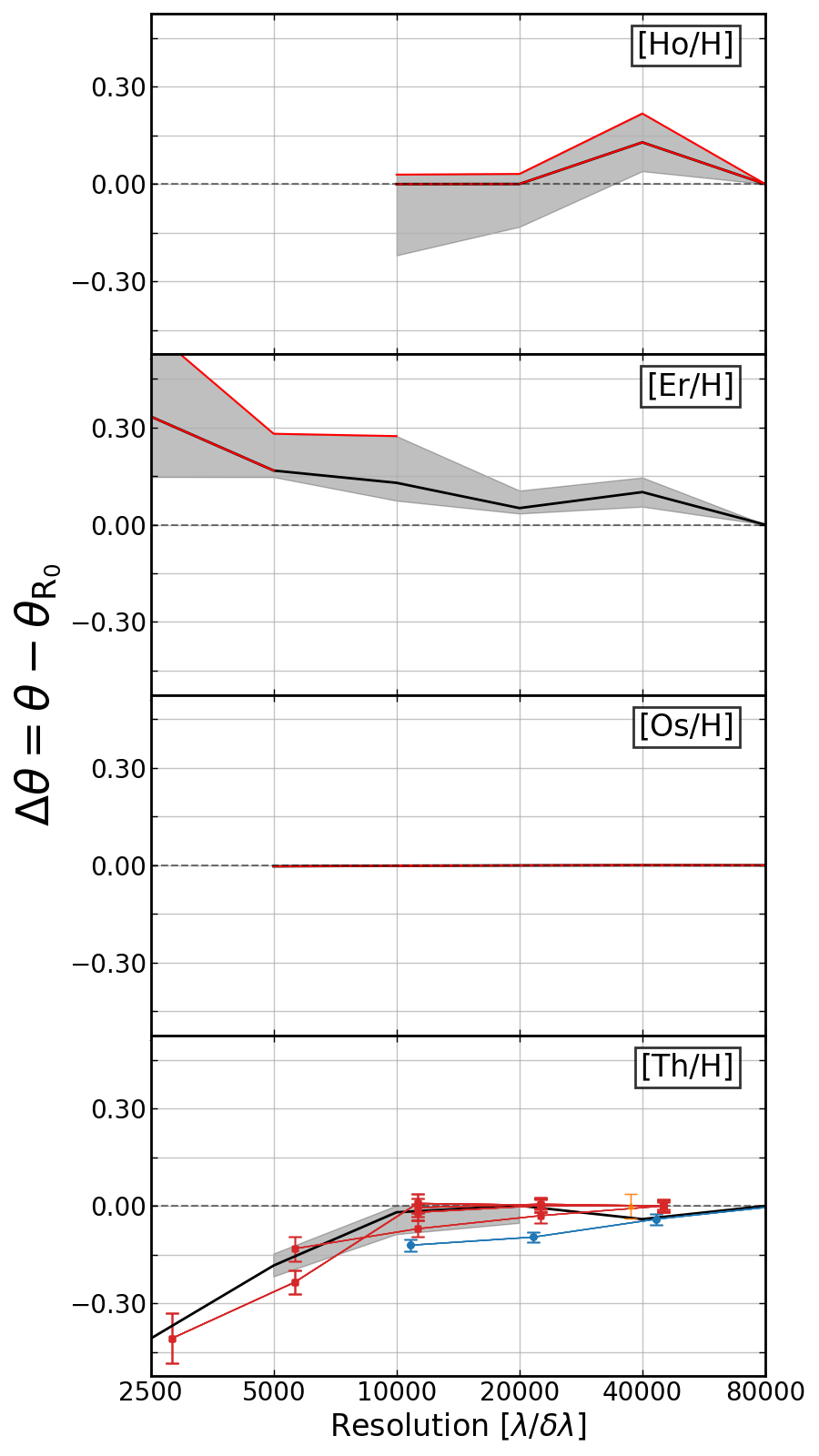}
    \caption{
        Same as Figure \ref{fig:4_atm1_v_res} except for neutron-capture elements Ho, Er, Os, and Th (from top to bottom). The recovery of [Ho/H], [Er/H], [Os/H], and [Th/H] are all made difficult due to a paucity of absorption lines in the observed spectra. Little can be said about the resolution-dependent measurement of [Ho/H] and [Os/H]. The recovery of [Er/H] exhibits a substantial bias towards larger values as the resolution is decreased. For the few instances in which [Th/H] can be recovered below $R\sim10000$, a negative bias is apparent.
        \label{fig:4_neutron4_v_res}
    }
\end{figure}

\paragraph{Holmium}
For the majority of stars in our sample, we recover [Ho/H] as lower limits at the boundary of our model ($\text{[Ho/Fe]} > 1$). As such, the impact of resolution-induced systematics is hard to robustly quantify. Only five Ho lines are present in our spectra and all are either weak or in poorly-fit portions of the spectrum. This leads to large 0.15--0.4 dex scatter from repeat observations in the three stars (K431, K462, and K969) where [Ho/H] is constrained within the prior bounds and the inability to measure [Ho/H] to better precision than 0.5 dex at resolutions below $R\lesssim10000$.

\paragraph{Erbium}
We find that the recovery of [Er/H] is biased high by $>$0.1 dex at all resolutions lower than the default. As the resolution is decreased to $R\sim2500$, the bias increases steadily to at least 0.3 dex. The positively skewed systematic uncertainty increases with decreasing resolution to $\gtrsim$0.2 dex. Due to the recovery of several lower limits ($\text{[Er/Fe]} > 1.0$), the bias and uncertainties for $R\lesssim5000$ may be underestimated. Compared to other neutron-capture elements, Er has far fewer lines in the observed spectra, and the lines that do exist are quite weak and blended. As a result, the recovery of [Er/H] is quite sensitive to model fidelity at all but the highest resolutions.

\paragraph{Osmium}
Owing to the lack of good Os absorption lines in the archival spectra, we struggle at all resolutions to recover [Os/H] within the bounds of our training set ($\text{[Os/Fe]} = 1.0$). This precludes any robust quantification of the systematic bias and uncertainty. 

\paragraph{Thorium}
We find [Th/H] to be consistently recovered for $R\gtrsim10000$. At lower resolutions, we measure [Th/H] to be $\sim$0.4 dex smaller. Systematic uncertainties are $\sim$0.05 dex, although the limited number of stars for which we can measure [Th/H], especially at low resolution, adds makes the uncertainty difficult to quantify across the full resolution range. Our ability to recover Th is limited by the small handful of Th II lines detectable in the observed spectra. Of the three strongest lines, \wave\wave{3676.6, 3742.2, 4020.3}, only the last is contained within spectral coverage of the C147Hr and C316Hr observations. Nearly all Th II lines are substantially impacted by blends at lower resolutions, leading to the observed bias.

\subsection{Label Recovery as a Function of Signal/Noise}
\label{sec:Abundance_v_SNR}
Here we present the change in the recovered stellar parameters as a function of S/N for our sample of stars fit at $R\sim10000$. Similar to the presentation in \S\ref{sec:Abundance_v_Resolution}, the change in stellar parameters, $\Delta\theta$, is reported relative to a fiducial measurement, in this case, the value recovered at the native S/N with the same resolution ($\Delta\theta=\theta_{\sigma}-\theta_{\sigma_0}$). In this analysis, we consider stellar label measurements from individual exposures rather than from the stacked posteriors so that they can be more easily mapped to a median S/N. The results of this analysis for each element are presented in Table \ref{tab:abundance_v_snr}.

\begin{deluxetable*}{cCCCCCC}
	\centerwidetable
	\caption{Trends in Stellar Label Recovery with S/N}
	\label{tab:abundance_v_snr}
	\tablehead{
         &
        \colhead{$\text{S/N}\sim5$} &
        \colhead{$\text{S/N}\sim10$} &
        \colhead{$\text{S/N}\sim20$} &
        \colhead{$\text{S/N}\sim40$} &
        \colhead{$\text{S/N}\sim75$} \\
        \colhead{$\theta$} &
        \colhead{$\Delta\theta\pm_{\sigma_\text{16th}}^{\sigma_\text{84th}}\pm\sigma_\text{stat}$} &
        \colhead{$\Delta\theta\pm_{\sigma_\text{16th}}^{\sigma_\text{84th}}\pm\sigma_\text{stat}$} &
        \colhead{$\Delta\theta\pm_{\sigma_\text{16th}}^{\sigma_\text{84th}}\pm\sigma_\text{stat}$} &
        \colhead{$\Delta\theta\pm_{\sigma_\text{16th}}^{\sigma_\text{84th}}\pm\sigma_\text{stat}$} &
        \colhead{$\Delta\theta\pm_{\sigma_\text{16th}}^{\sigma_\text{84th}}\pm\sigma_\text{stat}$} 
	}
	\startdata		
		\tefftext   & 0.38\pm_{0.79}^{1.04}\pm1.44          & -0.16\pm_{0.55}^{0.51}\pm0.99         & -0.41\pm_{0.51}^{0.19}\pm0.55         & -0.09\pm_{0.45}^{0.14}\pm0.30         & 0.00\pm_{0.00}^{0.03}\pm0.20          \\
		\loggtext   & -0.00\pm_{0.02}^{0.01}\pm0.01         & 0.00\pm_{0.01}^{0.01}\pm0.01          & 0.00\pm_{0.00}^{0.00}\pm0.00          & 0.00\pm_{0.00}^{0.00}\pm0.00          & 0.00\pm_{0.00}^{0.00}\pm0.00          \\
		\vmicrotext & 0.16^{*}\pm_{0.11}^{0.28^{*}}\pm0.22  & 0.11\pm_{0.06}^{0.18^{*}}\pm0.13      & 0.06\pm_{0.07}^{0.06^{*}}\pm0.07      & 0.01\pm_{0.03}^{0.05}\pm0.04          & 0.00\pm_{0.00}^{0.00}\pm0.02          \\
		\vmacrotext & 4.67\pm_{4.73}^{0.43}\pm0.83          & 4.42\pm_{4.08}^{0.36}\pm0.42          & 4.30\pm_{4.07}^{0.23}\pm0.22          & 4.11\pm_{4.11}^{0.21}\pm0.12          & 0.00\pm_{0.00}^{0.00}\pm0.01          \\
		$v_r$       & -0.22\pm_{0.12}^{1.17}\pm0.51         & -0.09\pm_{0.07}^{0.58}\pm0.26         & -0.05\pm_{0.03}^{0.25}\pm0.14         & -0.01\pm_{0.03}^{0.09}\pm0.08         & 0.00\pm_{0.00}^{0.03}\pm0.06          \\
		{[}C/H{]}   & -0.11\pm_{0.09}^{0.09}\pm0.10         & -0.10\pm_{0.13}^{0.07}\pm0.07         & -0.07\pm_{0.11}^{0.06}\pm0.05         & -0.02\pm_{0.09}^{0.02}\pm0.03         & 0.00\pm_{0.00}^{0.01}\pm0.01          \\
		{[}N/H{]}   & 0.05^{*}\pm_{0.15}^{0.02^{*}}\pm0.71  & -0.00^{*}\pm_{0.09}^{0.00^{*}}\pm0.19 & -0.03^{*}\pm_{0.07}^{0.03^{*}}\pm0.08 & -0.01^{*}\pm_{0.01}^{0.01^{*}}\pm0.04 & 0.00^{*}\pm_{0.00}^{0.00^{*}}\pm0.02  \\
		{[}O/H{]}   & \nodata                               & -0.07\pm_{0.41}^{0.01}\pm0.14         & -0.09\pm_{0.47}^{0.12^{*}}\pm0.22     & -0.00\pm_{0.29}^{0.01^{*}}\pm0.06     & 0.00^{*}\pm_{0.00}^{0.00^{*}}\pm0.04  \\
		{[}Na/H{]}  & \nodata                               & \nodata                               & -0.11^{*}\pm_{0.01}^{0.01^{*}}\pm0.36 & 0.00\pm_{0.08^{*}}^{0.13^{*}}\pm0.31  & 0.00\pm_{0.00^{*}}^{0.14^{*}}\pm0.20  \\
		{[}Mg/H{]}  & 0.45^{*}\pm_{0.14}^{0.19^{*}}\pm0.35  & 0.14\pm_{0.15}^{0.15}\pm0.20          & 0.01\pm_{0.10}^{0.22}\pm0.13          & 0.01\pm_{0.05}^{0.12}\pm0.08          & 0.00\pm_{0.00}^{0.04}\pm0.05          \\
		{[}Al/H{]}  & \nodata                               & \nodata                               & 0.04^{*}\pm_{0.11^{*}}^{0.10}\pm0.42  & 0.00^{*}\pm_{0.30^{*}}^{0.07}\pm0.31  & 0.00\pm_{0.04^{*}}^{0.00}\pm0.17      \\
		{[}Si/H{]}  & 0.13^{*}\pm_{0.08}^{0.09^{*}}\pm0.35  & 0.03\pm_{0.06}^{0.06^{*}}\pm0.12      & -0.02\pm_{0.04}^{0.04}\pm0.06         & -0.01\pm_{0.03}^{0.02}\pm0.03         & 0.00\pm_{0.00}^{0.00}\pm0.02          \\
		{[}K/H{]}   & \nodata                               & \nodata                               & \nodata                               & \nodata                               & 0.00^{*}\pm_{0.00^{*}}^{0.00^{*}}\pm0.30\\
		{[}Ca/H{]}  & -0.28^{*}\pm_{0.07^{*}}^{0.05}\pm0.07 & -0.28^{*}\pm_{0.07^{*}}^{0.06}\pm0.05 & -0.24^{*}\pm_{0.08^{*}}^{0.14}\pm0.05 & -0.13\pm_{0.07}^{0.15}\pm0.05         & 0.00\pm_{0.00}^{0.03}\pm0.03          \\
		{[}Sc/H{]}  & -0.06\pm_{0.22^{*}}^{0.04}\pm0.31     & -0.08\pm_{0.12}^{0.03}\pm0.17         & -0.06\pm_{0.04}^{0.02}\pm0.08         & -0.03\pm_{0.02}^{0.03}\pm0.05         & 0.00\pm_{0.02}^{0.00}\pm0.03          \\
		{[}Ti/H{]}  & 0.10\pm_{0.16}^{0.03}\pm0.08          & 0.03\pm_{0.06}^{0.02}\pm0.05          & 0.01\pm_{0.02}^{0.01}\pm0.03          & 0.00\pm_{0.01}^{0.01}\pm0.02          & 0.00\pm_{0.00}^{0.00}\pm0.01          \\
		{[}V/H{]}   & 0.13\pm_{0.04}^{0.03}\pm0.20          & 0.05\pm_{0.11}^{0.06}\pm0.11          & 0.01\pm_{0.06}^{0.04}\pm0.06          & -0.00\pm_{0.02}^{0.01}\pm0.03         & 0.00\pm_{0.01}^{0.00}\pm0.02          \\
		{[}Cr/H{]}  & 0.20\pm_{0.04}^{0.05}\pm0.21          & 0.09\pm_{0.03}^{0.03}\pm0.12          & 0.03\pm_{0.02}^{0.02}\pm0.07          & 0.01\pm_{0.02}^{0.01}\pm0.04          & 0.00\pm_{0.00}^{0.00}\pm0.03          \\
		{[}Mn/H{]}  & 0.17^{*}\pm_{0.14^{*}}^{0.11}\pm0.25  & 0.04^{*}\pm_{0.10^{*}}^{0.06}\pm0.15  & 0.00^{*}\pm_{0.10^{*}}^{0.02}\pm0.10  & -0.00^{*}\pm_{0.12^{*}}^{0.01}\pm0.09 & 0.00^{*}\pm_{0.05^{*}}^{0.00}\pm0.08  \\
		{[}Fe/H{]}  & -0.01\pm_{0.04}^{0.02}\pm0.04         & 0.00\pm_{0.03}^{0.02}\pm0.03          & 0.01\pm_{0.01}^{0.01}\pm0.02          & 0.00\pm_{0.01}^{0.01}\pm0.01          & 0.00\pm_{0.00}^{0.00}\pm0.01          \\
		{[}Co/H{]}  & -0.01\pm_{0.02}^{0.02^{*}}\pm0.31     & 0.00\pm_{0.04}^{0.02}\pm0.14          & -0.01\pm_{0.13}^{0.02}\pm0.07         & -0.01\pm_{0.10}^{0.02}\pm0.04         & 0.00\pm_{0.06}^{0.00}\pm0.02          \\
		{[}Ni/H{]}  & -0.05\pm_{0.04^{*}}^{0.05}\pm0.19     & -0.06\pm_{0.03}^{0.03}\pm0.12         & -0.05\pm_{0.02}^{0.03}\pm0.07         & -0.01\pm_{0.01}^{0.01}\pm0.04         & 0.00\pm_{0.00}^{0.01}\pm0.03          \\
		{[}Cu/H{]}  & \nodata                               & 0.00^{*}\pm_{0.00^{*}}^{0.00}\pm0.12  & 0.00^{*}\pm_{0.00^{*}}^{0.00}\pm0.14  & -0.00^{*}\pm_{0.30^{*}}^{0.00}\pm0.06 & 0.00^{*}\pm_{0.23^{*}}^{0.00}\pm0.05  \\
		{[}Zn/H{]}  & \nodata                               & -0.06\pm_{0.05^{*}}^{0.16}\pm0.38     & -0.06\pm_{0.16}^{0.14}\pm0.27         & 0.00\pm_{0.12}^{0.11^{*}}\pm0.17      & 0.00\pm_{0.00}^{0.08}\pm0.12          \\
		{[}Ga/H{]}  & \nodata                               & \nodata                               & \nodata                               & \nodata                               & 0.00^{*}\pm_{0.00}^{0.00^{*}}\pm0.05  \\
		{[}Sr/H{]}  & \nodata                               & -0.10^{*}\pm_{0.14^{*}}^{0.03^{*}}\pm0.22& -0.18\pm_{0.17^{*}}^{0.18^{*}}\pm0.20 & -0.03\pm_{0.12}^{0.03^{*}}\pm0.13     & 0.00\pm_{0.00}^{0.00^{*}}\pm0.09      \\
		{[}Y/H{]}   & 0.09\pm_{0.10}^{0.36^{*}}\pm0.24      & 0.02\pm_{0.04}^{0.18}\pm0.15          & -0.01\pm_{0.03}^{0.10}\pm0.08         & -0.01\pm_{0.02}^{0.03}\pm0.05         & 0.00\pm_{0.00}^{0.02}\pm0.04          \\
		{[}Zr/H{]}  & -0.05\pm_{0.14}^{0.12}\pm0.28         & 0.02\pm_{0.14}^{0.09^{*}}\pm0.18      & 0.01\pm_{0.10}^{0.06^{*}}\pm0.10      & 0.00\pm_{0.05}^{0.06}\pm0.06          & 0.00\pm_{0.00}^{0.05}\pm0.04          \\
		{[}Ba/H{]}  & \nodata                               & -0.14^{*}\pm_{0.53^{*}}^{0.10^{*}}\pm0.25& -0.16\pm_{0.52^{*}}^{0.16^{*}}\pm0.19 & -0.00\pm_{0.39}^{0.00^{*}}\pm0.11     & 0.00^{*}\pm_{0.00}^{0.00^{*}}\pm0.05  \\
		{[}La/H{]}  & -0.17^{*}\pm_{0.19}^{0.17^{*}}\pm0.35 & -0.10^{*}\pm_{0.22}^{0.12^{*}}\pm0.15 & -0.01^{*}\pm_{0.17}^{0.01^{*}}\pm0.07 & -0.00\pm_{0.09}^{0.00^{*}}\pm0.04     & 0.00\pm_{0.00}^{0.00^{*}}\pm0.03      \\
		{[}Ce/H{]}  & -0.16\pm_{0.09}^{0.07}\pm0.21         & -0.10\pm_{0.06^{*}}^{0.05}\pm0.12     & -0.06\pm_{0.05^{*}}^{0.03}\pm0.09     & -0.02\pm_{0.05^{*}}^{0.02}\pm0.05     & 0.00\pm_{0.00}^{0.01}\pm0.02          \\
		{[}Pr/H{]}  & -0.25\pm_{0.11}^{0.09}\pm0.31         & -0.16\pm_{0.03}^{0.06}\pm0.16         & -0.07\pm_{0.11}^{0.05}\pm0.09         & -0.04\pm_{0.12}^{0.07}\pm0.05         & 0.00\pm_{0.04}^{0.00}\pm0.04          \\
		{[}Nd/H{]}  & 0.13\pm_{0.33^{*}}^{0.07}\pm0.19      & 0.04\pm_{0.27}^{0.05}\pm0.11          & 0.01\pm_{0.17}^{0.02}\pm0.05          & -0.00\pm_{0.04}^{0.01}\pm0.03         & 0.00\pm_{0.01}^{0.00}\pm0.02          \\
		{[}Sm/H{]}  & -0.00^{*}\pm_{0.14}^{0.12^{*}}\pm0.35 & -0.00^{*}\pm_{0.04}^{0.09^{*}}\pm0.14 & 0.03^{*}\pm_{0.03}^{0.02^{*}}\pm0.07  & 0.02\pm_{0.02}^{0.05^{*}}\pm0.04      & 0.00\pm_{0.00}^{0.03^{*}}\pm0.03      \\
		{[}Eu/H{]}  & \nodata                               & -0.00^{*}\pm_{0.02}^{0.00^{*}}\pm0.23 & -0.00^{*}\pm_{0.03}^{0.02^{*}}\pm0.22 & 0.00^{*}\pm_{0.00}^{0.05^{*}}\pm0.13  & 0.00^{*}\pm_{0.00}^{0.00^{*}}\pm0.08  \\
		{[}Gd/H{]}  & 0.10^{*}\pm_{0.07}^{0.07^{*}}\pm0.38  & 0.01\pm_{0.05}^{0.08}\pm0.22          & -0.03\pm_{0.05}^{0.06}\pm0.15         & -0.01\pm_{0.04}^{0.02}\pm0.08         & 0.00\pm_{0.00}^{0.00}\pm0.05          \\
		{[}Dy/H{]}  & -0.69\pm_{0.01}^{0.01}\pm0.32         & -0.29\pm_{0.20}^{0.29^{*}}\pm0.28     & -0.09\pm_{0.18}^{0.13^{*}}\pm0.17     & -0.02\pm_{0.09}^{0.08^{*}}\pm0.11     & 0.00\pm_{0.00}^{0.05}\pm0.07          \\
		{[}Ho/H{]}  & \nodata                               & \nodata                               & \nodata                               & \nodata                               & 0.00^{*}\pm_{0.00}^{0.00^{*}}\pm0.08  \\
		{[}Er/H{]}  & -0.36^{*}\pm_{\text{NaN}}^{\text{NaN}}\pm0.49& -0.00\pm_{0.30}^{0.06^{*}}\pm0.41     & 0.01\pm_{0.18}^{0.03^{*}}\pm0.24      & 0.01\pm_{0.01}^{0.08^{*}}\pm0.13      & 0.00\pm_{0.00}^{0.01^{*}}\pm0.06      \\
		{[}Os/H{]}  & \nodata                               & \nodata                               & \nodata                               & 0.00^{*}\pm_{0.00}^{0.00^{*}}\pm0.08  & 0.00^{*}\pm_{0.00}^{0.00^{*}}\pm0.08  \\
		{[}Th/H{]}  & \nodata                               & 0.24^{*}\pm_{0.00}^{0.00^{*}}\pm0.26  & 0.18^{*}\pm_{0.17}^{0.06^{*}}\pm0.24  & 0.05\pm_{0.05}^{0.11^{*}}\pm0.15      & 0.00\pm_{0.00}^{0.00}\pm0.09 
	\enddata
	\tablecomments{
	    Asterisks denote instances where the reported quantities are impacted by the imposed boundaries of the training set. Median statistical uncertainties at each S/N are included for reference.
	}
\end{deluxetable*}

\begin{figure*}[ht!]
    \epsscale{1.1}
    \plotone{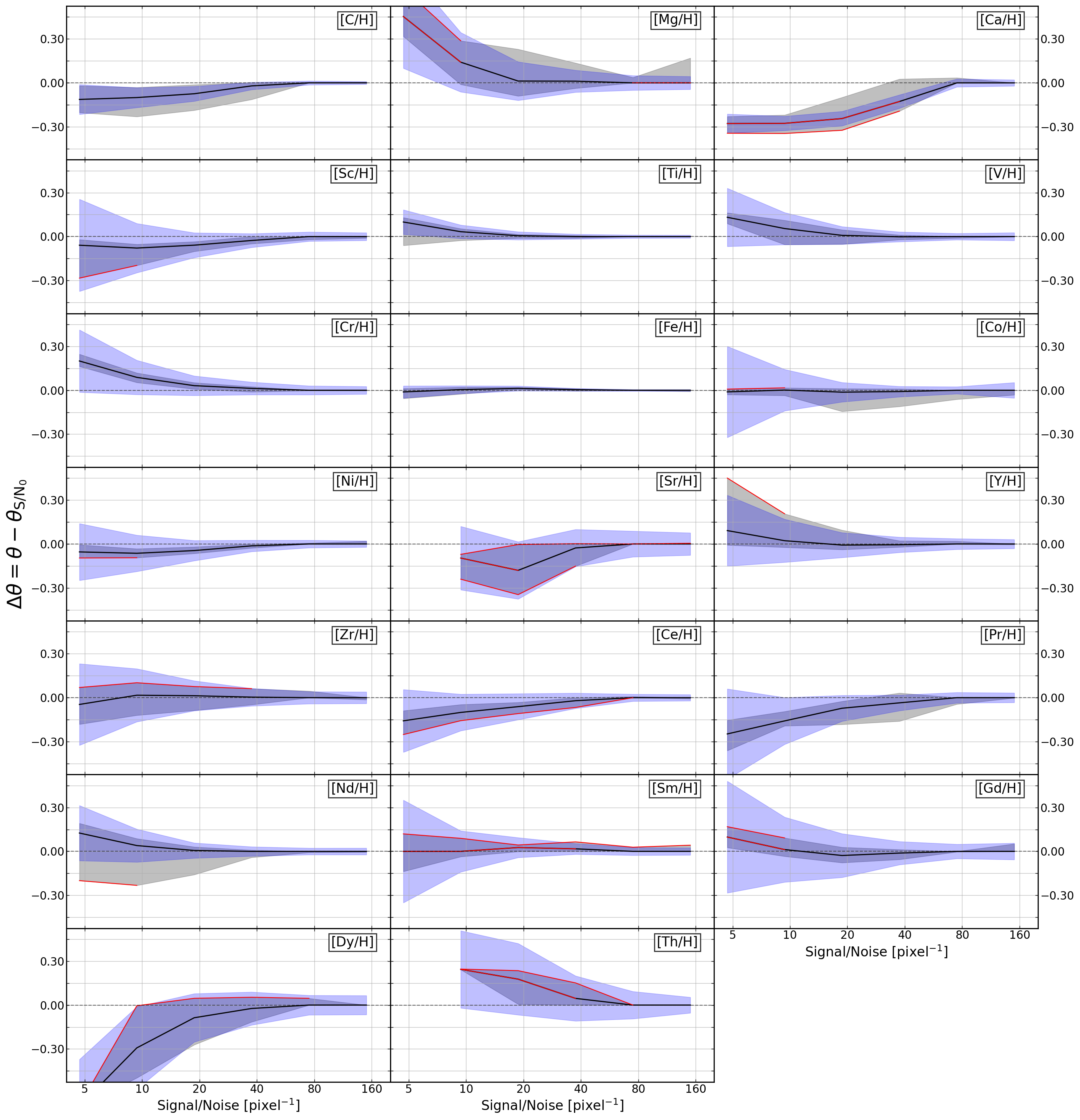}
    \caption{
        Systematic biases (solid black lines) and uncertainties (gray shaded regions) in the recovery of elements at $R\sim10000$ as a function of S/N. The median formal statistical uncertainties (blue shaded regions) are included for reference. Only elements that were found to have minimal resolution-dependent systematics in  \S\ref{sec:Abundance_v_Resolution} are included. Most of these elements display S/N-dependent  systematic effects that are small compared to the statistical uncertainties, though a few (C, Mg, Ca, Dy) are biased at low S/N.
        \label{fig:4_all_good_v_snr}
    }
\end{figure*}

Figure \ref{fig:4_all_good_v_snr}, illustrates the trends in recovery as a function of S/N for the 20 elements (C, Mg, Ca, Sc, Ti, V, Cr, Fe, Co, Ni, Sr, Y, Zr, Ce, Pr, Nd, Sm, Gd, Dy, and Th) that we found to have minimal resolution-dependent systematic bias in \S\ref{sec:Abundance_v_Resolution}. The presentation of these results follows the same conventions as Figures \ref{fig:4_atm1_v_res}--\ref{fig:4_neutron3_v_res} except that we also include the 1$\sigma$ statistical uncertainties inferred from MCMC sampling as blue shaded regions for reference.

We find that most of these elements show little to no dependence on the S/N down to $\text{S/N}\sim5$ \perpix.  While we find small differences between high and low S/N measurements, they are typically smaller than the 1$\sigma$ statistical uncertainties inferred from the posteriors. The scatter found in the trends between individual exposures is generally consistent with the statistical uncertainty.

The recovery of upper/lower limits at the model grid boundary impede robust characterization of the S/N-dependence for several elements across the full S/N range, including: Sm and Th below $\text{S/N}<40$ \perpix, Sr below $\text{S/N}<20$ \perpix, and Gd below $\text{S/N}<10$ \perpix. 

For two elements, Mg and Dy, we find that the low-S/N measurements become inconsistent with the high-S/N measurements below $\text{S/N}\lesssim10$ \perpix at which point the measurement precision is already quite poor ($\gtrsim$0.3 dex). For two other elements, C and Ca, we find more substantial trends as the S/N is decreased. For C, we find a negative bias that increases to $\sim$0.15 dex and a systematic uncertainty of $\sim$0.1 dex $\text{S/N}\lesssim40$ \perpix. For Ca, we find a much more striking trend with S/N. Below $\text{S/N}\lesssim40$ \perpix, [Ca/H] is recovered to be at least 0.3 dex lower than at the default S/N. The origin of these S/N-dependent systematics is challenging to ascertain and is worth of future investigation.

%

\section{Discussion}
\label{sec:discussion}

\subsection{Fidelity of Low-Resolution Abundance Measurements}
\label{sec:element_fidelity_res}
The primary motivation of this work is to identify which elements can and cannot be robustly measured from low-resolution stellar spectra and quantify the trends in abundance recovery exist as a function of resolution so that low- and high-resolution measurements can be integrated. Here we summarize our findings, referring to the resolution-dependent systematic biases and uncertainties for each element reported in Table \ref{tab:abundance_v_resolution}. We recommend the usage of these values for 22 of the 36 elements considered in this work: C, Mg, Ca, Sc, Ti, V, Cr, Fe, Co, Ni, Zn, Sr, Y, Zr, Ce, Pr, Nd, Sm, Gd, Dy, Er, and Th. We urge caution in the adoption of these factors for the remaining elements due to inconsistent agreement with literature measurements (O, Na, Al, Si, K, and Ho; see Appendix \ref{sec:lit_by_element}) or the limitations set by the extent of our training grid (N, Mn, Cu, Ga, Ba, La, Os, Eu).

For the broad optical wavelength coverage considered in this work, we highlight $R\sim10000$ as an inflection point in the trends of $\Delta\theta$ and $\sigma_\text{syst}$, below which both quantities increase more sharply. At $R\sim10000$, 20 elements (C, Mg, Ca, Sc, Ti, V, Cr, Fe, Co, Ni, Sr, Y, Zr, Ce, Pr, Nd, Sm, Gd, Dy, and Th) are recovered with $\Delta\theta\lesssim0.1$ dex and $\sigma_\text{syst}\lesssim0.15$ dex. This decreases to 14 elements (C, Mg, Ca, Sc, Ti, V, Fe, Ni, Y, Zr, Ce, Nd, Sm, and Gd) at $R\sim5000$ and 9 elements at $R\sim2500$ (C, Mg, Ca, Sc, Ti, Fe, Ni, Y, and Nd). With that said, that multiple individual elements---including at least one from each broad nucleosynthetic grouping no-less---can be robustly measured at $R\sim2500$ is very promising for low-resolution surveys in the MW and LG (e.g., LAMOST, DESI, and PFS, and the low-resolution modes of SDSS-V, 4MOST, and WEAVE).

Generally speaking, the fidelity of an element's recovery as a function of resolution is related primarily to the number (and secondarily to the strength) of its absorption features. Elements with many absorption lines spread across the entire spectrum tend to show the least sensitivity to model-data mismatch at low-resolution, while elements with only a few lines---especially a few weak lines---exhibit the strongest trends with resolution. This makes sense intuitively as the presence of many additional lines anchors the measurement even if some of the lines are contaminated by poorly-modeled neighboring features. When only a few lines are present, contamination of any single line can substantial bias the measurement. Similarly, we find that elements that are primarily constrained by their indirect and subtle effects on the lines of other elements (i.e., through changes to the atmospheric structure) are also sensitive to model fidelity. While these elements may still be able to be measured from low-resolution spectroscopy, much more careful treatment of the spectral features and the regions around these features is necessary. 

The qualitative conclusions (e.g., which elements are more/less robustly recovered as a function of resolution) of this analysis should be broadly applicable, though the exact systematic uncertainties and biases that are reported are likely to be a strong function of the observed wavelength coverage, the observed stars' parameters, and the adopted stellar models. This analysis must be extended to larger and broader datasets and spectroscopic configurations before any low-resolution ``corrections" from this work are naively applied to drastically different observations (e.g., solar-metallicity dwarf stars or NIR observations). Similarly, this study should be repeated for additional stellar models if models other than \atlas\ and \synthe\ are used.

\subsection{Fidelity of Low-S/N Abundance Measurements}
\label{sec:element_fidelity_snr}
A secondary motivation of this work was to evaluate the prospect of accurately measuring multi-element abundances from low-S/N data at $R\sim10000$. Here we summarize our findings, referring to the S/N-dependent systematic biases and uncertainties for each element reported in Table \ref{tab:abundance_v_snr}. We recommend the usage of these values for the 20 elements identified in \S\ref{sec:element_fidelity_res}, which show only small to modest bias and uncertainties at $R\sim10000$ ($\Delta\theta\lesssim0.1$ dex and $\sigma_\text{syst}\lesssim0.15$ dex): C, Mg, Ca, Sc, Ti, V, Cr, Fe, Co, Ni, Sr, Y, Zr, Ce, Pr, Nd, Sm, Gd, Dy, and Th. For nearly all of these elements, robust, albeit less precise, measurements can be made at S/N as low as 5 \perpix\ without the need to invoke additional systematic uncertainty. For four elements, C, Mg, Ca, and Dy, we find biases at low S/N in excess of the statistical and systematic uncertainties. The origin of these trends is difficult to identify and warrants additional investigation. We recommend a minimum S/N of $\sim$10 \perpix\ for Mg and Dy and $\sim$40 \perpix\ for C and Ca.

\subsection{Stellar Label Uncertainties and Correlations}
\label{sec:crlb}
\begin{figure*}[ht!]
    \epsscale{1.25}
    \plotone{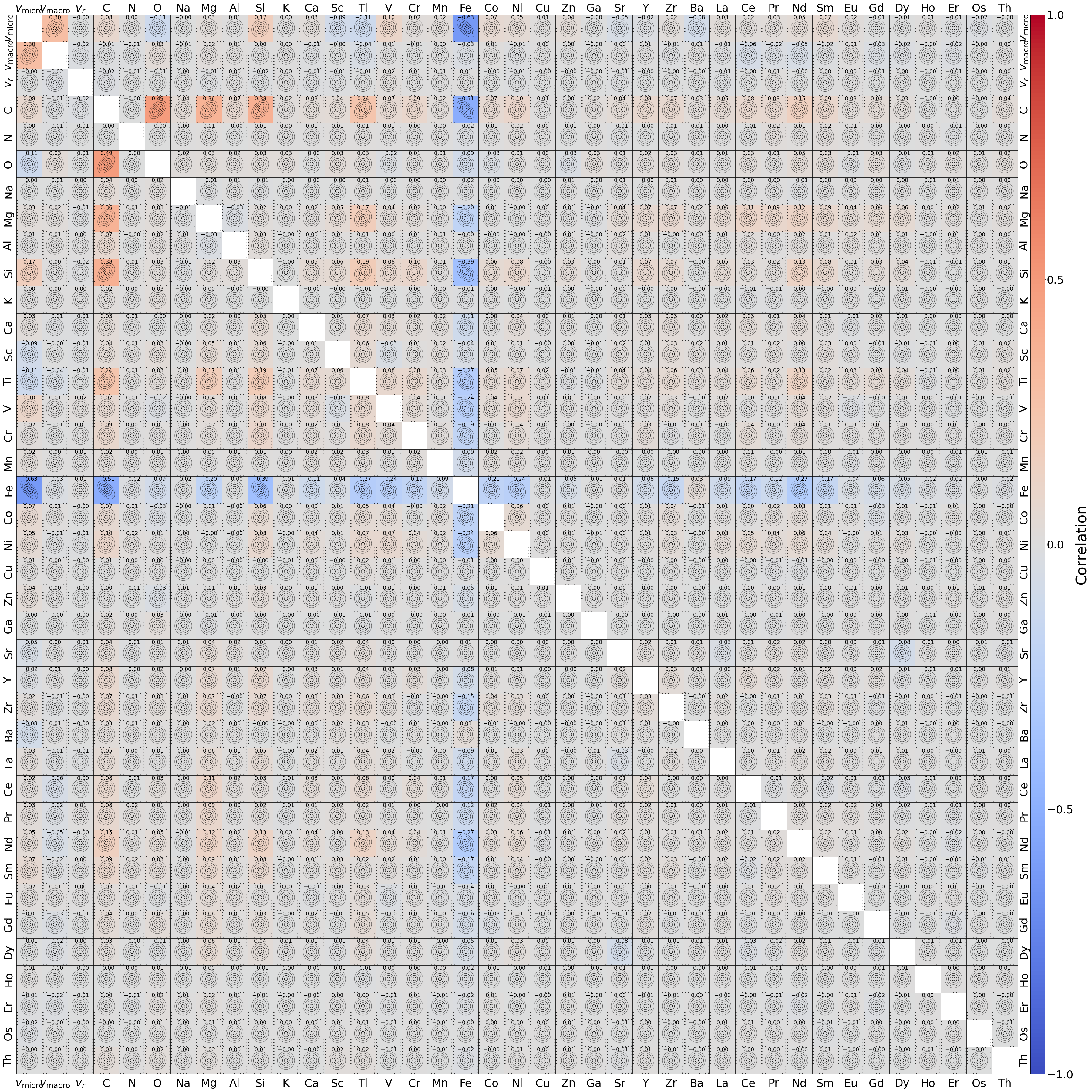}
    \caption{
        Median correlations in the measurements of all 36 elemental abundances, \vmicrotext, \vmacrotext, and $v_r$ at $R\sim10000$. Each panel depicts the correlation of a different pair of labels with the color of the panel indicating the strength and direction of the correlation. While most labels are not strongly correlated with one other, labels that contribute to a large number of pixels across the observed wavelength range like Fe, C, Mg, Si, and Ti exhibit modest correlations.
        \label{fig:5_measurement_correlation}
    }
\end{figure*}

The use of MCMC methods in our spectroscopic analysis enables us to robustly quantify the formal statistical uncertainties on measurements of [X/H] as well as the element-to-element measurement correlations. In Figure \ref{fig:5_measurement_correlation}, we present the median pairwise correlations found between the 36 elemental abundances, \vmicrotext, \vmacrotext, and $v_r$ at the convolved resolution of $R\sim10000$. \tefftext\ and \loggtext\ are omitted as they are 1-to-1 correlated with Fe. Each panel depicts the correlation of a label pair as measured from the MCMC posterior samples. To guide the eye, panels are shaded according to their Pearson correlation coefficient, $r$. 

Figure \ref{fig:5_measurement_correlation} shows that the majority of stellar labels are not strongly correlated ($r\lesssim0.05$) at $R\sim10000$. The strongest correlations belong to elements which have many absorption features across the observed wavelength range. Most obvious among these is Fe, which is strongly anti-correlated ($r\sim-0.1$ to $-0.6$) with $\sim20$ other stellar labels. C, and to a lesser extent Mg, Si, and Ti, also exhibit correlations of $r\gtrsim0.05$ with roughly a dozen other elements as a result of their contributions to stellar atmospheric structure. We also compare the correlations we infer at $R\sim10000$ with those that we infer at both lower and higher resolutions. We find that the pairwise correlation between elements at $R\sim40000$ is very similar to what we find at $R\sim10000$. At $R\sim2500$, the pairwise correlation increases in magnitude for most elements and for \vmacrotext, though it still remains below $r\lesssim0.2$ for most element pairs.

\subsubsection{Comparison of Uncertainties to CRLBs}
Calculating uncertainties and correlations using MCMC sampling is a computationally expensive undertaking, especially given the high-dimensionality of abundance measurements explored here. As a result, its application to the datasets of large spectroscopic surveys (e.g., APOGEE, GALAH, LAMOST) are intractable. Recently, the use of Cram\'er-Rao Lower Bounds (CRLBs; \citealt{frechet:1943, rao:1945, darmois:1945, cramer:1946}), the maximum precision predicted by a Fisher Information analysis, has been proposed as a fast and easy method to forecast the chemical abundance precision achievable from a given stellar spectral dataset \citep[e.g.,][]{ting:2019, sandford:2020b}. Here we take the opportunity to compare the statistical uncertainties we measure from our MCMC fitting technique to those forecasted by the CRLBs.

To calculate the CRLBs of our observations, we employ the \texttt{Chem-I-Calc}\footnote{https://chem-i-calc.readthedocs.io/en/latest/} Python package with a few minor adjustments \citep{sandford:2020a, sandford:2020b}. For each star in our sample, we generate gradient spectra using \thePayne\ and adopt the total S/N of the fit (model errors and observational masks included). Because \tefftext\ and \loggtext\ are inferred deterministically from each star's photometry and [Fe/H], we treat them as fixed parameters in the CRLB calculation.

\begin{figure*}[ht!]
    \epsscale{1.15}
    \plotone{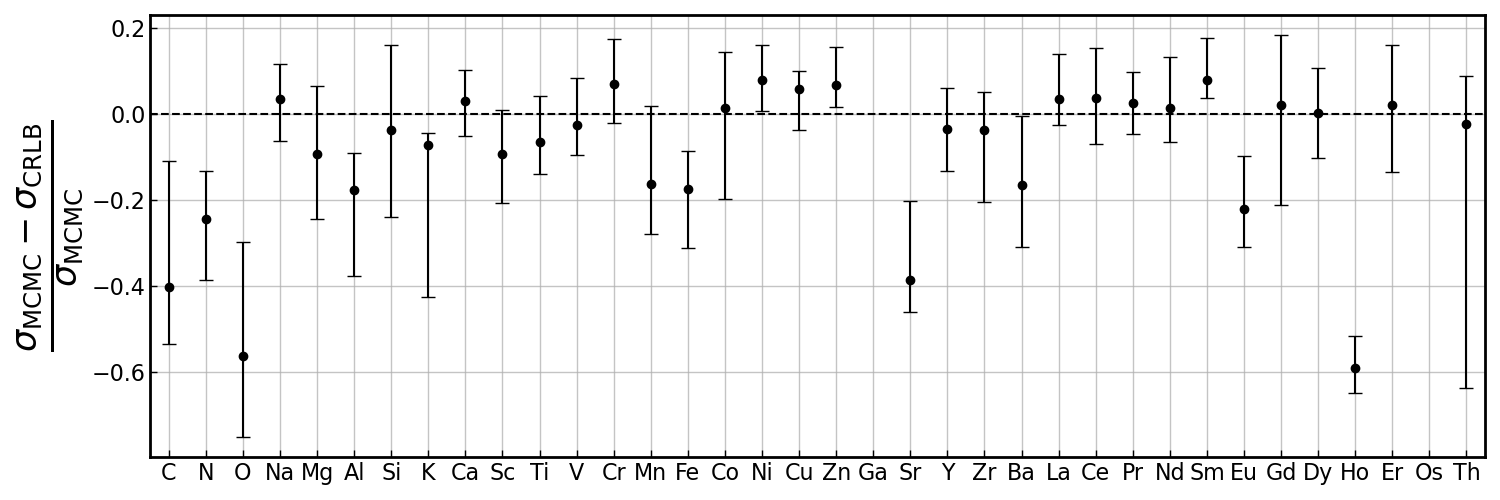}
    \caption{
        Fractional difference in the formal statistical uncertainty on [X/H] and the precision forecasted by the CRLB. Points and error bars represent the median and 16th and 84th percentiles across all individual exposure measurements performed in this study. For $\sim$75\% of the elements considered here, the uncertainties are in general agreement. Large deviations from zero are found in instances of non-Gaussian posteriors (e.g., C, N, and O) and/or substantial model-data mismatches (e.g., Sr and Eu).
        \label{fig:5_unc_vs_crlb}
    }
\end{figure*}

In Figure \ref{fig:5_unc_vs_crlb}, we present a comparison of the statistical uncertainty found through MCMC sampling $\sigma_\text{MCMC}$ and the uncertainty forecasted by the CRLB $\sigma_\text{CRLB}$ for each element. Points and error bars represent the median and 16th and 84th percentiles across all individual exposure measurements performed in this study, omitting measurements which are within 2$\sigma$ of the uniform prior bounds and measurements for which $\sigma_\text{CRLB}>0.5$. We have no suitable measurements for Ga and Os. For 28 (24) of the 36 chemical abundances, $\sigma_\text{CRLB}$ is within 20\% (10\%) of $\sigma_\text{CRLB}$. We find no trend in the (dis)agreement as a function of the resolution or S/N of the observation, nor as a function of the expected precision.

Because the CRLB represents the maximum theoretically achievable precision, we would expect $\sigma_\text{MCMC} \gtrsim \sigma_\text{CRLB}$. While this is the case for many elements and measurements, it is not universally true. For example, the statistical uncertainties on Fe and Eu are consistently $\sim$20\% smaller than forecasted by the CRLBs. C, N, O, Al, K, Sr, Ba, and Ho are also recovered to better precision than the CRLBs predict---in some cases by large margins. These deviations from the forecasted precision are driven by 1) non-Gaussian posteriors for which $\sigma_\text{MCMC}$ underestimates the true uncertainty and/or 2) mismatches between the model and observed spectra that invalidate the assumption of an un-biased estimator in the CRLB calculation \citep[e.g.,][]{ting:2017a, sandford:2020b}. In the case of C, N, and O, we believe that the better-than-expected precision is due in part to non-Gaussian posteriors and in part to over-estimation of the correlation between these three elements in the CRLBs. Indeed, if the correlation in CNO spectral features is ignored in the CRLB calculation, the agreement between the forecasted and realized statistical uncertainties is much better (though large variance remains).

Given the general agreement found in this comparison, and no instances of the CRLB drastically over-predicting the expected precision, we suggest that, going forward, CRLBs can be safely adopted as conservative forecasts of the statistical precision to the 10--20\% level. 

\subsection{Implications for Chemical Evolution Studies}
A primary use of stellar chemical abundance measurements is to constrain stellar and galactic physics by fitting models to a system's chemical enrichment history. Recovering meaningful constraints, however, requires accurate abundances with well-characterized uncertainties. Anything less will lead to biased or misleading conclusions. High-precision, inaccurate measurements are perhaps the most disastrous combination as they will strongly influence a chemical evolution model from the true solution. Less catastrophic, but still undesirable are accurate measurements with uncertainties that are under-predicted as these will lead to accurate model predictions but will overstate the constraining power of the data. Moreover, mischaracterized uncertainties will bias studies concerned with the intrinsic dispersion of stellar chemical abundances (e.g., to understand stochasticity in nucleosynthetic pathways or inhomogenous mixing of the ISM; \citealt{griffith:2022, ting:2022}).

For these reasons, it is important to fold in accurate estimates of the systematic uncertainties like those presented in this work. For most low-resolution stellar spectroscopic observations, this precludes the $\lesssim$0.1 dex precision on many elements that higher resolution surveys can achieve. As a result, the vast majority of chemical abundance measurements in the next decade, especially for stars outside the MW, will be systematics limited in precision. Nevertheless, low-precision (0.2--0.3 dex) measurements can still be incredibly informative as long as they are accurate and there are sufficient numbers of stars \citep[e.g.,][]{kirby:2011b, sandford:2022}. Even after appropriately accounting for the systematic uncertainties quantified in this work, the highly-multiplexed, low-resolution spectrographs of the next decade have the potential to reveal transformative new insight into galactic chemical evolution in the MW and throughout the LG and beyond.

\section{Conclusion}
\label{sec:conclusion}
We perform a completely self-consistent analysis of 40 Keck/HIRES observations of 8 RGB stars in M15, which have been degraded to a range of lower resolutions and S/N. We fit for 39 stellar labels (including 36 elemental abundances) and $\sim$100--200 nuisance parameters (mostly continuum coefficients) using full-spectrum fitting techniques and quantify the systematic biases and uncertainties that are introduced as the quality of the data is degraded. Our primary conclusions are as follows:

\begin{itemize}
    \item Observations at resolutions down to $R\sim10000$ can measure 20 elements (C, Mg, Ca, Sc, Ti, V, Cr, Fe, Co, Ni, Sr, Y, Zr, Ce, Pr, Nd, Sm, Gd, Dy, and Th) to within $\lesssim$0.1 dex of high-resolution observations with $\lesssim0.15$ dex systematic uncertainties.
    \item Nine elements (C, Mg, Ca, Sc, Ti, Fe, Ni, Y, and Nd) can be measured to this same level of consistency down to $R\sim2500$.
    \item Only four elements (C, Mg, Ca, and Dy), exhibit substantial S/N-dependent bias at $R\sim10000$ in excess of statistical uncertainties below $\text{S/N}\sim10$ \perpix.
    \item For $\sim$75\% of elements, the precision forecasted by the CRLBs provides a good estimate of the formal uncertainties computed with MCMC sampling.
    \item The predominant source of systematic bias and uncertainty at low-resolution is blending of poorly-modelled absorption features, which impacts elements with few and/or weak lines most strongly.
\end{itemize}

We conclude with an optimistic outlook. In this work we find that even with imperfect models, low-resolution measurements that are consistent with their high-resolution counterparts are possible for a representative sample of elements. As such, the next decade of highly-multiplexed low-resolution spectroscopic surveys and instruments are poised to dramatically increase our understanding of the MW and LG's chemical evolution. Furthermore, because we have adopted 1D-LTE stellar models in this analysis, the systematic effects we report represent a conservative estimate. Ongoing improvements to stellar models (e.g., 3D-NLTE physics, updated atomic
line data), will continue to alleviate these systematics and further increase the viability of high-precision accurate low-resolution spectroscopic chemical abundance measurements.

\section*{Acknowledgements}
We thank Bob Kurucz for developing and maintaining programs and databases without which this work would not be possible. We thank Jennifer Sobeck, Chris Sneden, Anish Amarsi, Thomas Nordlander, and Mikhail Kobal\"ev for their helpful insight on NLTE effects in stellar spectra.
NRS is grateful for the hospitality of the Research School of Astronomy and Astrophysics at the Australian National University at which a portion of this work was conducted.
NRS acknowledges financial support from the NSF GRFP under grants DGE 1752814 and DGE 2146752.
NRS and DRW also acknowledge support from HST-GO-15901 from the Space Telescope Science Institute, which is operated by AURA, Inc., under NASA contract NAS5-26555.
YST acknowledges financial support from the Australian Research Council through DECRA Fellowship DE220101520.

The computations in this paper were partially run on the Savio computational cluster resource provided by the Berkeley Research Computing Program at the University of California, Berkeley.

The data presented herein were obtained at the W. M. Keck Observatory, which is operated as a scientific partnership among the California Institute of Technology, the University of California and the National Aeronautics and Space Administration. The Observatory was made possible by the generous financial support of the W. M. Keck Foundation. Further, this data was made accessible by the Keck Observatory Archive (KOA), which is operated by the W. M. Keck Observatory and the NASA Exoplanet Science Institute (NExScI), under contract with the National Aeronautics and Space Administration. This research also made use of the VizieR catalogue access tool \citep{ochsenbein:2000} and the SIMBAD database \citep{wenger:2000}, both operated by the CDS, Strasbourg, France. 

The authors wish to recognize and acknowledge the very significant cultural role and reverence that the summit of Maunakea has always had within the indigenous Hawaiian community.  We are most fortunate to have the opportunity to conduct observations from this mountain.

\facilities{
Keck:I (HIRES), Gaia
}

\software{
astropy \citep{astropycollaboration:2013, astropycollaboration:2018},
Chem-I-Calc \citep{sandford:2020a},
emcee \citep{foreman-mackey:2013},
matplotlib \citep{Hunter:2007},
numpy \citep{walt:2011, harris:2020},
pandas \citep{mckinney:2010, reback:2022},
PyTorch \citep{paszke:2019},
Pytorch-Lightning\citep{falcon:2020},
}

\appendix

\section{\texttt{The Payne}: Technical Details and Training}
\label{app:payne}

\subsection{Neural Network Architecture}
\label{app:architecture}
As in previous implementations of \texttt{the Payne}, we adopt a fully-connected neural network with two hidden layers of $N_1=N_2=300$ neurons each.
The first hidden layer expects as input an array of $N_\theta=39$ stellar labels (\tefftext, \loggtext, \vmicrotext, and [X/H], where X includes the elements C, N, O, Na, Mg, Al, Si, K, Ca, Sc, Ti, V, Cr, Mn, Fe, Co, Ni, Cu, Zn, Ga, Sr, Y, Zr, Ba, La, Ce, Pr, Nd, Sm, Eu, Gd, Dy, Ho, Er, Os, and Th). The output of the model is an array of normalized flux values corresponding to each wavelength pixel of the ab initio spectra it is trained on. Employing a leaky ReLU activation function,
\begin{equation}
    \text{LReLU}(x) = 
    \begin{cases}
    x,      & \text{if } x > 0\\
    0.01x,  & \text{otherwise},
    \end{cases}
\end{equation}
the model architecture can be represented by the following equations:
\begin{align}
    f^{(1)}_{j}(\theta_*) &= \text{LReLU}\left(\sum_{i=1}^{N_{\theta}}\left[w^{(1)}_{i,j}\theta_{*,i}\right] + b^{(1)}_{j}\right) \\
    f^{(2)}_{k}(\theta_*) &= \text{LReLU}\left(\sum_{j=1}^{N_{1}}\left[w^{(2)}_{j,k}f^{(1)}_{j}(\theta_*)\right] + b^{(2)}_{k}\right) \\
    f^{(\text{out})}_\lambda(\theta_*) &= \sum_{k=1}^{N_{2}}\left[w^{(\text{out})}_{k,\lambda}f^{(2)}_{k}(\theta_*)\right] + b^{(\text{out})}_{\lambda},
\end{align}
where $w^{(1)}$, $b^{(1)}$, $w^{(2)}$, $b^{(2)}$, $b^{(\text{out})}$, and $w^{(\text{out})}$ are the weights and biases of the neurons in the first hidden layer, second hidden layer, and the output layer. Like later implementations of \texttt{the Payne} \citep[e.g.,][]{kovalev:2019, xiang:2022, straumit:2022}, this architecture capitalizes on the spectrum's continuity in the wavelength dimension and utilizes the information contained in adjacent pixels to better predict the flux of each pixel---in contrast to the architecture used originally in \citet{ting:2019}, which used an independent model for each pixel.

The total number of model parameters in a neural network with this architecture is given by
\begin{equation}
    N_\text{par} = (N_\theta+1) \times N_1 + (N_1+1)\times N_2 + (N_2+1)\times N_\text{pix},
\end{equation}
where $N_\text{pix}$ is the number of pixels in the model spectrum. Adopting $N_1=N_2=300$ and training on ab initio spectra with $N_\text{pix}=262144$ with $N_{\theta}=39$ as we do, requires a model with $N_\text{par}\sim7.9\times10^7$ parameters. Despite the large number of parameters to optimize, such a model can be optimized in a reasonable $\sim$150 hours on a NVIDIA A40 GPU.

\subsection{Training Set}
\label{app:payne_dataset}

Training \texttt{the Payne} requires a set of stellar spectra with known labels that span the parameter space of the observed stars. Because the stars considered in this work have been well studied, we could generate a dense training set around the literature values for these stars ($\text{[Fe/H]}\sim-2.5$ at the tip of the RGB). However, we choose to generate a much more ambitious training set that covers the entire RGB over a large range of metallicities. The reasons for this are twofold: 1) to avoid simply reproducing literature results by construction and 2) to generate a training set with applications beyond the RGB of M15.

We begin the construction of our training set by randomly drawing 25000 sets of $T_\text{eff}$, $\log g$, and [Fe/H] values from MIST isochrones \citep{paxton:2011, paxton:2013, paxton:2015, dotter:2016, choi:2016, paxton:2018} with $3500 \leq T_\text{eff}~\text{K} \leq 6000$, $0.0 \leq \log g \leq 4.0$, $-4.0 \leq \text{[Fe/H]} \leq -1.0$, and $10 \leq t_\text{age}~\text{[Gyr]} \leq 14$. Only RGB stars are included in this sample. For each sample, $v_\text{micro}$ is determined from the empirical relation found in \citet{holtzman:2015},
\begin{equation}
    v_\text{micro}=2.478 - 0.325\log g.
\end{equation}
To smooth over the discrete isochrone tracks and allow for $v_\text{micro}$ offset from the empirical relation, we add zero-mean Gaussian scatter to each of these labels with $\sigma_{T_\text{eff}}=250~\text{K}$, $\sigma_{\log g}=0.25$, $\sigma_\text{[Fe/H]}=0.25$, and $\sigma_{v_\text{micro}}=0.25~\text{km/s}$. Lastly, for each sample, we draw elemental abundances [X/H] from a uniform distribution with the condition $-1.0 \leq \text{[X}_1\text{/Fe]} \leq 1.0$ for $\text{X}_1$ = C, N, and O; $-0.5 \leq \text{[X}_2\text{/Fe]} \leq 0.5$ for $\text{X}_2$ = Na, Sc, V, Cr, Mn, Co, Ni, Cu, Zn, Ga, Sr, Y, Zr, Ba, and La; and $-0.25 \leq \text{[X}_3\text{/Fe]} \leq 1.0$ for $\text{X}_3$ = Mg, Al, Si, K, Ca, Ti, Ce, Pr, Nd, Sm, Eu, Gd, Dy, Ho, Er, Os, and Th. Summaries of our MIST isochrones and sampling scheme are presented in Tables \ref{tab:mist} and \ref{tab:sampling} respectively. We note that while 25000 ab initio may seem like a large training set, it is still orders of magnitude smaller than would be required for grid interpolation over the broad 39-dimensional parameter space.

\begin{table}[ht!] \label{tab:mist}
\begin{center}
	\caption{MIST Isochrone Set} 
    \begin{tabular}{lc}
    \hline \hline
        MIST version & 1.2 \\
		Initial $v/v_\text{crit}$ & 0.4 \\
		$t_\text{age}$ & 10 to 14 Gyr \\
		$\Delta\log t_\text{age}$ & 0.01 \\
		$\text{[Fe/H]}$ & $-4.0$ to $-1.0$ \\
		$\Delta\text{[Fe/H]}$ & 0.1 \\
		$[\alpha\text{/H]}$ & 0.0 \\
    \tableline
	\end{tabular}
	\tablecomments{
	    Characteristics of the MIST isochrone set from which \tefftext, \loggtext, and [Fe/H] are initially drawn.
	}
\end{center}
\end{table}

\begin{table}[ht!] \label{tab:sampling}
    \begin{center}
	\caption{Stellar Label Sampling Scheme}
    \begin{tabular}{ll}
	\hline \hline
	    Label & Distribution \\
	\hline
	    \multicolumn{2}{l}{Intermediate Samples from MIST Isochrones}\\
	\hline
	    $T_\text{eff,~iso}$ & $\mathcal{U}_\text{MIST}(3500~\text{K}, ~6000~\text{K})$ \\
	    $\log g_\text{iso}$ & $\mathcal{U}_\text{MIST}(0.0, ~0.4)$ \\
	    $\text{[Fe/H]}_\text{iso}$ & $\mathcal{U}_\text{MIST}(-4.0, ~-1.0)$ \\
	\hline
        \multicolumn{2}{l}{Final Samples with Scatter}\\
	\hline
	    $T_\text{eff}$ & $\mathcal{N}(T_\text{eff,~iso}, ~250~\text{K})$ \\
	    $\log g$ & $\mathcal{N}(\log g_\text{iso}, ~0.25)$ \\
	    $\text{[Fe/H]}$ & $\mathcal{N}(\text{[Fe/H]}_\text{iso}, ~0.25)$ \\
	    $v_\text{micro}$ & $\mathcal{N}(2.478 - 0.325\log g, ~0.25)$ \\
	    $\text{[X}_1\text{/Fe]}$ & $\mathcal{U}(-1.00, ~1.00)$ \\
	    $\text{[X}_2\text{/Fe]}$ & $\mathcal{U}(-0.50, ~0.50)$ \\
	    $\text{[X}_3\text{/Fe]}$ & $\mathcal{U}(-0.25, ~1.00)$ \\
	\hline
	\end{tabular}
	\tablecomments{
	    Distributions from which the training label sets are drawn. \tefftext, \loggtext, and [Fe/H] are drawn initially from the MIST isochrone set described in Table \ref{tab:mist} before additional scatter is applied.
	    $\text{X}_1$ includes C, N, and O. $\text{X}_2$ includes Na, Sc, V, Cr, Mn, Co, Ni, Cu, Zn, Ga, Sr, Y, Zr, Ba, and La. $\text{X}_3$ includes Mg, Al, Si, K, Ca, Ti, Ce, Pr, Nd, Sm, Eu, Gd, Dy, Ho, Er, Os, and Th.
	}
    \end{center}
\end{table}

Ab initio spectra are generated using the same method described in \citet{ting:2019}, which we summarize here. For each of the 25000 sets of stellar labels, we compute 1D LTE model atmospheres using the \atlas\ code maintained by R.\ Kurucz \citep{kurucz:1970, kurucz:1993, kurucz:2005, kurucz:2013, kurucz:2017, kurucz:1981}. We adopt Solar abundances from \citet{asplund:2009} and the standard mixing length theory with a mixing length of 1.25 and no overshoot for convection. After the model atmosphere converges, we use the \synthe\ radiative transfer code (also maintained by R.\ Kurucz) to produce its normalized spectrum at a nominal resolution of $R=300000$. 

For a little less than $\sim$20\% of the labels, the stellar atmosphere and/or spectrum fails to converge. These failed models predominantly belong to stellar atmospheres with very low metallicities ($\text{[Fe/H]}\lesssim-3.0$). It is possible that better initialization of low-metallicity atmospheres might improve convergence, but we leave this to a future study.

The $\sim$20500 successfully generated spectra are then continuum normalized using the theoretical continua from \synthe. Lastly, the normalized spectra are convolved and sub-sampled down to the highest spectral resolution and wavelength sampling present in our archival data ($R=86600$ and d$v=1.17$km/s \perpix).

\subsection{Training Procedure}
We implement our adoption of \thePayne\ using \texttt{PyTorch}, a powerful and flexible Python machine learning framework, and  \texttt{Pytorch Lightning}\footnote{\url{https://www.pytorchlightning.ai/}}, a lightweight wrapper designed to streamline the development and training of PyTorch models.

As with many machine learning techniques, it is helpful to scale the input labels so that they all share a similar dynamic range of order unity with zero mean. To do so, we normalize all stellar labels according to
\begin{equation}
    \theta_{*,i}^\prime = \frac{\theta_{*,i}-\theta_{*,i,\text{min}}}{\theta_{*,i,\text{max}}-\theta_{*,i,\text{min}}} - 0.5,
\end{equation}
where $\theta_{*,i,\text{min}}$ and $\theta_{*,i,\text{max}}$ are the minimum and maximum values of each label, $i$, included in the training set.
For clarity, we drop the prime notation throughout the rest of this work and convert back to physical units when reporting results. 

For each model/training set, we train directly on 80\% of the successfully generated spectra ($\sim$16000) and validate with the remaining 20\% ($\sim$4000). Training is performed iteratively in batches of 512 spectra using a rectified Adam optimizer (with a learning rate of 0.0001) to minimize the model's L1 loss (i.e., the mean absolute error). Though unknown to the optimization, the L1 loss is also calculated on the cross-validation dataset each epoch. Training is halted after 2000 epochs without improvement of the best L1 validation loss, at which point the model that minimized the L1 validation loss is chosen as the final model. Training of the model was completed in $\sim$150 GPU hours after $\sim$24000 epochs.

\subsection{Accuracy}
\label{app:payne_acc}
We determine the internal accuracy (i.e., the median interpolation error; MIE) of \thePayne\ as is in \citep{ting:2019}. Using the trained neural networks, we generate spectra for each set of stellar labels in the cross-validation dataset and compare to the original ab initio spectra generated with \atlas\ and \synthe. The median interpolation error is thus
\begin{equation}
    \sigma_\text{MIE} = \text{Med}(|f_\lambda(\theta_{*,\text{valid}})-f_{\lambda, \text{valid}}|).
\end{equation}

\begin{figure*}[ht!]
    \epsscale{1.25}
    \plotone{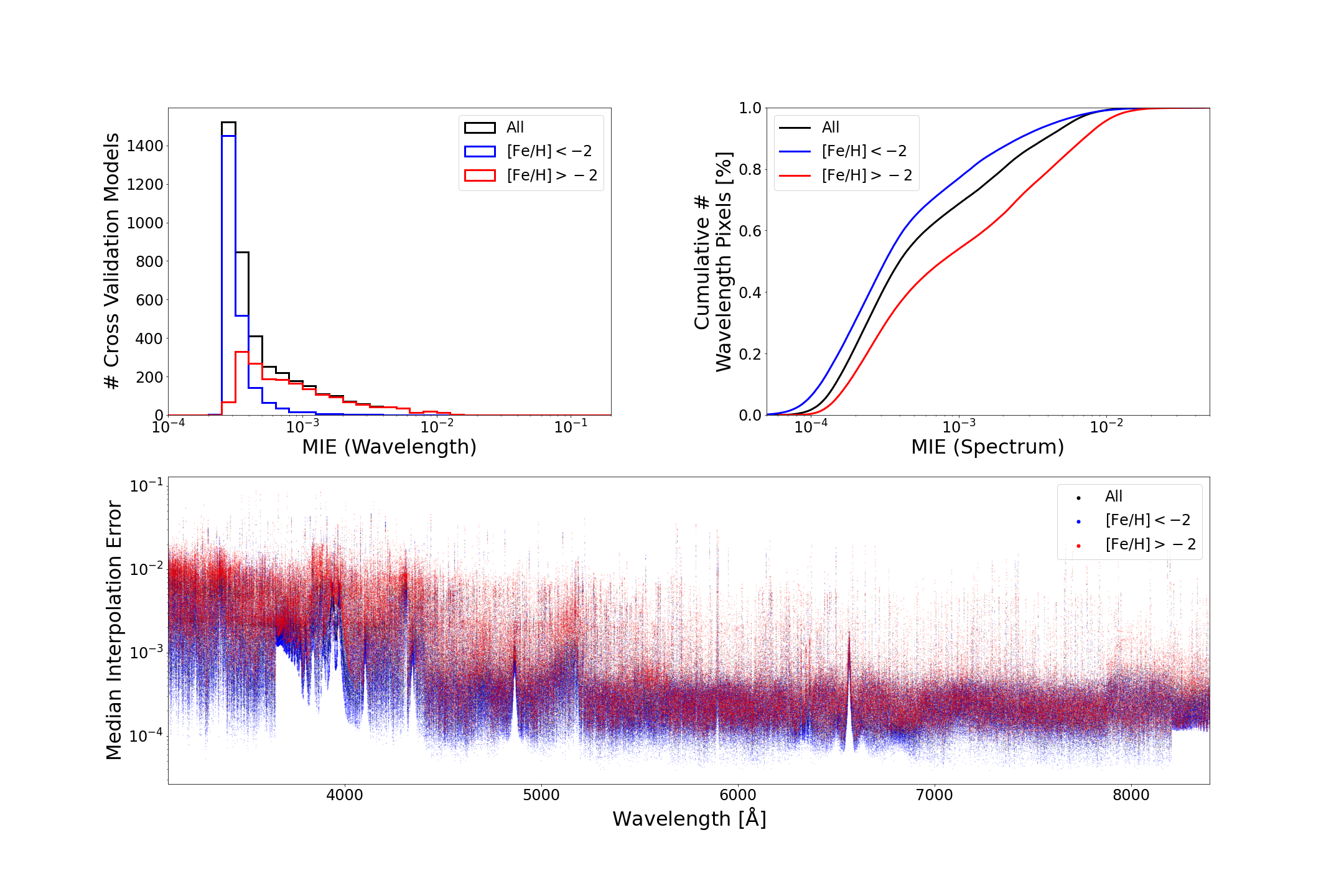}
    \caption{\textbf{Top Left}: Histogram of the median interpolation error of each model in the cross validation set. The median error is consistently larger for higher-metallicity stars ($\text{[Fe/H]}>-2$; red) compared to lower-metallicity stars ($\text{[Fe/H]}>-2$; blue). \textbf{Top Right}: Cumulative percentage of pixels in each spectrum as a function of the median interpolation error. Approximately 80\% of pixels have $\sigma_\text{MIE} < 0.001$, and 95\% of pixels have $\sigma_\text{MIE} < 0.006$. \textbf{Bottom}: The median interpolation error across the cross-validation set as a function of wavelength. Errors are largest in the proximity of strong H lines and complicated molecular features. \label{fig:6_payne_interpolation_errors}}
\end{figure*}

Figure \ref{fig:6_payne_interpolation_errors} graphically presents how accurately \thePayne\ interpolates the synthetic spectra. In the top left panel, we show the distribution of interpolation errors for our cross-validation set, taking the median over all wavelength pixels. We find that for $\sim$85\% of spectra, the MIE is $\lesssim$0.1\%, though the long tail of the distribution indicates that some spectra have errors as high as $\sim$1\%. Unlike \citet{ting:2019} who find larger MIE for cooler stars, we find that this long tail towards higher errors corresponds to stars with higher metallicities ($\text{Fe/H} > -2$; red histogram). Because the stars analyzed in this paper all have $\text{[Fe/H]}\lesssim-2.4$, the adopted MIE is likely on the conservative end.

In the bottom panel, we show the pixel-by-pixel MIE for the entire wavelength range of the model, taking the median over all cross-validation spectra; this is the $\sigma_\text{MIE}$ adopted in Equation \ref{eq:model_unc}. Typical pixel-by-pixel MIEs for the \thePayne\ are 0.01--0.1\% for $\lambda>4500$ and 0.1--0.5\% for $\lambda<4500$. The MIE is generally larger in the blue due to the higher density of absorption lines and the presence of complicated molecular features. The MIE is also larger in the proximity of strong absorption features like the the Balmer lines. We believe this to be the reason why higher-metallicity stars in our cross-validation set have larger MIE than the lower-metallicity stars. The results over all wavelength pixels is summarized in the top right panel, which shows the cumulative number of wavelength pixels as a function of MIE. Roughly 80\% of pixels have $\sigma_\text{MIE} < 0.001$, and 95\% of pixels have $\sigma_\text{MIE} < 0.006$.

\section{Fitting Routines}
\label{sec:fitting}
\subsection{Optimization}
\label{sec:optimization}

Using \texttt{Pytorch}'s automatic differentiation engine and the Adam optimization algorithm, we minimize the negative log-posterior, which is equivalent to maximizing Equation \ref{eq:posterior}. The optimization is performed 10 times with unique initializations. The parameters from the trial with the highest log-posterior value after convergence are taken as the best-fit optimization values. In the rest of this section, we present our choices of initialization, learning rates, and convergence criteria. These choices are summarized in Table \ref{tab:optimization}.

\paragraph{Initialization}
For each initialization, we begin by defining a fiducial model spectrum with the mean stellar labels of the training set (i.e., $\theta_{*,i}^\prime = 0$) and the appropriate resolution for the observations. No Doppler shift, macroturbulent broadening, or continuum correction (other than the blaze function) is applied . Using this fiducial spectrum, the radial velocity is then initialized via a grid search from -300 to 300 km/s in steps of 2 km/s to the value that minimizes the negative log-posterior. The polynomial continuum coefficients are then initialized by performing a polynomial fit with \texttt{np.polyfit} to the ratio of the observed spectrum and the (now Doppler-shifted) fiducial spectrum. All other labels are initialized by randomly sampling from their priors. 

\paragraph{Learning Rates}
We find learning rates of 0.1 to work well for all labels but the radial velocity, which requires a much smaller learning of 0.001 due to the sensitivity of $P(\Theta|D)$ to $v_r$ at high resolution.
To improve convergence, the learning rates are decayed every 10 step by a multiplicative factor of 0.9 for $v_r$ and 0.99 for all other labels.

\paragraph{Convergence}
Convergence of the optimization is achieved when the change in all model parameters is below a given threshold. We define this tolerance to be $10^{-5}$ for the scaled stellar labels, $\theta_{*}^\prime$, and $10^{4}$ for both $\log_{10}v_\text{macro}$ and $v_r$.

We define the convergence criteria for the continuum coefficients slightly differently. Instead of imposing a threshold on the change in $c_{n,o}$, we require that at every pixel (excluding masked pixels) the value of the continuum polynomial changes by less than 5\% of the observed flux uncertainty.

\begin{deluxetable}{lccccc}
	\centerwidetable
	\caption{Treatment of labels in the optimizer}
	\label{tab:optimization}
	\tablehead{
	    \colhead{Label} & \colhead{Initialization} &
	    \colhead{Learning Rate} &  \colhead{Decay} & \colhead{Timescale} &
	    \colhead{Tolerance}
	}
	\startdata
	    $\theta_*^\prime$ & Prior & 1e-1 & 0.99 & 10 & 1e-5 \\
	    $\log_{10}v_\text{macro}$ & Prior & 1e-1 & 0.99 & 10 & 1e-4 \\
	    $v_r$ & Grid Search & 1e-3 & 0.9 & 10 & 1e-4 \\
	    $c_{o,n}$ & np.polyfit & 1e-1 & 0.99 & 10 & 5e-2 \\
	\enddata
	\tablecomments{
	    Initialization procedure and optimization hyper-parameters for the stellar labels and nuisance parameters of our model.
	}
\end{deluxetable}

\subsection{MCMC Sampling}
\label{sec:mcmc}

Using the affine invariant MCMC methods introduced by \citet{goodman:2010} and implemented in \texttt{emcee}\footnote{\url{https://emcee.readthedocs.io/}}, we sample our log-posterior distribution (Equation \ref{eq:posterior}). As in the optimization routine, the log-posterior is evaluated using scaled stellar labels, $\theta_{*}^\prime$, which are converted to physical units when results are reported. In the rest of this section, we present the specifics of our sampling routine.

\paragraph{Initialization}
Before sampling the posterior in earnest, we begin by initializing 128 walkers at the maximum \textit{a posteriori} value of $\Theta$ found via our optimization algorithm. Gaussian scatter of 0.1 is applied to all labels except the continuum coefficients, which are held constant throughout MCMC sampling. During this burn-in phase, the walkers sample the posterior until 1) the mean value for each label of the walkers changes less than 0.5\% over the previous 100 steps and 2) the mean $\log P$ of the walkers has changed by less than 0.00001\%. After the burn-in phase is complete, 512 walkers are initialized around the location of the walker with the highest $\log P$ with a Gaussian scatter in each label equal to half the standard deviation of the burn-in walkers for that label. Now that the initialization is complete, and the walkers have had a chance to settle around the maximum \textit{a posteriori}, we begin the production run of our posterior sampling. 

\paragraph{Convergence}
We sample the posterior distribution until the following two convergence criteria have been met: 1) the auto-correlation time, $\tau$, has changed by $<$1\% over the previous 100 steps and 2) the sampler has run for $>$30$\tau$ steps. If these criteria have not been met after 15000 steps, walkers are re-initialized around the location of the walker with the highest $\log P$, and the sampling is restarted. Once convergence has been reached, we discard the first $5\tau$ samples from each walker and  thin each chain's samples by $\sim\tau/2$ to remove any residual effects of burn-in or correlated samples. Unthinned chains can be made available upon request.

\paragraph{Move Proposal}
The default move proposal of \texttt{emcee} is the ``stretch move" method of \citet{goodman:2010}, which is not well suited for the dimensional of our problem. Instead, we adopt a weighted mixture of 80\% differential evolution proposals (\texttt{emcee.moves.DEMove}; \citealt{terbraak:2006, nelson:2014}) and 20\% differential evolution snooker proposals (\texttt{emcee.moves.DESnookerMove}; \citealt{terbraak:2008}). We find that this combination of move proposals improves the convergence time by more than an order of magnitude over the default move proposal.

\section{Comparison with Literature Values}
\label{sec:literature}
As a check on our fitting procedure, we compare our default high-resolution high-S/N stellar label measurements to those measured from previous stellar spectral analyses of the same stars. Our high-resolution measurements and those included in the literature comparison are presented in Table \ref{tab:literature}. Their chemical abundances have been adjusted to place them on the \citet{asplund:2009} Solar abundance scale adopted in our analysis. This collection of literature measurements is a non-exhaustive, but representative sample of previous spectroscopic studies in M15 across a wide range of spectroscopic configurations, fitting techniques, and measured elements. 
\begin{deluxetable}{lRRRRRc}
	\caption{Literature Stellar Parameters}
	\label{tab:literature}
	\tablehead{Reference &\teff~\text{[K]} &\logg & \text{[Mg/Fe]} &\text{[Ca/Fe]} &\text{[Fe/H]} & ... }
	\startdata 
	\multicolumn{7}{c}{K341}\\
	\tableline
	This Study&4415	&0.78	&0.37	& 0.04	& -2.47	&... \\
	S+97	&4275	&0.45	&0.72	& 0.32	& -2.35	&... \\
	S+00b	&4275	&0.45	&...		& 0.56	& -2.45	&... \\
	S+06	&4275	&0.45	&...		& ...		& -2.46	&... \\
	R+09  	&...		&... 		&... 		& ...		& -2.32	&... \\
    	C+09b 	&4324	&0.69 	&0.49 	& ...		& -2.23	&... \\
    	S+11   	&4343 	&0.88 	&0.60 	& 0.22	& -2.53	&... \\
    	W+13  	&4324 	&0.69 	&... 		& 0.16	& -2.32	&... \\
    	K+18   	&4253 	&0.67 	&0.66 	& 0.28	& -2.49	&... \\
    	M+19   	&4545 	&0.80 	&0.27	& 0.17	& -2.08	&... \\
    	J+20    	&4377 	&0.64 	&0.39 	& 0.28	& -2.30	&... \\
	\tableline  \multicolumn{7}{c}{K386}\\ \tableline
	This Study&4390	&0.54	&0.24	&0.01	& -2.49	&... \\
	S+97    	&4200	&0.15 	&...		&0.19	& -2.43	&... \\
    	S+00b 	&4200	&0.15 	&...		&0.19	& -2.35	&... \\
    	O+06   	&4200 	&0.35 	&...		&...		& -2.40	&... \\
    	S+06   	&4200 	&0.15 	&...		&...		& -2.51	&... \\
    	C+09a   	&4313 	&0.65 	&...		&...		& -2.33	&... \\
    	W+13    	&4313 	&0.65 	&...		&0.10	& -2.33	&... \\
    	K+18    	&4263 	&0.65 	&0.15	&0.19	& -2.50	&... \\
    	M+19    	&4548 	&0.81 	&0.28	&0.51	& -2.14	&... \\
    	J+20    	&4449 	&0.90 	&...		&...		& -2.08	&... \\
\tableline  \multicolumn{7}{c}{K431}\\ \tableline
	This Study&4489	&0.78	&0.16	& 0.15	& -2.45	&... \\
	S+97    	&4375	&0.50 	&0.38	& 0.28	& -2.43	&... \\
    	L+06   	&4350 	&0.50 	&0.33	& 0.32	& -2.36	&... \\
    	S+06   	&4375 	&0.50 	&...		& ...		& -2.50	&... \\
    	W+13    	&4377 	&0.77 	&...		& ...		& -2.34	&... \\
    	K+18    	&4351 	&0.84 	&...		& 0.12	& -2.49	&... \\
    	M+19    	&4670 	&1.09 	&0.22	& 0.15	& -2.20	&... \\
    	J+20    	&4543 	&1.02 	&...		& ...		& -2.09	&... \\
\tableline  \multicolumn{7}{c}{...}\\ \tableline
	\enddata
	\tablecomments{
		Non-exhaustive compilation of literature measurements of stellar parameters for stars in our sample. All chemical abundances have been scaled to the \citet{asplund:2009} Solar abundance scale for consistency in comparison. Reference abbreviations are as follows: S+97 = \citet{sneden:1997}, S+00b = \citet{sneden:2000b}, L+06 = \citet{letarte:2006}, O+06 = \citet{otsuki:2006}, S+06 = \citet{sobeck:2006}, R+09 = \citet{roederer:2009}, C+09a = \citet{carretta:2009a}, C+09b = \citet{carretta:2009b}, S+11 = \citet{sobeck:2011}, W+13 = \citet{worley:2013}, K+18 = \citet{kirby:2018}, M+19 = \citet{masseron:2019}, J+20 = \citet{jonsson:2020}.\\
		(This table is available in its entirety in machine-readable form online.)
	}
\end{deluxetable}

\begin{figure*}[ht!]
    \epsscale{1.15}
    \plotone{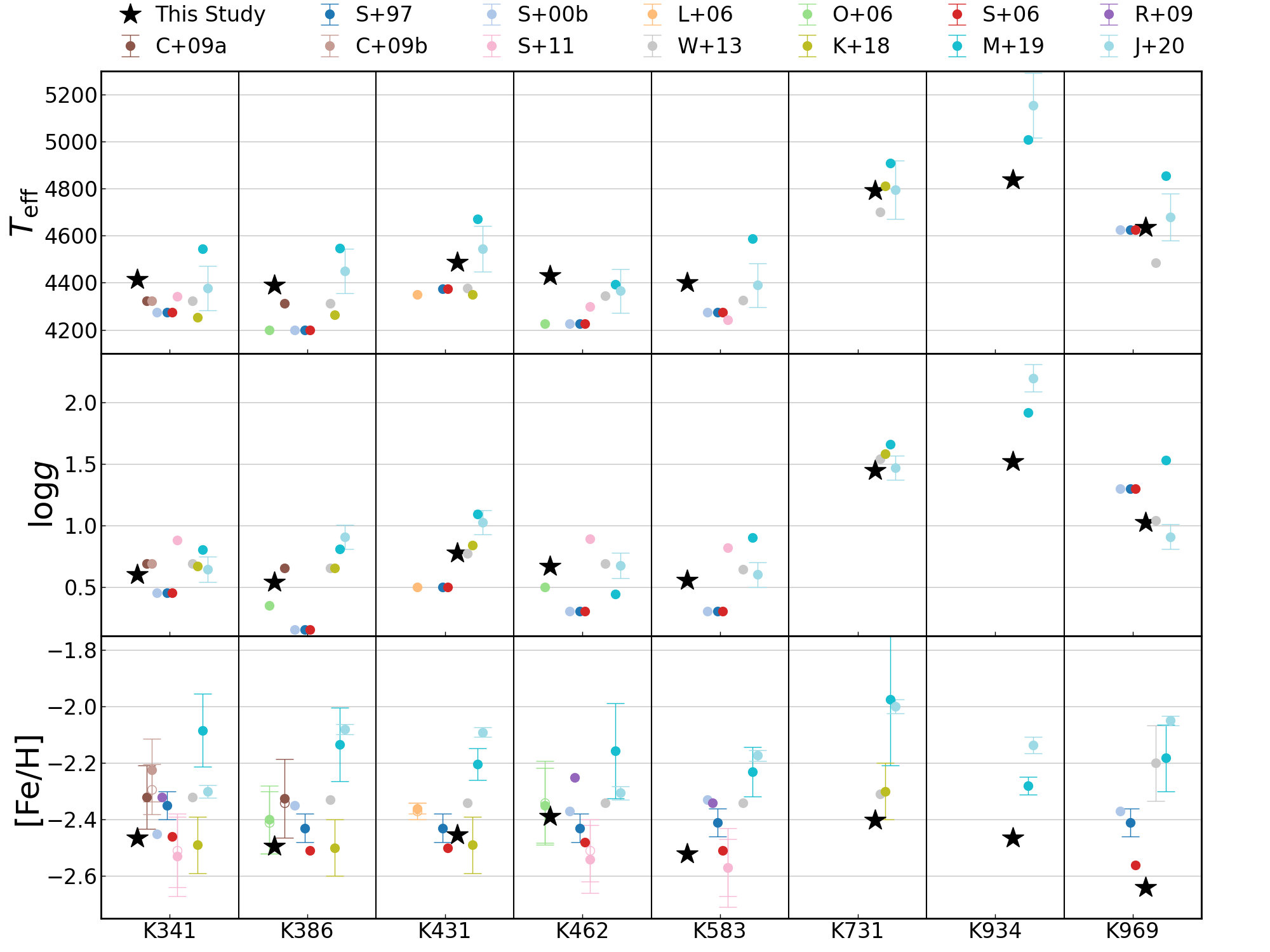}
    \caption{
    	The \tefftext\ (top), \loggtext\ (middle), and [Fe/H] (bottom) measured for each of the 8 stars in our sample using the full-spectrum fitting techniques presented in this paper (black stars). For comparison, we also plot the values for \tefftext, \loggtext, and [Fe/H] reported in a representative sample of literature studies of the same stars (colored circles). In instances where studies report separate values for neutral and ionized atomic species, the ionized value is represented by an open circle. Error bars represent $1\sigma$ uncertainties when provided and are too small to be visible for our own measurement uncertainties. Scatter is added in the x-dimension for clarity; points are ordered from left to right in order of increasing mean observed wavelength. A key to the abbreviated references is provided in Table \ref{tab:literature}.
        \label{fig:literature_comparison0}
    }
\end{figure*}

In Figure \ref{fig:literature_comparison0} we present our measurements of \tefftext, \loggtext, and [Fe/H] (black stars) alongside literature measurements (colored circles) for the 8 stars in our sample. Error bars represent $1\sigma$ uncertainties where available and are too small to be visible for the statistical uncertainty of our own measurements. 

Overall, our measurements fall nicely amidst the locus of literature measurements. On average, our measurements differ from the median literature value for \tefftext, \loggtext, and [Fe/H] by roughly 100 K, $-0.1$ dex, and $-0.15$ dex respectively. The most outlying atmospheric parameters are recovered for star K934, for which we recover values of \tefftext, \loggtext, and [Fe/H] to be $\sim$300 K, 0.5 dex, and 0.3 dex lower respectively than measured from APOGEE spectra by \citet{masseron:2019} and \citet{jonsson:2020}. We believe that the proximity of K934 to K731 in the CMD (Figure \ref{fig:m15_cmd}) justify our measurements, which would place the two stars similarly close in \tefftext-\loggtext\ space. We also note that two APOGEE studies also consistently recover [Fe/H] higher than most other studies, perhaps owing to the difference in wavelength coverage (NIR vs.\ optical).


\begin{figure*}[ht!]
    \epsscale{0.95}
    \plotone{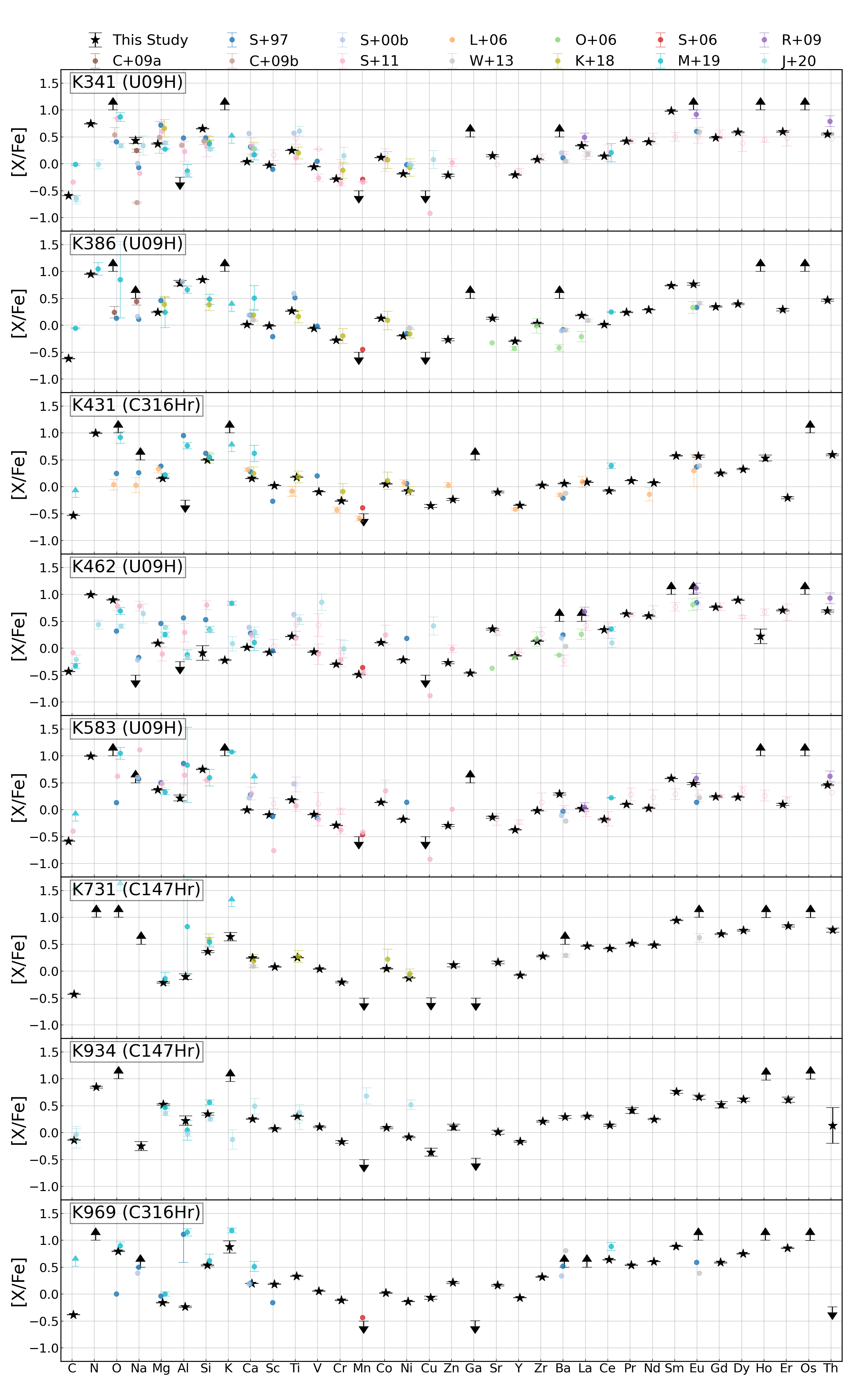}
    \caption{
        The detailed chemical composition of the 8 M15 stars as measured in this study (black stars) and as reported in the literature (colored circles; the same as in Figure \ref{fig:literature_comparison0}). 95\% upper and lower limits are plotted where appropriate (see \S\ref{sec:methods_stack}) and when reported by the literature. As in Figure \ref{fig:literature_comparison0}, when separate abundances are provided for neutral and ionized atomic species, the ionized values are represented with open circles. Scatter is added in the x-dimension for clarity; points are ordered from left to right in order of increasing mean observed wavelength. 
        \label{fig:literature_comparison1}
    }
\end{figure*}

In Figure \ref{fig:literature_comparison1}, we present a comparison of our measurements with measurements from the same literature studies for the remaining 35 chemical abundances. The same symbol schema is adopted as in Figure \ref{fig:literature_comparison0}. When separate abundances are provided for neutral and ionized atomic species (e.g., [Ti I/Fe I] vs.\ [Ti II/Fe II]), the abundances for the ionized species are represented with open circles. 

In brief, we find generally good agreement with the literature for measurements of C, N, O, Mg, Ca, Ti, V, Cr, Mn, Co, Ni, Cu, Y, Zr, La, Ce, Pr, Nd, Gd, Dy, Er, and Th with a few caveats. For 5 elements, our measurements are systematically offset $\sim$0.25 dex higher (Sc, Ba, Sm, and Eu) or lower (Zn) compared to literature measurements. Mixed or poor agreement with values was found for (Na, Al, Si, K, and Ho). No literature measurements were found to compare to for Ga and Os. 
In general, differences between our abundance measurements and those from the literature can be attributed to differences in the stellar models, line lists, oscillator strengths, and/or wavelength coverage employed.
Across all elements, our measurements agree best with literature measurements made over similar wavelength ranges (e.g., the optical; \citealt{sneden:1997, sobeck:2011}) and less well with those made over non-overlapping spectral coverage (e.g., the NIR; \citealt{meszaros:2015, masseron:2019, jonsson:2020}). 
As in \S\ref{sec:results}, we present a detailed comparison with the literature for each element loosely grouped by nucleosynthetic origin.

\subsection{Literature Comparison by Element Group}
\label{sec:lit_by_element}

\paragraph{Iron-Peak Elements}
In addition to Fe, we find generally good agreement with the literature for iron-peak elements V, Cr, Co, and Ni. While we recover only upper limits on Mn, these upper limits are generally consistent with literature values.
The one exception to our general agreement with the literature is for K934, for which \citet{jonsson:2020} reports $\text{[Mn/Fe]} = 0.68\pm0.15$. Because this is the only reported measurement of [Mn/Fe] in K934, and \citet{jonsson:2020} do not report [Mn/Fe] for any of the other stars in this sample, it is difficult to identify the source of this discrepancy. Given the ubiquity of low [Mn/Fe] abundances in metal-poor stars \citep[see][]{sobeck:2006}, we are inclined to trust our measurement in this instance. We note that large NLTE offsets ($\sim$0.2--0.4 dex) have been calculated for [Mn/Fe] in low-metallicity stars \citep[e.g.,][]{bergemann:2008, larsen:2022}, which have not been accounted for in either this study or any of the referenced studies.

Only three stars in our sample have literature [Cu/Fe] measurements and all come from one of two studies, \citet{sobeck:2011} and \citet{jonsson:2020}. The values reported by these two studies are discrepant by as much as 1 dex, an effect we attribute to the difference in wavelength coverage of the two surveys: \citet{sobeck:2011} used optical Keck/HIRES spectra (the same archival spectra, in fact, as analyzed in this paper), and \citet{jonsson:2020} used NIR APOGEE spectra. \citet{jonsson:2020} also urges caution in adopting the [Cu/Fe] measurements for metal-poor stars with $\text{[Fe/H]} < -1$ as they find systematically higher [Cu/Fe] at lower [Fe/H] in contrast to previous studies and the expectations of Cu's nucleosynthetic origin \citep[e.g.,][]{sneden:1991a, cayrel:2004, ishigaki:2013}. It is possible, however, that NLTE effects are responsible for the low [Cu/Fe] values found from optical spectroscopy \citep[e.g.,][]{roederer:2018}.
For these three stars, we recover upper limits on [Cu/Fe] that are consistent with the lower abundances of \citet{sobeck:2011}. This is in line with expectations given the same underlying observations and LTE assumptions.

We routinely recover [Zn/Fe] to be $\sim$0.25 dex smaller than reported in the literature by \citet{letarte:2006} and \citet{sobeck:2011}. We believe this offset is driven by the Zn I line at \wave{4681.4}, which is blended with a poorly modelled Fe I line at \wave{4681.6} and excluded from the analysis of \citet{letarte:2006} and \citet{sobeck:2011}. The other two Zn lines at \wave\wave{4723.5, 4811.9} are slightly underestimated in our fits, consistent with the $\sim$0.25 dex lower measurement of [Zn/Fe].

Owing to the paucity of quality Ga lines in the archival spectra---there are only two ($\lambda\lambda$40341a and 4173.2) and both are weak, heavily blended, and within NLTE masks---we are unable to precisely constrain [Ga/Fe] within our model bounds for all but two stars, K462 and K479. Because no literature values of [Ga/Fe] exist for these stars, we cannot confirm the fidelity of these measurements and urge caution in their adoption.

\paragraph{$\alpha$ Elements}
In general, we find good agreement with the literature for $\alpha$ elements Mg, Ca, and Ti, though [Ca/Fe] measurements are on the low end of reported values for a few stars. This is most likely a result of differences in the handling of NLTE effects. 

We find good agreement with the literature for Si for the C147Hr and C316Hr programs, but the blue-only U09H program observations yield somewhat more mixed agreement, in some cases off by 0.25--0.5 dex. This can be traced to Si's role as an electron donor in stellar atmospheres. When the full optical spectrum is available, Si is primarily constrained through its isolated absorption lines near \wave{5700} and \wave{7400}. However, when only the blue-optical spectrum is available, Si is primarily constrained through its indirect influence on other absorption features through changes to the atmospheric structure. While indirect measurement of elements is feasible \citep[e.g.,][]{ting:2018b}, it relies heavily on the accuracy of the stellar atmospheric models. We know the 1D-LTE models employed in this work to be imperfect, thus explaining the inconsistencies with the literature seen for the blue-only [Si/Fe] abundances.

\paragraph{C, N, O}
There exists large (0.5--1.0 dex) scatter in the literature values measured for C, N, and O abundances, owing to the complicated nature of their molecular features and the varying wavelength coverage and methods of these studies (see for example the analysis of [C/Fe] measurements across surveys by \citealt{arentsen:2022}). We find good agreement with the literature for [C/Fe] with the exception of few measurements from \citet{masseron:2019}, which are substantially higher than our values.

Only three of our stars have previously measured [N/Fe]. For K386, our measurement agrees with the measurement of \citet{masseron:2019}, but for K341 and K462, we measure [N/Fe] to be $>$0.5 dex larger than reported by \citet{jonsson:2020}.

For nearly all stars in our sample, literature values span a large range from $\text{[O/Fe]}\sim0.0$--1.0. Our measurements, either lower limits of $\text{[O/Fe]}\gtrsim1.0$ or in the range of $\text{[O/Fe]}\sim0.75$--1.0, are most consistent with the high end of the reported literature values.

\paragraph{Light-Odd Elements}
The light-odd elements Na, Al, and K, similar to C, N, and O, exhibit large (0.5--1.0 dex) scatter in the reported literature values, which is due to the combination of the limited absorption features available for these elements as well as their sensitivity to NLTE effects \citep[e.g.,][]{asplund:2005, asplund:2009}. We recover [Na/Fe] values that either fall among the literature values or lie slightly above the literature measurements except for stars K462, for which we measure [Na/Fe] $\sim$ 0.25 lower than previously reported.
For roughly half of the stars in our sample we measure [Al/Fe] in agreement with literature values, while we recover substantially lower [Al/Fe] for the others.
In general, measurements of [K/Fe] (or lower limits) are in coarse agreement with measurements (or lower limits) from \citet{masseron:2019}. As with Na, the exceptions to this is K462, for which we recover much lower [K/Fe].
%

Literature measurements of [Sc/Fe] come from \citet{sneden:1997} and \citet{sobeck:2011}, which largely analyzed the same archival spectra as in this paper. Our measurements of [Sc/Fe] are in good agreement with those from \citet{sobeck:2011} except for the sole Sc I measurement in K583, which is itself highly discrepant from the Sc II measurement from the same study. Agreement is good with \citet{sneden:1997} for the 4 stars observed as part of the U09H program, while for the remaining stars our values are higher by 0.25--0.5 dex. Not coincidentally, the 4 stars for which agreement is best are the 4 stars for which the same spectra are analyzed in both studies.

\paragraph{Neutron-Capture Elements}
We find good agreement for neutron-capture elements Y, Zr, La, Pr, Nd, Gd, Dy, Er, and Th. We recover values for [Sr/Fe] that are in good agreement with the values reported by \citet{sobeck:2011}, but are $\sim$0.5 dex higher than reported by \citet{otsuki:2006}. This discrepancy was previously identified by \citet{sobeck:2011} and attributed to uncertainties in measuring abundances from the Sr II resonance lines.

Similarly, we find good agreement in [Ce/Fe] with \citet{sobeck:2011}, but mixed agreement with \citet{masseron:2019}. For stars K341 and K462, we recover [Ce/Fe] values that match \citet{masseron:2019} quite well, but for stars K386, K431, K583, and K969, we recover [Ce/Fe] values that are $\sim$0.25--0.50 dex smaller. This is of order the systematic error that \citet{masseron:2019} reports between [Ce/Fe] values measured using differently derived atmospheric parameters.

Our recovered abundances for Ba, Sm, and Eu are consistently $\sim$0.25--0.50 dex larger than the values reported in the literature. We link these offsets to a combination of factors, including line saturation, NLTE effects, and hyperfine splitting that result in inaccurately modeled line profile shapes (see \citealt{roederer:2008, eitner:2019}).

In the case of Ho, our measurements disagree by $>$0.5 dex from the only available measurements of \citet{sobeck:2011}. Four of the five Ho II lines in our spectrum (\wave\wave{3797.8, 3811.8, 3892.1, 4046.6}) are heavily blended in poorly modelled portions of the the spectrum while the fifth (\wave{4153.8}) is quite weak. As a result, we believe our [Ho/Fe] measurements to be in err.

Owing to the dearth of Os lines in the archival spectra, we are unable to precisely constrain [Os/Fe] within our model bounds for any star in our sample. Because no literature values of [Os/Fe] exist for this star, we cannot confirm the fidelity of these measurements and urge caution in their adoption.


\bibliography{GC_validation}{}
\bibliographystyle{aasjournal}



\end{document}